\documentclass[aps,prd,preprint,showpacs,amsmath,amssymb,superscriptaddress]{revtex4}

\usepackage{graphicx}% Include figure files
\usepackage{dcolumn}% Align table columns on decimal point
\usepackage{bm}% bold math

\begin{document}
\title{Measurement of the Top Quark Mass with the Dynamical Likelihood 
Method using Lepton plus Jets Events with $b$-tags in $p\bar p$ Collisions 
at $\sqrt{s}$ = 1.96 TeV}
\date{\today}

\begin{abstract}
This report describes a measurement of the top quark mass, $M_{top}$, with the
dynamical likelihood method (DLM) using the CDF II detector at the
Fermilab Tevatron.  The Tevatron produces top/anti-top ($t\bar t$)
pairs in $p\bar p$ collisions at a center-of-mass energy of 1.96
TeV. The data sample used in this analysis was accumulated from March
2002 through August 2004, which corresponds to an integrated
luminosity of 318 pb$^{-1}$. We use the $t \bar t$ candidates in the
``lepton+jets'' decay channel, requiring at least one jet identified
as a $b$ quark by finding a displaced secondary vertex.  The DLM
defines a likelihood for each event based on the differential cross
section as a function of $M_{top}$ per unit phase space volume of the
final partons, multiplied by the transfer functions from jet to parton
energies. The method takes into account all possible jet combinations
in an event, and the likelihood is multiplied event by event to derive
the top quark mass by the maximum likelihood method. Using 63 $t \bar
t$ candidates observed in the data, with 9.2 events expected from
background, we measure the top quark mass to be $173.2$
$^{+2.6}_{-2.4}$ (stat.) $\pm$ 3.2 (syst.) GeV/$c^2$, or 173.2
$^{+4.1}_{-4.0}$ GeV/$c^2$.
\end{abstract}

\affiliation{Institute of Physics, Academia Sinica, Taipei, Taiwan 11529, Republic of China} 
\affiliation{Argonne National Laboratory, Argonne, Illinois 60439} 
\affiliation{Institut de Fisica d'Altes Energies, Universitat Autonoma de Barcelona, E-08193, Bellaterra (Barcelona), Spain} 
\affiliation{Baylor University, Waco, Texas  76798} 
\affiliation{Istituto Nazionale di Fisica Nucleare, University of Bologna, I-40127 Bologna, Italy} 
\affiliation{Brandeis University, Waltham, Massachusetts 02254} 
\affiliation{University of California, Davis, Davis, California  95616} 
\affiliation{University of California, Los Angeles, Los Angeles, California  90024} 
\affiliation{University of California, San Diego, La Jolla, California  92093} 
\affiliation{University of California, Santa Barbara, Santa Barbara, California 93106} 
\affiliation{Instituto de Fisica de Cantabria, CSIC-University of Cantabria, 39005 Santander, Spain} 
\affiliation{Carnegie Mellon University, Pittsburgh, PA  15213} 
\affiliation{Enrico Fermi Institute, University of Chicago, Chicago, Illinois 60637} 
\affiliation{Joint Institute for Nuclear Research, RU-141980 Dubna, Russia} 
\affiliation{Duke University, Durham, North Carolina  27708} 
\affiliation{Fermi National Accelerator Laboratory, Batavia, Illinois 60510} 
\affiliation{University of Florida, Gainesville, Florida  32611} 
\affiliation{Laboratori Nazionali di Frascati, Istituto Nazionale di Fisica Nucleare, I-00044 Frascati, Italy} 
\affiliation{University of Geneva, CH-1211 Geneva 4, Switzerland} 
\affiliation{Glasgow University, Glasgow G12 8QQ, United Kingdom} 
\affiliation{Harvard University, Cambridge, Massachusetts 02138} 
\affiliation{Division of High Energy Physics, Department of Physics, University of Helsinki and Helsinki Institute of Physics, FIN-00014, Helsinki, Finland} 
\affiliation{University of Illinois, Urbana, Illinois 61801} 
\affiliation{The Johns Hopkins University, Baltimore, Maryland 21218} 
\affiliation{Institut f\"{u}r Experimentelle Kernphysik, Universit\"{a}t Karlsruhe, 76128 Karlsruhe, Germany} 
\affiliation{High Energy Accelerator Research Organization (KEK), Tsukuba, Ibaraki 305, Japan} 
\affiliation{Center for High Energy Physics: Kyungpook National University, Taegu 702-701; Seoul National University, Seoul 151-742; and SungKyunKwan University, Suwon 440-746; Korea} 
\affiliation{Ernest Orlando Lawrence Berkeley National Laboratory, Berkeley, California 94720} 
\affiliation{University of Liverpool, Liverpool L69 7ZE, United Kingdom} 
\affiliation{University College London, London WC1E 6BT, United Kingdom} 
\affiliation{Massachusetts Institute of Technology, Cambridge, Massachusetts  02139} 
\affiliation{Institute of Particle Physics: McGill University, Montr\'{e}al, Canada H3A~2T8; and University of Toronto, Toronto, Canada M5S~1A7} 
\affiliation{University of Michigan, Ann Arbor, Michigan 48109} 
\affiliation{Michigan State University, East Lansing, Michigan  48824} 
\affiliation{Institution for Theoretical and Experimental Physics, ITEP, Moscow 117259, Russia} 
\affiliation{University of New Mexico, Albuquerque, New Mexico 87131} 
\affiliation{Northwestern University, Evanston, Illinois  60208} 
\affiliation{The Ohio State University, Columbus, Ohio  43210} 
\affiliation{Okayama University, Okayama 700-8530, Japan} 
\affiliation{Osaka City University, Osaka 588, Japan} 
\affiliation{University of Oxford, Oxford OX1 3RH, United Kingdom} 
\affiliation{University of Padova, Istituto Nazionale di Fisica Nucleare, Sezione di Padova-Trento, I-35131 Padova, Italy} 
\affiliation{LPNHE-Universite de Paris 6/IN2P3-CNRS} 
\affiliation{University of Pennsylvania, Philadelphia, Pennsylvania 19104} 
\affiliation{Istituto Nazionale di Fisica Nucleare Pisa, Universities of Pisa, Siena and Scuola Normale Superiore, I-56127 Pisa, Italy} 
\affiliation{University of Pittsburgh, Pittsburgh, Pennsylvania 15260} 
\affiliation{Purdue University, West Lafayette, Indiana 47907} 
\affiliation{University of Rochester, Rochester, New York 14627} 
\affiliation{The Rockefeller University, New York, New York 10021} 
\affiliation{Istituto Nazionale di Fisica Nucleare, Sezione di Roma 1, University of Rome ``La Sapienza," I-00185 Roma, Italy} 
\affiliation{Rutgers University, Piscataway, New Jersey 08855} 
\affiliation{Texas A\&M University, College Station, Texas 77843} 
\affiliation{Istituto Nazionale di Fisica Nucleare, University of Trieste/\ Udine, Italy} 
\affiliation{University of Tsukuba, Tsukuba, Ibaraki 305, Japan} 
\affiliation{Tufts University, Medford, Massachusetts 02155} 
\affiliation{Waseda University, Tokyo 169, Japan} 
\affiliation{Wayne State University, Detroit, Michigan  48201} 
\affiliation{University of Wisconsin, Madison, Wisconsin 53706} 
\affiliation{Yale University, New Haven, Connecticut 06520} 
\author{A.~Abulencia}
\affiliation{University of Illinois, Urbana, Illinois 61801}
\author{D.~Acosta}
\affiliation{University of Florida, Gainesville, Florida  32611}
\author{J.~Adelman}
\affiliation{Enrico Fermi Institute, University of Chicago, Chicago, Illinois 60637}
\author{T.~Affolder}
\affiliation{University of California, Santa Barbara, Santa Barbara, California 93106}
\author{T.~Akimoto}
\affiliation{University of Tsukuba, Tsukuba, Ibaraki 305, Japan}
\author{M.G.~Albrow}
\affiliation{Fermi National Accelerator Laboratory, Batavia, Illinois 60510}
\author{D.~Ambrose}
\affiliation{Fermi National Accelerator Laboratory, Batavia, Illinois 60510}
\author{S.~Amerio}
\affiliation{University of Padova, Istituto Nazionale di Fisica Nucleare, Sezione di Padova-Trento, I-35131 Padova, Italy}
\author{D.~Amidei}
\affiliation{University of Michigan, Ann Arbor, Michigan 48109}
\author{A.~Anastassov}
\affiliation{Rutgers University, Piscataway, New Jersey 08855}
% ...
\collaboration{CDF Collaboration}
\noaffiliation
                       
\maketitle
%===========================================================================
\section{\label{sec:chap1} Introduction}
The top quark mass is an important quantity in particle
physics.  Its precise value not only serves for setting basic
parameters in calculations of electroweak processes, but also provides
a constraint on the mass of the Higgs boson. Therefore it is desirable
to have a measurement with a precision comparable to that of other
relevant electroweak parameters, typically of the order of $0.1\% -
1\%$, the latter corresponding to about 2 GeV$/c^2$ in $M_{top}$.
Based on Run I Tevatron data at a center-of-mass energy of 1.8 TeV
(1992--1996), the CDF and D$\O$ collaborations published several
direct experimental measurements of $M_{top}$ with all decay
topologies arising from $t\bar t$ production: the dilepton
channel~\cite{Ref:RunIdilCDF,Ref:RunIdilD0}, the lepton+jets
channel~\cite{Ref:RunIljCDF,Ref:RunIljD0}, and the all-jets
channel~\cite{Ref:RunIallj,Ref:RunIalljD0}. Including a recent
reanalysis of D$\O$ Run I data~\cite{Ref:D0runIana}, the Run I world
average top quark mass is $178.0\pm4.3$ GeV/$c^2$~\cite{Ref:RunIcomb}.
A global standard model fit using this updated value gives the most
likely value of the Higgs boson mass of 129 $^{+74}_{-49}$ GeV/$c^2$,
and the 95$\%$ C.L. upper limit of 285 GeV/$c^2$
\cite{Ref:HiggsLEP}.

In this paper we present a measurement of the top quark mass in $p
\bar p$ collisions at $\sqrt{s}$ = 1.96 TeV at the Fermilab
Tevatron. The data were obtained with the upgraded Collider Detector
at Fermilab (CDF II) operated during Run II. The integrated luminosity
of the data sample, collected from March 2002 through August 2004, is
318 pb$^{-1}$. This is the total dataset for which all detectors
including the silicon tracker were operating.
The method employed is the Dynamical Likelihood Method (DLM)
\cite{Ref:dlmKK1}--\cite{Ref:dlmKK4}, which uses the differential cross 
section for the $t\bar t$ process as a function of $M_{top}$ in the
likelihood definition. The permutations in assigning jets to the
primary quarks from the $t$ and $\bar{t}$ decays and the quadratic
ambiguity in the $z$-component of the neutrino momentum are
incorporated in the likelihood for an event, and the likelihood is
multiplied event by event to extract $M_{top}$ by the maximum
likelihood method. Similar but not identical techniques were
introduced and employed during Run
I~\cite{Ref:D0runIana,Ref:RunIWhelD0,Ref:DGM1}.  Using Run II data,
CDF recently produced the best single top quark mass measurement using
the template method with $in$ $situ$ jet energy calibration
~\cite{Ref:TMassTempII}. That analysis and this one are summarized
together in~\cite{Ref:TMassPRL}.

In this method, we assume that the standard model (SM) accurately describes
$t\bar t$ production and decay. This assumption is justified for two
reasons: (1) Since the discovery of the top quark was established in
Run I~\cite{Ref:topdiscov}, its properties have been investigated
using both Run I and Run II data, but no significant discrepancies
between experimental data and the SM have been found
\cite{Ref:RunIIXsecKin}--\cite{Ref:RunIIBR}. (2) The SM
neither predicts the top quark mass directly, nor explains why it is
approximately 40 times more massive than the $b$ quark, the isodoublet
partner of the top quark. Therefore it is reasonable to use the SM in
the likelihood definition (through the differential cross section) for
the mass measurement.

\begin{figure}[htbp]
\includegraphics[width=1\columnwidth]{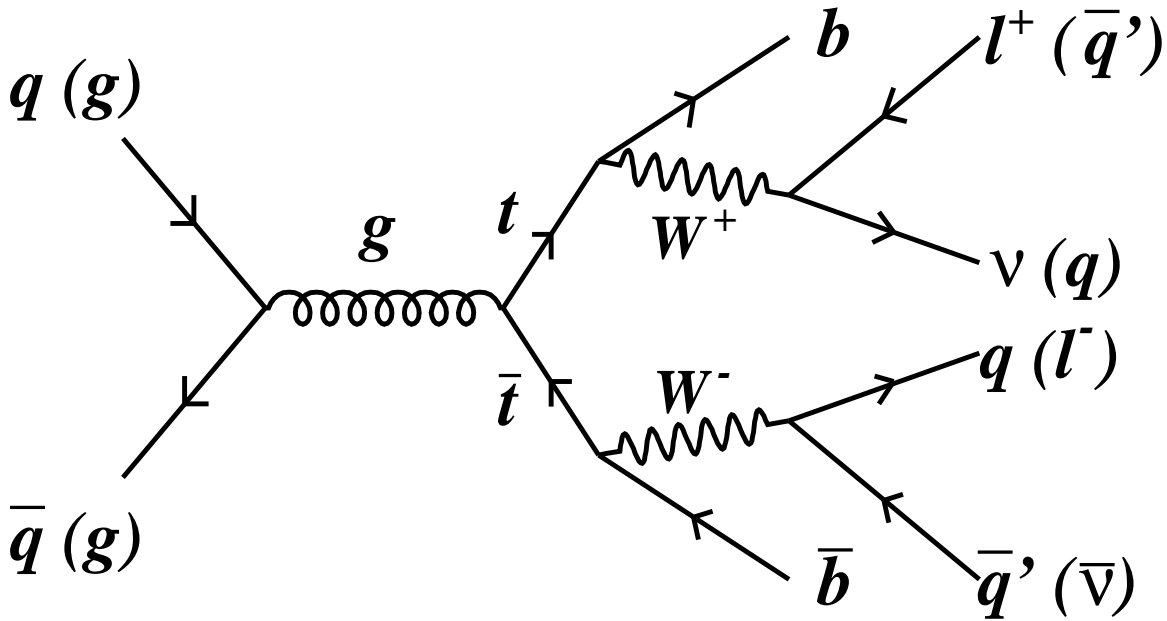}
\caption{Tree-level Feynman diagram of standard model $t\bar t$ production and
decay in the lepton+jets mode.}
\label{Fig:Feynman}
\end{figure}

According to the SM, the top quark decays
approximately 100$\%$ of the time into a $W$ boson and a $b$
quark. The $W$ then decays to a quark-antiquark or lepton-neutrino
pair. The measurement presented here uses events with $t\bar t$
decaying in the ``lepton+jets'' channel, $t\bar t\to W^{+}W^{-}b\bar b
\to l\nu q \bar q' b\bar b$, as shown in Fig.~\ref{Fig:Feynman},
which provided the most accurate mass measurement in Run I because of
higher statistics than the dilepton channel and lower background than
the all-jets channel. This channel is characterized by a single high
$p_{T}$ lepton (electron or muon) and missing transverse energy from a
$W$ $\to$ $l\nu$ decay, plus four jets, two from the hadronically
decaying $W$ boson and two $b$ quarks from the top decays. The $b$
quarks may be identified (``$b$-tagged'') by reconstructing secondary
vertices from the decay of $B$ hadrons with the silicon vertex
detector (SECVTX tagging), as described in Section~\ref{sec:btagalg}.

This paper is organized as follows. In Section~\ref{sec:chap2}, we
present a brief description of the most important detector subsystems
to this analysis. Section~\ref{sec:chap3} describes the data samples
that are used in the top quark mass
measurement. Section~\ref{sec:chap4} presents particle identification
and event selection.  The background estimates are described in
Section \ref{sec:chap5}. After a brief overview of the DLM procedure
in Section~\ref{sec:AO}, the definition of the dynamical likelihood
function and the reconstruction procedure are discussed in Section
\ref{sec:chap6}.  The transfer functions between jet and parton
kinematics, which play a key role in the method, are presented in
Section~\ref{sec:chap7}. Section~\ref{sec:chap8} describes top quark
mass determination studies using Monte Carlo for both the $t\bar t$
signal and background events. The effect of background on the
likelihood distribution is also investigated. Section~\ref{sec:chap9}
presents the final top quark mass result after correcting for the
mass-pulling effect of the background.  Section \ref{sec:chap10}
discusses further checks on this analysis. The systematic
uncertainties are presented in Section~\ref{sec:chap11}.  Conclusions
are summarized in Section~\ref{sec:chap12}.

%===========================================================================
\section{\label{sec:chap2}The CDF Detector Overview}
The CDF II detector, a general purpose detector with azimuthal and
forward-backward symmetry, is composed of independent subsystems
designed for distinct tasks relating to the study of $p\bar p$
interactions.  

The CDF coordinate system consists of the $z$-axis along the proton 
beam direction, the azimuthal angle $\phi$ defined in the plane 
transverse to the $z$-axis, and the polar angle $\theta$ from the proton 
direction (usually expressed as the pseudorapidity 
$\eta=-\ln(\tan(\theta /2))$). The $x$- and $y$-axes point outward and
upward from the Tevatron ring, respectively. Transverse energy ($E_T$) and 
momentum ($p_T$) are defined in this plane, perpendicular to the 
$z$-axis.

The three most relevant
subsystems to $t \bar t \to$ lepton+jets event detection are the
tracking chambers, the calorimeters and the muon chambers.  These
subsystems are briefly described below. A complete description of the
CDF II detector can be found elsewhere~\cite{Ref:CDF}.

%Tracking Systems
The tracking system consists of a large open-cell drift chamber and
silicon microstrip detectors.  These lie inside a superconducting
solenoid of length 5 m and diameter 3.2 m, which produces a 1.4 T
magnetic field aligned coaxially with the beampipe, and are
used for measuring charged particle momenta.  The outermost system,
the Central Outer Tracker (COT) is a 3.1 m long open-cell drift
chamber which provides 96 position measurements in the radial region
between 0.43 and 1.32 m~\cite{Ref:COT} and in the pseudorapidity 
region $|\eta| \leq$ 1.0. Sense wires are arranged in 8
alternating axial and $\pm$2$^{\circ}$ stereo superlayers with 12
wires each. The position resolution of a single drift time measurement
is approximately 140 $\mu$m. Between the interaction region and the
COT, there are three separate silicon detectors. In combination,
silicon detectors provide high resolution position measurements for
charged particles out to $|\eta|=2.0$.  The innermost device, Layer
00~\cite{Ref:L00}, is a single-sided layer of silicon microstrip
detectors mounted directly on the beampipe at a radius of 1.6 cm that
provides an axial measurement as close to the collision point as
possible.  Between the COT and Layer 00, a five layer double-sided
silicon detector (SVXII) covers the radial region between 2.4 and 10.7
cm~\cite{Ref:SVX}.  Three separate SVX barrel modules are located
along the beamline, covering a length of 96 cm. Three of the five
layers combine an $r$-$\phi$ measurement on one side and a 90$^\circ$
stereo measurement on the other, and the remaining two layers combine
$r$-$\phi$ with small-angle stereo at $\pm$1.2$^\circ$.  The typical
hit resolution is 11 $\mu$m.  Three additional layers of double-sided
silicon strips, the Intermediate Silicon Layers (ISL), are located at
larger radii, between 19 and 30 cm, and provide good linking between
tracks in the COT and SVXII~\cite{Ref:ISL}.

%Calorimetry
Outside of the tracking systems and the solenoid, segmented
electromagnetic (EM) and hadronic (HAD) sampling calorimeters are used
to reconstruct electromagnetic showers and jets in the pseudorapidity
interval $|\eta| <$ 3.6~\cite{Ref:elemagCal}--\cite{Ref:Cal}.  The
calorimeters are segmented into projective towers of size
7.5--15$^\circ$ in $\phi$ and 0.1 in $\eta$. At the front of each
tower, a lead-scintillator sampling electromagnetic calorimeter, 18
radiation lengths deep, records the energy of electromagnetic showers.
In the central region ($|\eta| <$ 1.0), a layer of multiwire
proportional chambers (CES) measures the transverse shower profile at
a depth of the maximum shower development. Behind the electromagnetic
calorimeter is the hadronic calorimeter with roughly 5 absorption
lengths of alternating layers of steel and scintillator.

%Muon systems
High $p_T$ muons used in this analysis are detected in three separate
subdetectors. Two separated drift chambers cover the region $|\eta|
\leq$ 0.6: Directly outside of the hadron calorimeter, four-layer
stacks of planar drift chambers (CMU) detect muons with $p_T >$ 1.4
GeV/$c$ which penetrate the five absorption lengths of the
calorimeter~\cite{Ref:CMU}.  Behind another 60 cm of steel, an
additional four layers (CMP) detect muons with $p_T >$ 2.0
GeV/$c$~\cite{Ref:CMP}. An additional system with 4 drift chamber
layers and scintillation counters occupies the region 0.6 $\leq |\eta|
\leq 1.0$ (CMX), completing the muon coverage over the full fiducial
region of COT tracking, $|\eta| \leq 1.0$.

%===========================================================================
\section{\label{sec:chap3}Data samples}
\subsection{Luminosity and Triggers}
The results reported here are based on the data recorded during the
period March 2002--August 2004, when the average instantaneous
Tevatron luminosity was approximately 4 $\times$ 10$^{31}$
cm$^{-2}$s$^{-1}$, and the highest was about 10 $\times$ 10$^{31}$
cm$^{-2}$s$^{-1}$. The recorded integrated luminosity for this period
is 318 $\pm$ 19 pb$^{-1}$ for electron and CMU/CMP muon analysis, and
305 $\pm$ 18 pb$^{-1}$ for CMX muon analysis.

CDF employs a three-level trigger system. We describe only the
triggers important for this analysis, which select events containing a
high momentum electron or muon. For electron candidates, the first
level (L1) trigger requires a track with $p_T \geq$ 8 GeV/$c$ matched
to an EM calorimeter cell with $E_T \geq$ 8 GeV, and a ratio of
hadronic to electromagnetic energy
($E_{\mathrm{Had}}/E_{\mathrm{EM}}$) less than 0.125. Calorimeter
clustering is done in the second level (L2) trigger, which requires a
track with $p_T \geq$ 8 GeV/$c$ matched to an EM cluster with $E_T
\geq$ 16 GeV. At the third level (L3), a reconstructed electron with
$E_T>$ 18 GeV is required.  For muon candidates, a track with $p_T
\geq$ 8 GeV/$c$ matched to muon stubs in the muon chambers (CMU, CMP,
or CMX) is required for L1 and L2; the L3 trigger requires a $p_T
\geq$ 18 GeV/$c$ track.

\subsection{Monte Carlo Programs}
The generation of $t\bar t$ events relies mainly on the HERWIG v6.505
\cite{Ref:Herwig} and PYTHIA v6.216~\cite{Ref:Pythia} Monte Carlo
programs, which employ leading order QCD matrix elements for the hard
process, followed by parton showering to simulate gluon radiation and
fragmentation. The CTEQ5L~\cite{Ref:CTEQ5L} parton distribution
functions are used. For heavy flavor jets, the decay algorithm QQ v9.1
\cite{Ref:QQv91} is used to provide proper modeling of bottom and charm hadron
decays. The ALPGEN v1.3 program~\cite{Ref:Alpgen}, which generates
high multiplicity parton final states using exact leading-order matrix
elements, is used in the study of backgrounds. The parton level events
are then passed to HERWIG and QQ for additional QCD radiation,
fragmentation and $B$ hadron decay.

The CDF II detector simulation~\cite{Ref:CDFsim} reproduces the
response of the detector to particles produced in $p\bar p$
collisions. Tracking of particles through matter is performed with
GEANT3~\cite{Ref:Geant3}.  Charge deposition in the silicon detectors
is calculated using a parametric model tuned to the existing data. The
drift model for the COT uses the GARFIELD package~\cite{Ref:COT}, with
the default parameters tuned to match COT data. The calorimeter
simulation uses the GFLASH~\cite{Ref:Gflash} parameterization package
interfaced with GEANT3. The GFLASH parameters are tuned to test-beam
data for electrons and pions, and are checked by comparing the
calorimeter energy of isolated tracks in $p\bar p$ collision data to
their momenta as measured in the COT.

%==================================
\section{\label{sec:chap4}Particle Identification and Event Selection}
\subsection{\label{sec:leptonID} Lepton Identification}
The identification of charged leptons produced by $W$ decay provides
the initial selection of the $t\bar t\to$ lepton+jets sample.  After
passing the trigger requirements, electron candidates are identified
by requiring the electrons to be in the central pseudorapidity region
of the detector ($|\eta|\leq$ 1) and to have an EM cluster with $E_T
\geq$ 20 GeV and a track with $p_T \geq$ 10 GeV/$c$.  Several
variables are used to discriminate against charged hadrons and photon
conversions.  We require that the extrapolated track match the shower
location as measured in the CES, that the ratio of hadronic to
electromagnetic energy in the calorimeter cluster,
$E_{\mathrm{Had}}/E_{\mathrm{EM}}$, be less than 0.055 + 0.00045
$\times$ $E_{\mathrm{EM}}$, and that the ratio of cluster energy to
track momentum, $E/p$, be less than 2.0 (unless $p_T >$ 50
GeV/$c$, in which case this cut is not applied).  The isolation variable,
defined as the ratio of the additional energy deposited in a
cone of radius $\Delta R \equiv
\sqrt{\Delta \eta^{2}+\Delta \phi^{2}}=0.4$ around the electron
cluster to the electron energy, is required to be less than
0.1. Conversion electrons are removed by rejecting events that have a
pair of opposite electric charge tracks (one of them the electron) in
which the distance $\Delta(xy)$ between the tracks in the $r$-$\phi$
plane (at the conversion point) is less than 0.2 cm, and the
difference between the polar angle cotangent of the two tracks,
$|\Delta\cot\theta|$, is less than 0.04. Fiducial cuts on the
electromagnetic shower position in the CES ensure that the shower is
located in a well-understood region of the calorimeter.  For isolated
high momentum electrons from $W$ decay, the tracking efficiency is
measured to be 99.93$^{+0.07}_{-0.35}\%$~\cite{Ref:WZcross}.  
The transverse energy can be measured from the electromagnetic cluster 
with a precision $\sigma/E_T$ = 13.5$\%$/$\sqrt{E_T\mathrm{(GeV})}$ 
$\oplus$ 2 $\%$~\cite{Ref:elemagCal}.

Muon candidates are identified by extrapolating COT tracks to the muon
detectors.  Two types of high-$p_T$ muon samples are used in this
analysis. CMUP muons ($|\eta|\leq0.6$) have a COT track linked to
track segments in both CMU and CMP. A CMX muon
($0.6\leq|\eta|\leq1.0$) has a COT track linked to a track segment in
the CMX.  For both CMUP and CMX muons, we require that the COT track
has $p_T \geq20$ GeV/$c$, and that the energy in the calorimeter tower
containing the muon is consistent with the deposit expected from a
minimum ionizing particle. The latter rejects secondary particles in
calorimeter hadron showers that produce tracks in the muon chambers.
An isolation variable is defined as the ratio of the total energy deposited in a
cone of radius $\Delta R =0.4$ around the muon track candidate
(excluding the towers the muon passed through) to the track
momentum, and is required to be less than 0.1.  Backgrounds from cosmic
rays are removed by requiring that the distance $d$ of closest
approach of the reconstructed track to the beam line be less than 0.2
cm. For high momentum COT tracks, the resolution at the origin is
$\delta z \approx$ 0.5 cm along the beamline and $\delta d \approx$ 350 
$\mu$m ($\approx$ 40 $\mu$m with SVXII) for the impact parameter in the
transverse plane. Additionally, the distance between the extrapolated
track and the track segment in the muon chamber is required to be less
than 3, 5 and 6 cm for CMU, CMP and CMX respectively. COT tracks are
required to have at least 3 axial and 2 stereo layers with at least 5
hits each for both electron and muon candidates. From the COT, the transverse
momentum resolution for high momentum particles is found to be $\delta p_T/p_T$
$\approx$ 0.15$\%$ $\times$ $p_T$(GeV/$c$).

\subsection{\label{sec:JetIDCorr} Jet Corrections and Systematics} 
Jet reconstruction in this paper employs a cone cluster algorithm with
cone radius $\Delta R \equiv \sqrt{\Delta \eta^{2}+\Delta \phi^{2}}
=0.4$~\cite{Ref:JetCone}. We measure the transverse energy $E_T =
E\sin
\theta$, where $\theta$ is the polar angle of the centroid of the
cluster's towers, calculated using the measured $z$ position of the
event vertex. The total energy $E$ is the sum of the energy deposited
in calorimeter towers within the cone. Jets are identified as isolated
clusters that contain significant hadronic energy.  Jet measurements
make the largest contribution to the resolution of the top quark mass
reconstruction due to their relatively poor energy resolution,
approximately ($0.1\times E_T+1.0$) GeV~\cite{Ref:JetRes}. Additionally,
the uncertainty arising from the jet energy scale is the dominant
source of systematic uncertainty for the top quark mass.  In contrast,
we assume the angles of the quarks are well measured from the jet
angles. They are therefore directly used in the mass reconstruction
without correction. We briefly describe the jet energy corrections and
their systematic uncertainties in this section.  More details on the
CDF jet energy response are available elsewhere~\cite{Ref:JESnim}.

\subsubsection{\label{sec:JetCorr} Jet Corrections}
\begin{figure}[htbp]
\includegraphics[width=1\columnwidth]{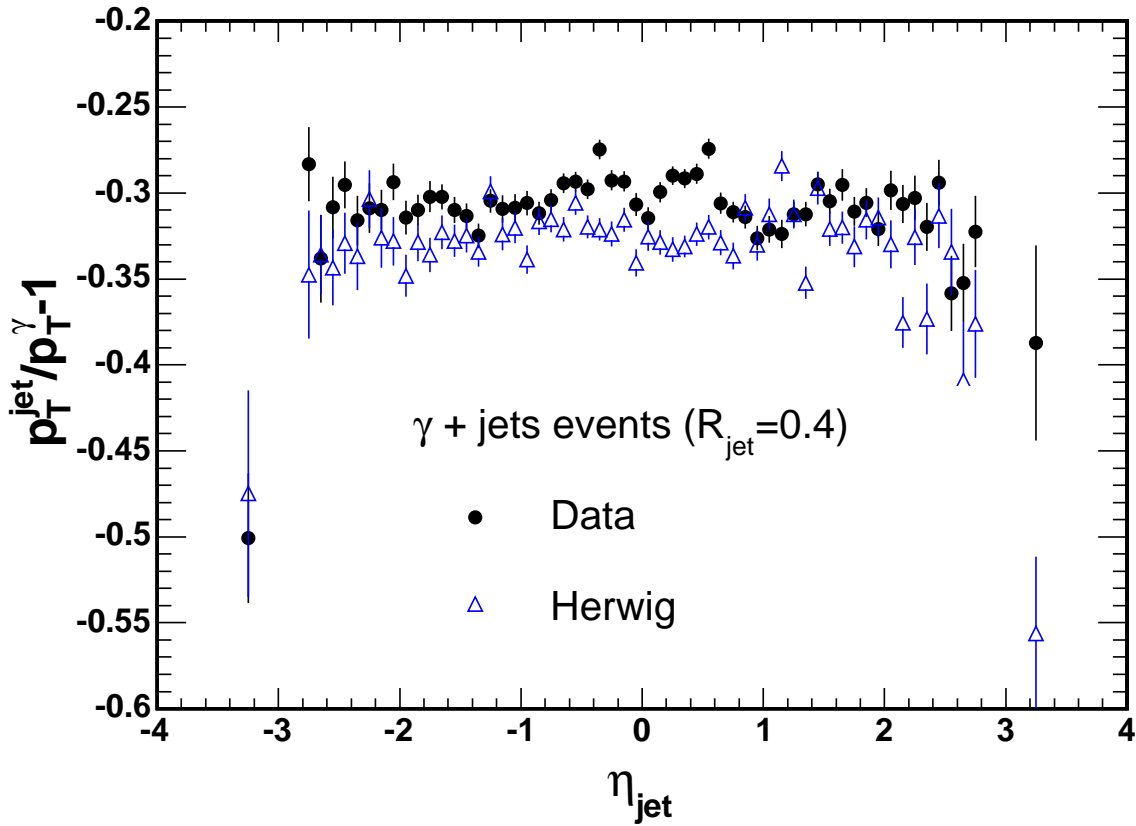}
\caption{$p_T$ balance, ~$p_T^{jet}/p_T^{\gamma}-1$, in $\gamma$-jet
events as a function of jet $\eta$ after relative corrections
are applied.  Circles are data and HERWIG Monte Carlo simulation is
plotted as triangles.}
\label{Fig:phojetcheck}
\end{figure}

\begin{figure}[htbp]
\includegraphics[width=1\columnwidth]{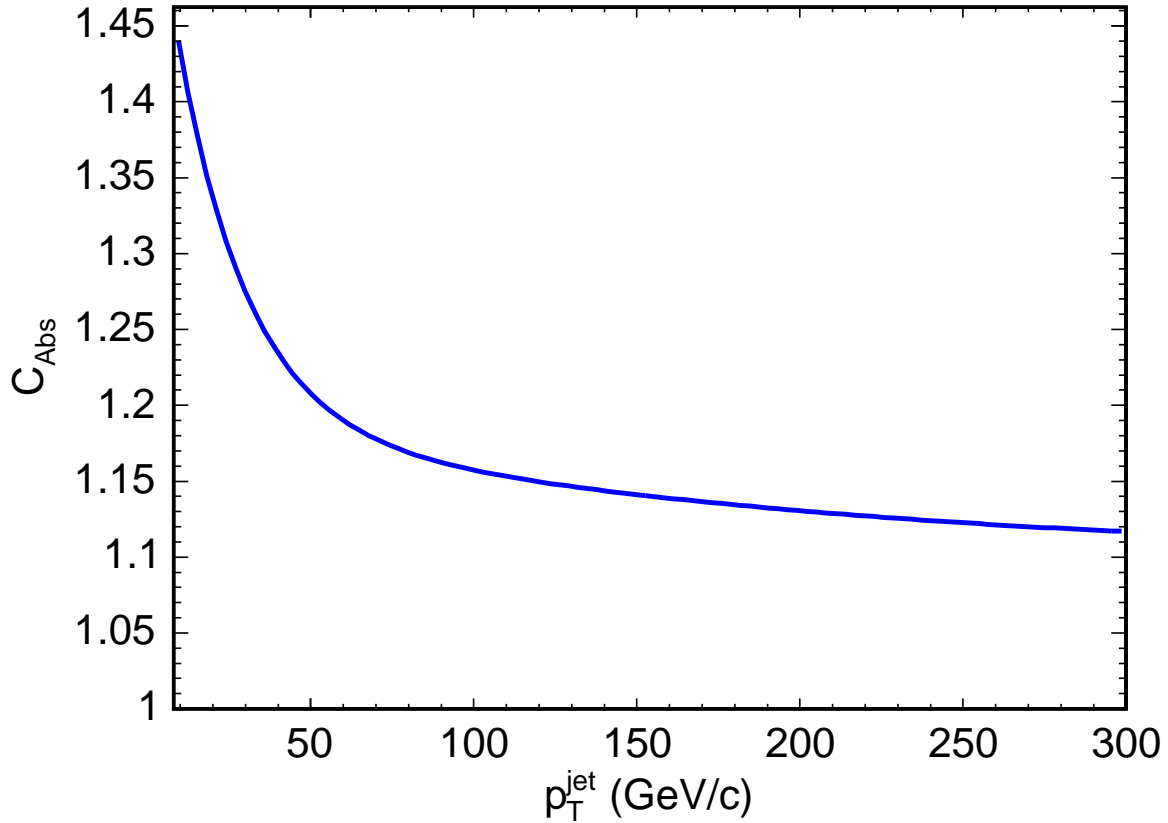}
\caption{The size of the absolute (hadron level) correction, $C_{\mathrm{Abs}}$, 
as a function of $p_T$ of the jet for cone size 0.4.}
\label{Fig:abscorr}
\end{figure}

To be used for top quark mass reconstruction, measured jet energies
are first corrected with a set of ``flavor-independent'' or
``generic'' corrections, so called because they are extracted mainly
from dijet and minimum bias samples.  These corrections are made in
several steps. A first correction scales the forward calorimeters to
the central calorimeter ($0.2 < |\eta| < 0.6$) scale for data and
Monte Carlo separately.  A dijet balancing procedure is used based on
the equality of the transverse energies of the two jets in a $2 \to 2$
process.  The correction is obtained as a function of $\eta$ and the
transverse momentum, $p_T$ of the jet.  The relative correction ranges
from about $-10\%$ to $+15\%$.  The corrections are tested by
comparing $E_T$ balance in $\gamma$+jet events in data and Monte Carlo
simulation. As shown in Fig.~\ref{Fig:phojetcheck}, after corrections
the response of the calorimeter is almost flat with respect to $\eta$
for both data and Monte Carlo simulations. A second correction is for
multiple $p \bar p$ interactions due to high-luminosity operation of
the Tevatron.  The energy from additional $p\bar p$ interactions
during the same accelerator bunch crossing can fall inside a jet
cluster, increasing the energy of the measured jet. The correction 
associated with this effect is derived from
minimum bias data and is parameterized as a function of the number of
identified interaction vertices in the event. This effect is corrected
on average and is very small (less than 1$\%$). A third correction is
called the absolute energy correction. This correction is applied to
account for calorimeter non-linearity and is based on the response of the
calorimeter to individual hadrons as measured by $E/p$ of single
tracks in the data. After this, the jet energy corresponds to the
energy of the hadrons incident on the jet cone. The absolute
correction varies between +10$\%$ and +40$\%$, depending on the jet
$p_T$ as shown in Fig.~\ref{Fig:abscorr}. The accuracy of this
correction depends on the Monte Carlo correctly modeling jet
fragmentation into hadrons, for example the charged to neutral particle 
ratio, and the particle multiplicity and $p_T$ spectrum. 
This has been checked by comparing the jet charged particle multiplicity 
distributions in data and Monte Carlo. 

After these generic jet energy corrections, we use the transfer
functions described in Section~\ref{sec:chap7} to account for the
fraction of the quark energy deposited outside the jet cone as well as
differences between light quark jets from $W$ boson decay and the $b$
jets coming directly from the top quark decay. 
Since the transfer functions are evaluated as a function of $E_T$, the 
resulting top quark mass is insensitive to the difference in the $E_T$ 
distributions of the dijet and top quark events.

\subsubsection{Systematics Uncertainty on the Jet Energy Scale}
The systematic uncertainty on the jet energy scale comes from a number
of sources.  The uncertainty in the calorimeter response relative to
the central calorimeter (relative response) is determined by varying
the dijet event selection criteria and the fitting procedure. This
uncertainty is typically between 0.5$\%$ and 1.0$\%$ for most jets
used in the top quark mass analysis. A second systematic uncertainty
comes from the hadron jet modeling used in the absolute energy scale
correction. The main sources here are uncertainties in the calorimeter
response to single hadrons ($E/p$) and jet fragmentation (charged to
neutral particle ratio).  Smaller contributions come from the Monte
Carlo modeling of the calorimeter response close to tower boundaries
in azimuth, and from the stability of the calorimeter calibration with
time.  In total, this uncertainty ranges from 1.5$\%$ to 3.0$\%$,
depending on jet $p_T$.

A third systematic uncertainty arises from modeling the energy that is
deposited outside the jet cone (out-of-cone correction).  This
uncertainty, which ranges from 2$\%$ to 6$\%$ depending on jet $p_T$,
is determined from the difference between data and Monte Carlo in
$\gamma+$jet events. The jet correction systematic uncertainties from
two other sources, the extra energy from multiple $p \bar p$
collisions and the underlying event, the energy associated with the
spectator partons in a hard collision event, are negligible for this
analysis.

\begin{figure}[htbp]
\includegraphics[width=1\columnwidth]{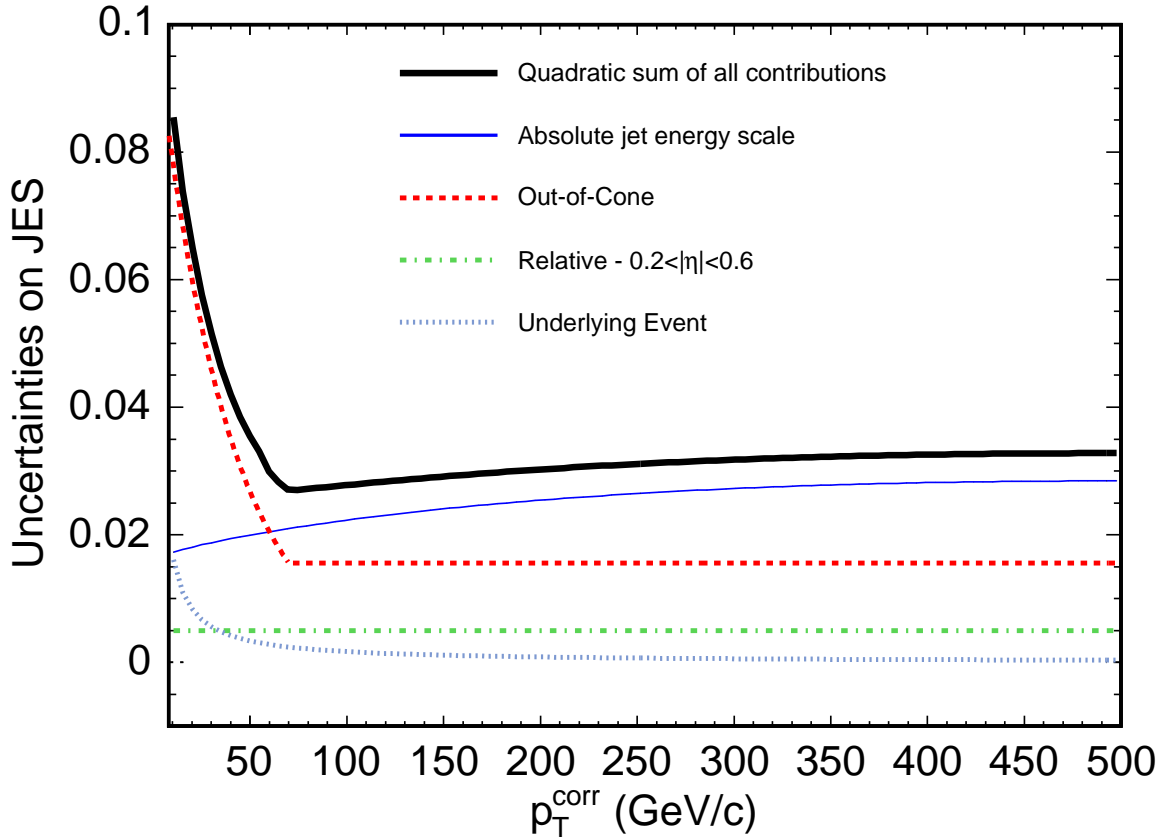}
\caption{The systematic uncertainties as a function of the corrected jet 
$p_T$ in the central calorimeter ($0.2 < |\eta| < 0.6$).}
\label{Fig:JESsys}
\end{figure}

\begin{figure}[htbp]
\includegraphics[width=1\columnwidth]{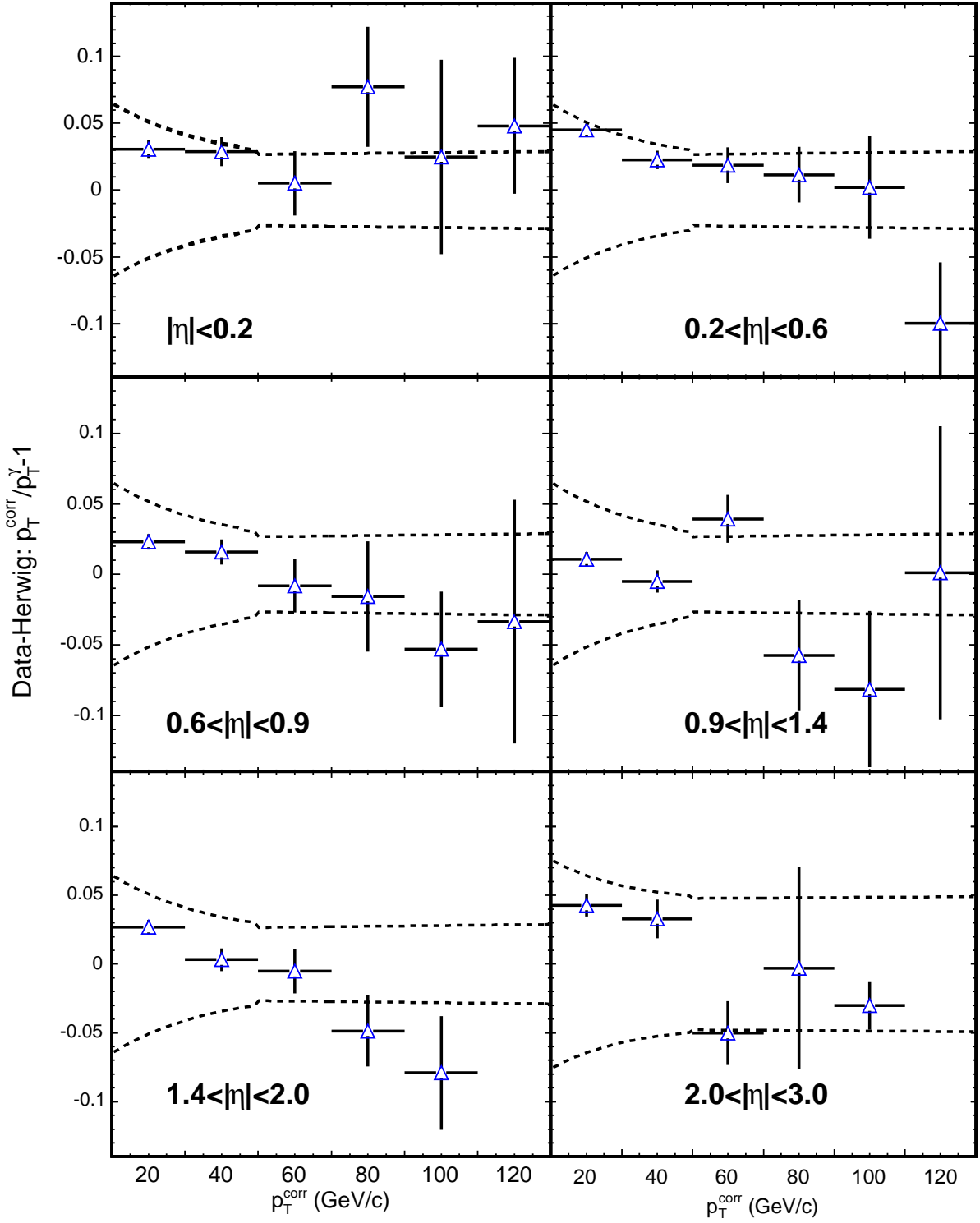}
\caption{The fractional difference between the jet and photon
transverse momenta in $\gamma$-jet events are calculated after all
jet corrections are applied. Plotted here is the difference between this
quantity in data and simulation as a function of photon $p_T$, for
different $\eta$ ranges. The dashed lines show $\pm 1\sigma$ from
the jet energy systematics.}
\label{Fig:jescheck}
\end{figure}

In summary, the systematic uncertainties on jet energy measurements
for jets in the central calorimeter ($0.2 < |\eta| < 0.6$) are shown
in Fig.~\ref{Fig:JESsys}.  The black line corresponds to the total
uncertainty, obtained by adding in quadrature all the sources
described above. Typically, it is 3$\%$ to 4$\%$ for jets with $p_T$
$>$ 40 GeV/c. In order to check the energy corrections and systematic
uncertainties, $\gamma$-jet events are used since the jet $p_T$ range
in this sample is similar to that in $t\bar{t}$
events. Figure~\ref{Fig:jescheck} shows the difference between
$\gamma$-jet balancing, defined as $p_T^{jet}/p_T^{\gamma}-1$, in data
and Monte Carlo after all jet corrections are applied. The $\pm$
1$\sigma$ range adequately covers the spread in these data points.
These jet energy uncertainties are propagated to the top quark mass
measurement as described in Section~\ref{sec:chap11}.  Additional
process-specific uncertainties are also considered in that
section. The most important of these for the top quark mass
measurement is the $b$-jet energy scale.

\subsection{\label{sec:btagalg}$b$-Jet Tagging using Secondary Vertex Identification}
The identification of $b$ jets from top quark decay plays an important
role in this analysis. Since most of the selected $W$+jets events
coming from non-$t\bar t$ processes do not contain bottom or charm
quarks in the final state, requiring the presence of $b$ jets provides
significant background reduction.

The SECVTX silicon vertex $b$-jet tagging algorithm searches within a
jet in the central region for a displaced secondary vertex due to the
decay of a $B$ hadron~\cite{Ref:btagRunI,Ref:RunIIXsecCount}.  It uses
tracks that are within $\Delta R <$ 0.4 of the jet axis and have hits
in the silicon detector. A set of cuts involving the transverse
momentum, the number of hits, and the $\chi^2/ndf$ of the track fit
are imposed to select good quality tracks in a jet.  Then the
algorithm is performed as follows: (1) Find at least three good tracks
with $p_T>$ 0.5 GeV/$c$ and an impact parameter significance
$|d_0/\sigma_{d_0}| >$ 2, where $d_0$ is the impact parameter of the
track relative to the accelerator beamline (measured on average for
each store of $p\bar p$ collisions) and $\sigma_{d_0}$ is the
uncertainty coming from both the track and beamline positions. At
least one of the tracks must have $p_T >$ 1 GeV/$c$.  (2) Reconstruct
a secondary vertex using the selected tracks.  (3) Calculate the
two-dimensional decay length of the secondary vertex ($L_{2D}$) from
the primary vertex. (4) Require $L_{2D}/\sigma_{L_{2D}} >$ 7.5, where
$\sigma_{L_{2D}}$ is the estimated uncertainty on $L_{2D}$, typically
190 $\mu$m, to reduce the background from false secondary vertices
(mistags).  If a secondary vertex is not found, a second pass of the
algorithm is carried out with tighter track requirements, demanding at
least two tracks with $p_T >$ 1 GeV/$c$ and $|d_0/\sigma_{d_0}| >$
3.5, including at least one track with $p_T > $ 1.5 GeV/$c$. The cut
on $L_{2D}/\sigma_{L_{2D}}$ is the same as in the first pass.

Based on simulation of the $b$-tagging algorithm, requiring at least
one $b$-tagged jet keeps 60$\%$ of top quark events while removing
more than 90$\%$ of background events. The difference between the
efficiency in the simulation and that in the data is measured using a
$b$-enriched dijet sample in which a non-isolated electron is found in
one jet. We find a data to Monte Carlo tagging efficiency scale factor
of 0.91 $\pm$ 0.06~\cite{Ref:RunIIXsecCount}, which is used with the
Monte Carlo in estimating the expected background (see
Section~\ref{sec:chap5}).  The uncertainty includes both systematic
and statistical contributions.  The main cause of the scale factor
being less than 1.0 is the difference in track resolution between the
data and Monte Carlo simulation.

\subsection{Missing Transverse Energy: $\not \!\!\bm{E}_T$ } 
The presence of neutrinos is inferred from transverse energy imbalance
in the detector. The missing transverse energy is calculated as
\begin{equation}
\not \!\!\bm{E}_T = - \sum_i {E_T^i \vec{n_i}},
\label{eq:metdef}
\end{equation}
where $E_T^i$ is the magnitude of the transverse energy contained in
calorimeter tower $i$, and $\vec{n_i}$ is the unit vector from the
interaction vertex to the tower in the plane transverse to the beam
direction. If isolated high-$p_T$ muon candidates are found in the
event, the $\not \!\!\bm{E}_T$ is corrected by subtracting the energy
deposited by the muon in the calorimeter, and adding the muon $p_T$ to
the vector sum. 
The typical $\not \!\!\bm{E}_T$ resolution in $t\bar t$ Monte Carlo
events is approximately 20 GeV.
Further corrections to $\not \!\!\bm{E}_T$ 
related to jet energy corrections and the transfer functions are described 
in Section~\ref{sec:chap7}.

\subsection{Event Selection}
The final state of the $t\bar t$ lepton+jets mode contains a
high-momentum lepton candidate, missing transverse energy that
indicates the presence of a neutrino from $W$ leptonic decay, and four
hadronic jets, of which two jets are expected to be $b$ quarks. We
summarize the selection criteria below:

Exactly one isolated electron (muon) candidate is required, having
$E_T\geq$ 20 GeV ($p_T\geq$ 20 GeV/$c$) and $|\eta| \leq$ 1.0. Any
event with two leptons satisfying the lepton criteria (see
Section~\ref{sec:leptonID}) is removed. We also remove events where
the second lepton candidate is an electron in the plug calorimeter or
a muon that fails the CMUP requirement but has one CMU or CMP muon
segment, to remove top dilepton events ($t\bar t\to l^+\nu l^- \bar
\nu b \bar b$). The missing transverse energy, $\not \!\! E_T$, is
required to be greater than 20 GeV. Events with $Z$ boson candidates
are removed by requiring that there be no second object that forms an
invariant mass with the primary lepton candidate within the window
76--106 GeV/$c^2$. Here, the second object is an oppositely-signed
isolated track with $p_T >10$ GeV/$c$ for primary muons; for primary
electrons it may be a track, an electromagnetic cluster, or a jet with
$E_T>$ 15 GeV and $|\eta|\leq$ 2.0 that has fewer than 3 tracks and a
high electromagnetic energy fraction.  The primary vertex of the event
must have its $z$ coordinate within 60 cm of the center of the CDF II
detector.  The jets are clustered after removing towers within
electron clusters and correcting each tower $E_T$ for the location of
the primary vertex $z$ coordinate. We select the events that have
exactly four jets with $E_T
\geq$ 15 GeV and $|\eta| \leq$ 2.0 to better match the leading order 
matrix element that is used in this analysis. This helps to reduce the
contamination by initial and final state radiation by 10$\%$ compared
to events with four or more jets.  Finally, at least one SECVTX tagged
$b$ jet is required. The above selection yields 63 $b$-tagged events
in which 39 events contain an electron and 24 a muon. Of these 63,
sixteen double $b$-tagged events are observed. The overall selection
efficiency for these criteria including the branching ratio, estimated
from $t\bar t$ Monte Carlo simulation, is approximately 1.94 $\pm$
0.01$\%$ for the electron channel, 1.22 $\pm$ 0.01$\%$ for muons in
the CMUP, and 0.41 $\pm$ 0.01$\%$ for muons in the CMX.

%===========================================================================
\section{\label{sec:chap5}Background Estimate}
It is important to this analysis that we have an accurate estimate of
the background level in the final event sample, because to extract the
top quark mass, we shift the measured mass to correct for background
contamination (see Section~\ref{sec:mapping}).  We use the technique
employed in the $t\bar t$ production cross section measurements
described in~\cite{Ref:btagRunI, Ref:RunIIXsecCount}.  The major
backgrounds come from misidentification, for example a fake lepton and
large missing $E_T$ in events not containing a $W$ boson, and events
in which a light quark or gluon jet is mistagged as a $b$ jet. The
major physics background is the production of a $W$ boson along with
heavy flavor quarks.

\subsection{Non-$W$ (QCD) Background}
\begin{figure}[htbp]
\includegraphics[width=1\columnwidth]{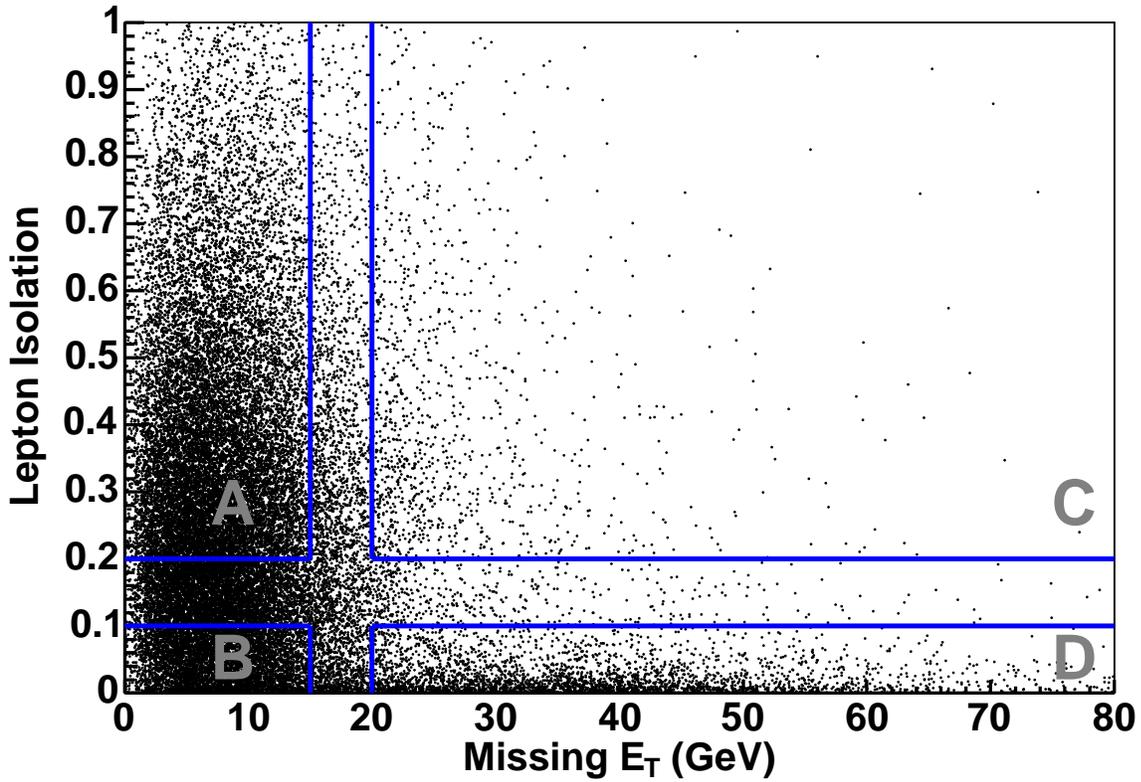}
\caption{Plot of missing $E_T$ vs lepton isolation for events that contain an 
electron candidate (without an isolation cut) and two or more jets
with $E_T >$ 15 GeV and $|\eta|\leq$ 2.0 before the $b$ tagging
requirement is applied.}
\label{Fig:MetvsIso}
\end{figure}

The non-$W$ background (QCD multijets), events that do not contain a
$W$ boson, is estimated directly from the data, separately for
electrons and muons. These events include fake leptons and missing
energy as well as semi-leptonic $B$ decays.  An isolated primary
lepton and large $\not \!\! E_T$ due to the neutrino are
characteristics of real $W$ events, not shared by most non-$W$
events. To estimate the number of non-$W$ events in the sample, we use
a 2-dimensional plot of $\not \!\! E_T$ vs lepton isolation, defining
four regions:
\begin{itemize}
\item[A:] isolation $>$ 0.2 and  $\not \!\! E_T <$ 15 GeV,
\item[B:] isolation $<$ 0.1 and  $\not \!\! E_T <$ 15 GeV,
\item[C:] isolation $>$ 0.2 and  $\not \!\! E_T >$ 20 GeV,
\item[D:] isolation $<$ 0.1 and  $\not \!\! E_T >$ 20 GeV,
\end{itemize}
as shown in Fig.~\ref{Fig:MetvsIso}. Region D contains the real $W$
events.  For the non-$W$ background, these two variables are assumed
to be uncorrelated; therefore $N_B/N_A$, the ratio of the numbers of
low $\not \!\! E_T$ events at low and high isolation, is the same as
$N_D/N_C$, the ratio at high $\not \!\!  E_T$.  The amount of non-$W$
contamination in region D is then calculated as $N_D$(non-$W$) =
$\frac{N_B\times N_C}{N_A}$.
Since backgrounds from $W$ + heavy flavor ($W b\bar b$, $W c\bar c$,
and $Wc$) and $W +$ a mistagged jet are estimated by normalizing to
the number of the ``pretagged'' events, those found prior to applying
the $b$ tagging algorithm, (see Section~\ref{chap:bkgWhf}), the
contributions of non-$W$ background to both the pretagged and the
tagged samples have to be measured, even though we use only the tagged
sample estimate directly in the mass measurement. To evaluate the
expected number of events in the tagged sample, we use two
methods. One is to estimate $N_D$ directly from the tagged
sample. However, this is limited by low statistics, hence we lower the
isolation boundary from 0.2 to 0.1 for regions A and C. A second
method is to scale the estimate in the pretagged sample by the tagging
rate for non-$W$ events. The number of events in region B with 2 or
more jets is used to obtain a reliable tag
rate~\cite{Ref:RunIIXsecCount}. These two estimates are found to be
consistent within the statistical uncertainty.  The final estimate is
obtained from the weighted average of the two methods.

\subsection{Mistags}
A SECVTX tag in a jet without a heavy flavor quark is called a
``mistag''. The mistag rate per jet is measured using a large
inclusive-jet data sample, without relying on the detector simulation.
It is parameterized as a function of the number of tracks in the jet,
the jet $E_T$ before energy corrections, the $\eta$ and $\phi$ of the
jet, and the sum of the $E_T$'s of all jets in the event with $E_T >$
10 GeV and $|\eta| <$2.4. To estimate the size of the mistag
background, each jet in the pretag sample is weighted by its mistag
rate, and then the sum of the weights over all jets in the sample is
computed, after correcting for the fraction of pretagged events that
are due to non-$W$ background ($\sim$ 10 $\%$ for electron 
and $\sim$ 5 $\%$ for muon channel) to avoid double counting these two
background sources. Using the number of mistagged jets as the number
of mistagged events is a good approximation because the mistag rate
per jet is sufficiently low, typically 1$\%$. This method is tested
using samples of pure mistagged jets in which the jet and
reconstructed secondary vertex are on opposite sides of the primary
vertex.  We find good agreement between the predicted and observed
numbers of jets in the pretagged sample as a function of jet
$E_T$~\cite{Ref:RunIIXsecCount}.

\subsection{\label{chap:bkgWhf} $W$ + Heavy Flavor ($W$+HF) Backgrounds}
The production of $W$ bosons accompanied by QCD production of heavy
flavor quarks in the processes $W b\bar b$, $W c\bar c$, and $Wc$
produces a signature very similar to $t\bar t$ events in the final
state, and is a significant part of the background for the tagged
sample. These contributions, $N_{HF}$, are evaluated by
$N_{HF}=N_{pretag}\times f_{HF}\times \epsilon_{btag}$, where
$N_{pretag}$ is the number of pretagged events in the data, $f_{HF}$
is the fraction of pretagged events containing $Wb\bar b$, $Wc\bar c$,
and $Wc$, estimated using the Monte Carlo models
ALPGEN+HERWIG, and $\epsilon_{btag}$ is
the $b$ tagging efficiency of each background source. The heavy flavor
fractions are found to be approximately 2--3$\%$ for $W b\bar b$ and
$W c\bar c$, and 6$\%$ for $Wc$ events, and were calibrated by
comparing dijet Monte Carlo events with data. Details of these
calculations can be found in~\cite{Ref:RunIIXsecCount}.

\subsection{Other Backgrounds}
The $WW$, $WZ$, and $ZZ$ background, $Z\to \tau\tau$, and electroweak
single top production by both $s$-channel $q\bar q$ fusion and
$t$-channel $W$-gluon fusion processes are evaluated based on
predictions from Monte Carlo simulation by multiplying the acceptances
for these processes, as determined from the PYTHIA Monte Carlo
program, by their production cross
sections~\cite{Ref:BkgXsec1,Ref:BkgXsec2} and the integrated
luminosity for the data sample. The Monte Carlo acceptance is
corrected for the differences between Monte Carlo and data for lepton
identification and trigger efficiencies. The $b$-tagging efficiency is
also scaled by the MC/data tagging scale factor which was described in
Section~\ref{sec:btagalg}.

\subsection{Background Summary}
\begin{figure}[htbp]
\includegraphics[width=1.0\columnwidth]{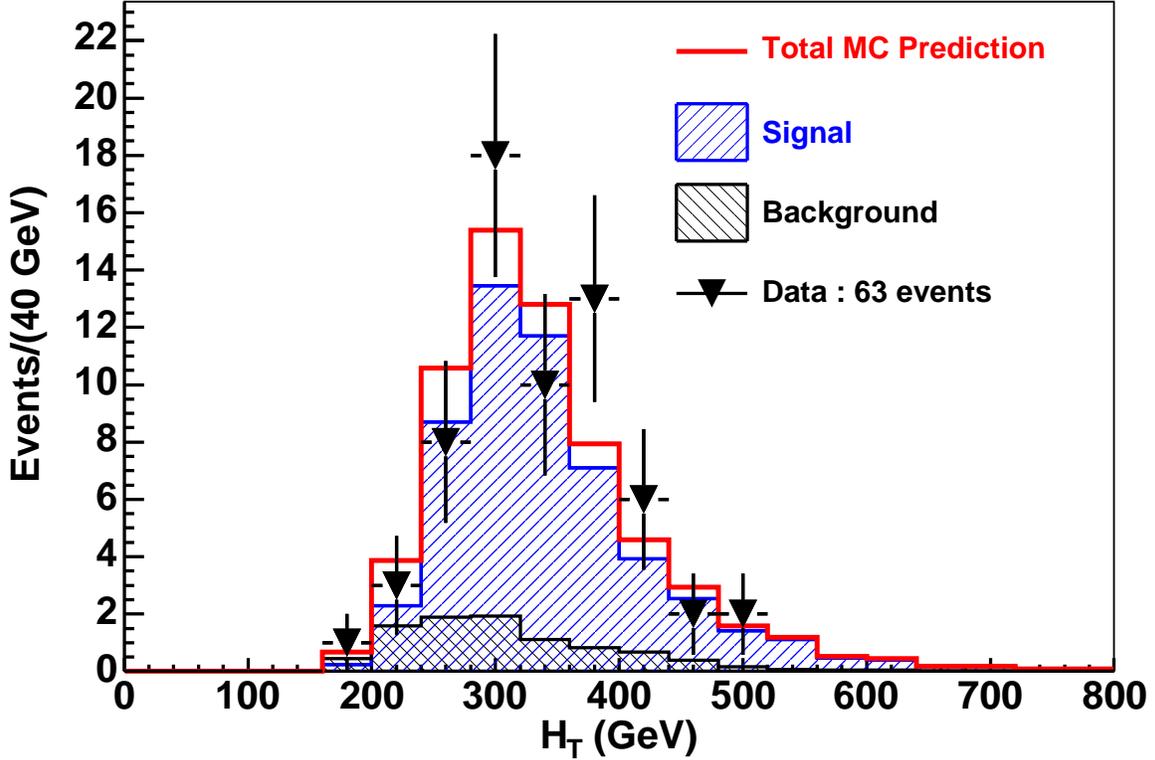}
\caption{$H_T$ distribution for signal ($M_{top}$ = 178 GeV/$c^2$) and background, 
normalized to 53.8 and 9.2 events, respectively. The total 63 data
events are shown as the triangles.}
\label{Fig:Htdist}
\end{figure}

Events having a leptonic $W$ decay plus 1 or 2 jets are used to test
the background estimation procedure.  We find agreement between the
data and Monte Carlo predictions within their uncertainties.  The
results provide confidence that we can estimate the number of
background events in the four jet topology.  The background
contributions to the W+4jets sample are summarized in Table
\ref{tab:expectedSummary}. We estimate the total number of background
events to be 9.2 $\pm$ 1.8. The expected number of signal events for
the predicted $t\bar t$ cross section ranges from 46 $\pm$ 5 events
for $M_{top}$ = 170 GeV/$c^2$ (7.8 pb) to 37 $\pm$ 4 events for
$M_{top}$ = 178 GeV/$c^2$ (6.1 pb).  The relative uncertainty on each
cross section value is roughly 10$\%$, mainly coming from the parton
distribution functions~\cite{Ref:ttXsecRef}.  However, the estimate of
9.2 background events has been extracted with little dependence on the
theoretical prediction of the $t\bar t$ cross section. We find that a
5 GeV/$c^2$ difference in $M_{top}$ (corresponding to about a 1.0 pb
difference in $t\bar t$ cross section) alters the background estimate
by roughly 1$\%$, corresponding to a negligible $\sim 0.1$ event.
Therefore in this analysis, 53.8 events are assumed to be from signal
$t\bar t$ events (9.2 background events subtracted from the observed
63 events). This is supported by the 16 double $b$ tagged events in
the data, where the expected number of events estimated by scaling the
63 observed events is 16.8 $\pm$ 1.8 events, including an expected 1.4
background events.  For a kinematic comparison, the $H_T$ distribution
is shown in Fig.~\ref{Fig:Htdist}.  $H_T$ is defined as the scalar sum
of the lepton $E_T$, the $\not \!\!\bm{E}_T$ and the $E_T$'s of the leading
four jets. We find good agreement between the data and the Monte Carlo
for both the double $b$ tag ratio and the kinematic distribution.

\begin{table}
\begin{ruledtabular}
\caption{\label{tab:expectedSummary}The expected number of background
events from individual sources and the fractions with respect to the
63 observed events.}
\begin{tabular}{lcr}
Source  & Number of Events & Fraction ($\%$)\\
\hline
Non-W (QCD)       	& 3.07 $\pm$ 1.06 & 4.87 \\
Mistag           	& 2.27 $\pm$ 0.45 & 3.60 \\
$W b\bar b$       	& 1.70 $\pm$ 0.79 & 2.70 \\
$W c\bar c$       	& 0.81 $\pm$ 0.40 & 1.28 \\
$Wc$               	& 0.51 $\pm$ 0.23 & 0.81 \\
$WW/WZ/ZZ$         	& 0.39 $\pm$ 0.08 & 0.62 \\
Single Top       	& 0.41 $\pm$ 0.09 & 0.65 \\
\hline
Background Total	& 9.2 $\pm$ 1.8  & 14.5\\
\hline
Observed events 	& 63             & 100 \\
\end{tabular}
\end{ruledtabular}
\end{table}

\section{\label{sec:AO}Analysis Overview}
The analysis proceeds as follows. For each event, a likelihood as a
function of top quark mass is calculated by the dynamical likelihood
method (DLM), described in Section~\ref{sec:chap6}. The DLM defines a
likelihood for each event based on the differential cross section per
unit phase space of the final partons in the elementary process.
It does not however use the number of observed events to constrain $M_{top}$ 
based on the theoretical ttbar cross section.
To infer the parton momenta, we employ transfer functions that relate the
observed jet energies to the corresponding parton energies: four jets
to four quarks ($q\bar{q}'$ from the $W$, $b$ and $\bar{b}$). The
transfer functions are obtained from HERWIG Monte Carlo $t\bar t$
signal samples.  Section~\ref{sec:chap7} describes the details and
performance checks of the transfer functions.  There are 6 or 2
possible assignments of the four jets to the individual partons,
depending on whether 1 or 2 jets are $b$-tagged, and for each
assignment 2 solutions for the $z$-component of the neutrino
momentum. Instead of selecting one particular assignment (e.g., the
one giving the maximum likelihood), we average the likelihoods for all
possible jet assignments and neutrino solutions in an event, and such
event likelihoods are multiplied together to obtain the joint
likelihood function for the entire data sample. We take the average
rather than the sum in order not to give greater weight to single
$b$-tag events with their larger number of jet assignments.  After
calculating the top quark mass under the assumption that all events
are $t\bar t$, the effect of the background is corrected by using a
mapping function that provides a mass-dependent correction factor. The
mapping functions are extracted using Monte Carlo pseudo-experiments
in which the numbers of signal and background events are Poisson
distributed around the expected means. The mean number of signal
events is not changed for different top quark mass samples. Finally,
we extract the measured value of the top quark mass using the expected
background fraction estimated in Section~\ref{sec:chap5}.

%===========================================================================
\section{\label{sec:chap6}Dynamical Likelihood Method}

The DLM was originally proposed in 1988~\cite{Ref:dlmKK1} and
developed in~\cite{Ref:dlmKK2,Ref:dlmKK3}; details of the latest
formulation are described in~\cite{Ref:dlmKK4}.  In DLM, we generate
the parton kinematics from the observed quantities, and the likelihood
of the reconstructed parton state is defined by the differential cross
section per unit phase space of the final partons in the elementary
process.

\subsection{Definition of the likelihood}
\subsubsection{Differential cross section}
The elementary parton process in a $p \bar p$ collision can be written
as,
\begin{equation}
\label{eq:process}
        a_1/p+a_2/\bar{p}\rightarrow \cdots \rightarrow 
C, \hspace{1.0cm}C \equiv \sum _{i=1}^n c_i.
\end{equation}
 where $a_1$ and $a_2$ are the initial partons -- quark, anti-quark,
or gluon -- in the proton and anti-proton, respectively, and
$c_{1},c_{2},\ldots,c_{n}$ are final state partons and leptons. These
are defined after initial-state radiation but before final-state
radiation.  In the case of the $t\bar{t}$ lepton+jets channel, the
initial parton set $(a_1,a_2)$ is $(q,\bar{q})$, $(\bar{q},q)$ or
$(g,g)$, and the final leptons and partons are
$l,\nu,q,\bar{q'},b,\bar{b}$ or their anti-particles, where
$(q,\bar{q^{'}})$ are quarks from $W$ decay, and $l=e$ or $\mu$.
Throughout this paper, a particle 4-momentum and its 3-momentum are
represented by a small letter in italics and in bold, respectively:
e.g., a symbol ``$p$'' represents the proton's 4-momentum, and $\bm p$
its 3-momentum.  The final partons are assumed to have their pole
masses (4.8 GeV/$c^2$ for the $b$ jets and 0.5 GeV/$c^2$ for the $W$
daughter jets), so that their 3-momenta define their states unambiguously.

The hadronic cross-section for process ($\ref{eq:process}$) is 
given by
\begin{eqnarray}
\label{eq:dsigma}
d\sigma =  dz_{a_1} dz_{a_2} d^2\bm{p}_T
f_{a_1/p}(z_{a_1}) f_{a_2/\bar{p}}(z_{a_2}) f_T(p_T) \nonumber
\\ \times d\hat{\sigma} (a_1+a_2\rightarrow C;\bm{\alpha}),
\end{eqnarray}
where $d\hat{\sigma}$ is the parton level cross section~\cite{Ref:Hagi},
\begin{widetext}
\begin{equation}
d\hat{\sigma}(a_1+a_2 \rightarrow C;\bm{\alpha}) =\frac{(2\pi )^4
  \delta^4(a_1+a_2-C)}{4\sqrt{(a_1\cdot a_2)^2 -m^{2}_{a_{1}} m^{2}_{a_{2}}}} |
{\cal{M}}(a_1+a_2\rightarrow C;\bm{\alpha}) |^2 d\Phi_{n}^{(f)}(a_1+a_2;C).
\label{eq:dsighat}
\end{equation}
\end{widetext}
In Eq.~(\ref{eq:dsigma}), the symbol $\bm{\alpha}$ represents a set of
dynamical constants to be measured, e.g.,\ masses, decay widths and
coupling constant ratios. In this analysis, $\bm{\alpha}$ is simply
the top quark mass $M_{top}$. The variables $z_{a_1}$ and $z_{a_2}$
are the energy fractions of $a_1$ and $a_2$ in hadrons $p$ and
$\bar{p}$ respectively, and $m_{a_{1}}$ and $m_{a_{2}}$ are their masses
that are assumed to be zero in this analysis. $\bm{p}_T$ is the total transverse
momentum of the initial and final systems in the plane transverse to
the beam axis. Functions $f_{a_1/p}(z_{a_1})$ and
$f_{a_{2}/\bar{p}}(z_{a_2})$ denote the parton distribution functions
(PDF's), while $f_{T}(p_{T})$ is the probability density function for
the total transverse momentum of the system acquired by initial state
radiation. In this analysis, we use the leading order PDF,
CTEQ5L~\cite{Ref:CTEQ5L}. Other PDF sets are used to calculate the
systematic uncertainty. The function $f_{T}(p_{T})$ is obtained by
running the PYTHIA generator.

In Eq.~(\ref{eq:dsighat}), ${\cal{M}}$ is the matrix element of the
process that is being studied (in this case, $t\bar t$ production and
decay described in Section~\ref{sec:matrix}), and $d\Phi_n^{(f)}$ is
the Lorentz invariant phase space element,

\begin{equation} 
d\Phi_{n}^{(f)} = \prod _{i=1} ^n \frac{d^3\bm{c}_i}{(2\pi)^3 2E_i}.
\end{equation}

We use Eq.~(\ref{eq:dsighat}) to formulate the parton level
likelihood.  The basic postulate is that final partons occupy an
$n$-dimensional unit phase space volume in the neighborhood of
$\bm{c}=(c_{1}, \ldots, c_{n})$.  When a momentum set $\bm{c}$ is
given, the total probability for this final state to occur is obtained
by integrating Eq.~(\ref{eq:dsigma}) over initial state variables
$z_{a_1}$, $z_{a_2}$ and $\bm{p}_T$, as
\begin{equation}
\label{eq:dsdphi}
\frac{d\sigma}{d\Phi_{n}^{(f)}}
= I(a_1,a_2) \left|{\cal{M}}(a_1+a_2\rightarrow C;\bm{\alpha})\right|^2 , 
\end{equation}
where
\begin{equation}
I(a_1,a_2) = \frac{(2\pi)^4}{4\sqrt{(a_1\cdot a_2)^2}} 
f_{a_1/p}(z_{a_1}) f_{a_2/\bar p}(z_{a_2}) f_{T}(p_{T})
\end{equation} 
is the integration factor for the initial state. Because of the
$\delta$-function in Eq. (\ref{eq:dsighat}), the initial parton
momenta $a_1$ and $a_2$ are uniquely defined by that of $C$.

For a given set of $\bm{c}=(c_{1}, \ldots, c_{n})$, we define the
parton level likelihood for $\bm{\alpha}$ by
\begin{equation}
\label{eq:plkhd}
L_{1}^{(p)}(\bm{\alpha}|\bm{c}) = l_{0} \frac{d\sigma}{d\Phi_{n}^{(f)}},
\end{equation}
where $l_{0}$ is given by  
\begin{equation}
\label{eq:lum0}
l_{0}=\frac{1}{\epsilon(M_{0})\sigma_{T}(M_{0})}
\end{equation}
In Eq.~(\ref{eq:lum0}), $\sigma_{T}(M_{0})$ and $\epsilon(M_{0})$ are
the total cross section and the detection efficiency for the true
(pole) top quark mass of the sample, respectively. Thus $l_{0}$ is the
integrated luminosity per event in the sample. This method does not
make use of any constraint from the theoretical $t\bar t$ cross
section as a function of $M_{top}$. Since $l_{0}$ only depends on the
true (pole) top mass, it does not vary event by event in the
sample~\cite{Ref:dlmKK4} and only changes the absolute value of the
likelihood, i.e., it has no effect on the final result.  In this sense,
this is not a real likelihood and any bias has to be corrected by the
mapping function. The statistical uncertainty is also corrected by
checking the pull distribution as described in
Section~\ref{sec:chap8}.

\subsubsection{Propagator factors}
When process (\ref{eq:process}) includes internal lines of the Feynman
graph, for example $r$ in
\begin{equation}
\label{eq:resprod}
a_1/p + a_2/\bar p \rightarrow r+c_{j+1}+\cdots + c_{n},
\end{equation}
\begin{equation}
\label{eq:resdec}
r \rightarrow c_{1}+\cdots + c_{j},
\end{equation}
we have to consider the propagator factor for a particle $r$.  We
treat, in this channel, $t$, $\bar{t}$, $W^{+}$ and $W^{-}$ as
internal lines ($r$) as illustrated in Fig.~\ref{Fig:Feynman}.

We factorize the matrix element as 
\begin{equation}
|{\cal{M}}(a_1+a_2\rightarrow C;\bm{\alpha})|^2
 =|{\cal{M}}_{prod}|^2{\cal{P}}(s_{r})|{\cal{M}}_{dec}|^2, 
\end{equation}
where ${\cal{M}}_{prod}$ and ${\cal{M}}_{dec}$ are the matrix elements
for the production process and decay respectively, and $s_{r}$ is the
virtual mass squared of $r$, which satisfies
\begin{equation}
\label{eq:qsq}
s_{r} = (\sum_{i=1}^{j} c_{i})^{2}.
\end{equation}
For the propagator factor ${\cal{P}}(s_{r})$, we assume the
Breit-Wigner form,
\begin{equation}
\label{eq:propfac}
{\cal P}(s_{r})  = \frac{1}{(s_{r}-M_{r}^2)^2+M_{r}^2\Gamma_{r}^2}.
\end{equation}
In the reconstruction of $\nu_{z}$, the unmeasured $z$-component of
the neutrino momentum, we generate the $W$ mass squared $s_{W}$
according to $\Pi (s_{W})$, where
\begin{equation}
\label{eq:normprop}
\Pi (s)={\cal P}(s)  \Big/ \int {\cal P}(s)  ds,
\end{equation}
and solve Eq. (\ref{eq:qsq}) for $\nu_{z}$ (quadratically ambiguous).
The function $\Pi (s)$ satisfies
\begin{equation}
\label{eq:propfaacnorm}
\int_{0}^{\infty} \Pi (s) ds = 1.
\end{equation}

\subsubsection{Transfer functions for observables}
Final quarks and gluons are not directly observed; they undergo
hadronization, are observed by detectors with finite resolution, and
are reconstructed as jets.  Jet energies are generally calibrated
using generic QCD jets, so we need additional corrections for $b$ jets
and $W$ daughter jets in the $t\bar t$ processes.  To describe the
relation between the parton and observed quantities (observables), we
introduce the transfer function (TF) $w(\bm {y}|\bm{x})$, where $\bm
{y}$ represents a set of observables and $\bm{x}$ is a parton variable
set that corresponds to $\bm {y}$. In the $l$+jets process, $\bm y$
consists of the momenta of the $e$ or $\mu$ and of the 4 jets, and the
missing transverse energy ($\not
\!\!\bm{E}_T$). In the present analysis, we use the TF only for quarks
and jets.  Electrons and muons are measured well in the detector, and
$\not
\!\!\bm{E}_T$ is calculated from other observed quantities in an event
(see Section~\ref{sec:met}).

The differential probability for the parton variables $\bm{x}$ to be
observed as $\bm{y}$, $dP(\bm{y};\bm{x})$, is defined by the TF
$w(\bm{y}|\bm{x})$ as
\begin{equation}
        dP(\bm{y};\bm{x})=w(\bm{y}|\bm{x})d\bm{y}.
\end{equation}
The TF for a single quark, $w(y|x)$, is obtained from the $(x,y)$
distribution of the $t \bar t$ Monte Carlo events.  The event
selection criteria are applied to these events.  
The effect of the detection efficiency for the variable set
$(x,y)$ is thus included in the determination of $w(y|x)$, and
the normalization condition,
\begin{equation}
        \int w(y|x)dy = 1,
\end{equation}
holds.

\subsubsection{Likelihood for a single path, a single event and multiple events}
\underline{\it Single path reconstruction and its likelihood}\hspace*{5mm}
The single path likelihood is defined for each complete set of parton
kinematics and calculated as follows:

\begin{description}
\item[(1)]{ We assume that the momentum of the $e$ or $\mu$ is precisely 
measured.}

\item[(2)]{ The four jets are assigned to the four final state quarks. 
We call such an assignment a ``topology'' denoted by $I_t$
($I_t=1,\cdots, N_t$).  Therefore $N_t$ represents 6 or 2 possible
topologies in an event, depending on whether 1 or 2 jets are
$b$-tagged.}

\item[(3)]{ Once a topology is specified, we randomly generate the parton kinematics 
($b,\bar{b},q,\bar{q}'$) according to the transfer functions. We
identify the momentum direction of each jet with that of the assigned
quark, and transfer variables $\bm{x}$
$(E_{Tb},E_{T\bar{b}},E_{Tq},E_{T\bar{q}'})$ are chosen using as input
$\bm{y}$, the transverse energies of the corresponding jets. More
details are given in Section~\ref{sec:chap7}.  Each random generation
is denoted by $k$.}

\item[(4)]{ After $\bm{(1)}$, $\bm{(2)}$ and $\bm{(3)}$, the transverse momentum of 
the neutrino $(\nu_{x},\nu_{y})$ is identified with the measured value
of $\bm{\not\!\! E_T}$ and then corrected using both jet corrections
and ($k$-th) jet transfer functions. Details of this correction are
discussed in Section~\ref{sec:chap7}. Then the parton momenta are
defined except for $\nu_{z}$, the unmeasured $z$ component of the
neutrino momentum. To get $\nu_{z}$ we choose $s_{W}$ according to
$\Pi(s_{W})$ in Eq.~(\ref{eq:normprop}), and $\nu_{z}$ is obtained by
solving Eq.~(\ref{eq:qsq}).  A quadratic ambiguity results in two
solutions ($\nu_{z1}, \nu_{z2}$) that are specified by an integer
$I_{s}$ (=1 or 2), which is treated separately from ``topology'' as
defined in (2).}

\item[(5)]{ From procedures $\bm{(1)}$, $\bm{(2)}$, $\bm{(3)}$ and $\bm{(4)}$, 
an event configuration ($I_t$ and $I_s$) and parton momenta ($k$ for a
generation by the transfer functions) are uniquely specified. The
likelihood of a single path is then
\begin{equation}
\label{eq:singleL}
\bm{L}^{(k)}_{1}(I_{t},I_{s},\bm{x}_{k};M_{top}|\bm{y}^{(i)})
= l_{0} \frac{d\sigma}{d\Phi_{6}^{(f)}} (I_{t}, I_{s}, k, i; M_{top}),
\end{equation}
where $i$ is the event number, and $d\Phi_{6}^{(f)}$ is the phase
space for $(l,\nu,b,\bar{b},q,\bar{q}')$. In this context, when we use
``a single path'' the likelihood (the differential cross section) can
be calculated without any ambiguity, since all information such as
assignments and parton momenta are determined.  Then for each path, we
make a parameter scan of $M_{top}$ uniformly in its search region
(typically 155--195 GeV/$c^2$).}
\end{description}

\underline{\it Likelihood for a single event}\hspace*{5mm}
All possible paths (configurations), each labeled by $k$, $I_t$ and $I_s$, 
are mutually exclusive, and we define the likelihood of the $i$-th event as the 
average of the likelihoods for all paths,
\begin{widetext}
\begin{equation}
\label{eq:evlikedef1}
\bm{L}(M_{top}|\bm{y}^{(i)}) = 
\frac{l_{0}}{2 K N_{t}} 
\sum_{k=1}^{K}\sum_{I_t=1}^{N_{t}}\sum_{I_s=1}^{2}
\bm{L}^{(k)}_{1}(I_{t},I_{s},\bm{x}_k;M_{top}|\bm{y}^{(i)}). 
\end{equation}
\end{widetext}
This definition of the event likelihood thus contains the correct set
of $(I_{t}, I_{s})$ (if the event is $t\bar{t}\rightarrow l+$jets).
The sum over $k$ corresponds to the numerical integration of the
parton kinematics according to the transfer functions. Therefore we
repeat the procedure $\bm{(2)}$--$\bm{(5)}$ a large enough number of
times ($K$) so that the value of $\bm{L}(M_{top}|\bm{y}^{(i)})$
converges, which is typically 50,000 times.  In summary, each time a
parton configuration and set of momenta are selected ($I_t$, $I_s$ and
$k$), we calculate the likelihood (single path) and in order to obtain
the event likelihood, we average all possible single path likelihoods
by numerical integration.\\

\underline{\it Likelihood for multiple events}\hspace*{5mm}
The single event likelihood is a function of $M_{top}$. For multiple
events, we get mutually independent functions of $M_{top}$.  Hence to
obtain the top quark mass from a total of $N_{ev}$ events, we form the
product of all the event likelihoods, and take negative two times the
logarithm of this product,
\begin{equation}
\label{eq:multilike}
\bm \Lambda(M_{top}) = -2\ln\left( \prod _{i=1}^{N_{ev}}
\bm{L}(M_{top}|\bm{y}^{(i)})\right).
\end{equation}
Then we obtain the top quark mass as the maximum likelihood estimate 
of $M_{top}$,
\begin{equation}
\label{eq:alphaOb}
\hat{M}_{top} = M_{top}\mbox{~at the minimum of ~}
\bm \Lambda(M_{top}), 
\end{equation}
and its uncertainty from the points where $\Delta \bm \Lambda = 1$.

\subsection{\label{sec:matrix} Matrix Element Calculation in the lepton+jets channel}
The matrix element squared $|{\cal M }|^{2}$ is factorized into 3
parts: (1) $t\bar t$ production ($| {\cal M}_{t \bar t} |^{2}$), (2)
the propagators of the top and anti-top (${\cal P}_{tl}$ and ${\cal
P}_{th}$), and (3) the decay matrices, $| {\cal M}_{tl} |^{2}$ and $|
{\cal M}_{th} |^{2}$, for leptonic and hadronic top decays,
respectively. Namely,
\begin{equation}
| {\cal M } |^{2} = | {\cal M}_{t \bar t}|^{2} {\cal
P}_{tl} {\cal P}_{th} | {\cal M}_{tl}
|^{2} | {\cal M}_{th} |^{2}.
\end{equation}
The production matrix element for the $q\bar q$ initial state at
leading order~\cite{Ref:ME1}--\cite{Ref:ME3} is
\begin{equation}
| {\cal M}_{t \bar t}(q\bar q \to t\bar t) |^{2} =
\frac{2g_s^4}{9}(2-\beta^2 \sin^2 \theta^*),
\end{equation}
where $\theta^*$ is the angle between the top quark and the incident
quark in the proton in the $t \bar t$ center of mass system, $\beta$
is the velocity of the top quark and $g_s$ is the strong coupling
constant.

For the $gg$ initial state~\cite{Ref:ME1}--\cite{Ref:ME3}, the matrix
element can be expressed as
\begin{equation}
| {\cal M}_{t \bar t}(gg \to t\bar t) |^{2} =
g_s^4(\frac{1}{6\tau_1\tau_2}-\frac{3}{8})(\tau_1^2+\tau_2^2+\rho-
\frac{\rho^2}{4\tau_1\tau_2}
),
\end{equation}
where
\begin{equation}
\tau_1 = \frac{2(g_1 \cdot t)}{\hat s}, \tau_2 = \frac{2(g_2 \cdot
t)}{\hat s}, \rho = \frac{4M_{top}^2}{\hat s}, {\hat s} = (g_1 + g_2)^2,
\end{equation}
 $g_1$ and $g_2$ are the incident gluon momenta in the proton and
 anti-proton, and $M_{top}$ is a free parameter for the top quark mass.
 In these equations, the $t\bar t$ spin correlations have been
 ignored.  This effect is included in the mapping functions described
 in Section~\ref{sec:mapping} since spin correlations are included in
 the HERWIG Monte Carlo samples that are used to
 make the mapping functions.  Since we don't know what the initial
 state was, the likelihoods for the two processes ($q\bar q$ and $gg$)
 are summed up in the event likelihood with the appropriate PDF
 weights obtained from CTEQ5L.

The propagators for the top and anti-top quarks are as specified by
Eq.~(\ref{eq:propfac}) in which $M_r$ corresponds to $M_{top}$ and
$s_r$ is the invariant masses of the leptonically ($tl$) or hadronically
decaying top quark ($th$).

The decay matrix elements for the leptonic and hadronic channels
are given by
\begin{eqnarray}
\label{eq:decaylep}
| {\cal M}_{tl} |^2 &=& 4g_w^4\frac{(t\cdot l)(b\cdot
\nu)}{(S_{l \nu}-M^2_W)^2+M^2_W\Gamma^2_W}, \\
\label{eq:decayhad}
| {\cal M}_{th} |^2 &=&
4g_w^4\frac{1}{2}\sum_{i\leftrightarrow j}^2\frac{(\bar
t\cdot q_i)(\bar b\cdot q_j)}{(S_{2j}-M^2_W)^2+M^2_W\Gamma^2_W},
\end{eqnarray}
where $S_{l\nu}$ and $S_{2j}$ represent the invariant masses squared
of the lepton+neutrino and the two quarks from the $W$ respectively.
For the mass and decay width of the $W$, we assume the world average
values, $M_{W}$ = 80.4 GeV/$c^2$ and $\Gamma_{W}$ = 2.1 GeV/$c^2$. In
Eq.~(\ref{eq:decaylep}), the dot product of $b$ and $\nu$ can be
calculated because the $z$-component of the neutrino momentum,
$\nu_z$, has already been determined in step (4) above.  In
Eq.~(\ref{eq:decayhad}), we make both possible assignments of the two
jets to $q$ and $\bar q'$ from the $W$, and the likelihoods
corresponding to the two possibilities are averaged.

%===============================================================================
\section{\label{sec:chap7}Transfer Functions (TF)}
As described in the preceding section, the transfer functions deal
with the relation between parton and corrected jet energies.  This
allows us to use the full distribution, including tails, of the
fraction of quark energy deposited outside of the jet cone.  Also,
since the generic jet corrections are based on the QCD dijet process,
the transfer functions can correct for $t\bar{t}$-specific $b$ jets
and $W$ daughter jets.

\subsection{Definition and Performance}
The transfer variable set $(x,y)$ we use in this analysis is the
transverse energy of a parton (quark) and the corresponding jet,
\begin{equation}
x=E_{T}(\mbox{parton}),
\;\;\;y=E_{T}(\mbox{jet}),
\end{equation} 
where $E_{T}$(jet) has been corrected with the CDF generic corrections
described in Section~\ref{sec:JetCorr}.
TF's are obtained for $b$ jets and $W$ daughter jets separately and
are applied only to the four highest $E_T$ jets in an event, which are
assumed to come from the $t$ and $\bar{t}$ decay.
 
The TF's are obtained with the following procedure.  We generate
events with the HERWIG and PYTHIA Monte Carlo event generators and a
full detector simulation, and select events with the same criteria as
applied to real data.  From the accepted events, we select those jets
that are within a distance $\Delta R<0.4$ from a final state quark.
Using these ``matched'' jets, we obtain a 2-dimensional density
function of the number of events at $(x,y)$, $D(x,y:M_{top})$.

The number of events in a $dx\,dy$ bin is given by
\begin{equation}
D(x,y:M_{top})dx\,dy = L_{int} \left( \frac{d\sigma}{dx} \right) dx
\times w(y|x||M_{top}) dy
\end{equation}
where $L_{int}$ is the integrated luminosity of the sample.  The
transfer function is obtained by removing the cross section factor
from $D(x,y:M_{top})$, i.e.,
\begin{equation}
\label{eq:wxy}
w(y|x||M_{top})=\frac{1}{n_{x}} D(x,y:M_{top}),
\end{equation}
where 
\begin{equation}
\label{eq:nx}
n_{x}=L_{int} \frac{d\sigma}{dx}=\int D(x,y:M_{top}) dy.
\end{equation}
Values of $n_x$ and $w(y|x||M_{top})$ are numerically obtained from
$D(x,y:M_{top})$ by Eqs.~(\ref{eq:nx}) and (\ref{eq:wxy}),
respectively.

The TF $w(y|x||M_{top})$ depends on $M_{top}$, most significantly
through the event selection criteria. The total selection efficiency
is about 2-3$\%$ depending on an input $M_{top}$; approximately 4,000
events are accepted from the 200,000 events that is the typical size
of the Monte Carlo sample for each top quark mass point. Thus the
statistics of the Monte Carlo samples are not sufficient to obtain an
$M_{top}$-dependent TF. Therefore as an approximation, we use TF's
averaged over the $M_{top}$ search region (130--230 GeV/$c^2$,
sampled every 5 GeV/$c^2$),

\begin{equation}
w(y|x) = \langle w(y|x||M_{top}) \rangle_{M_{top}}.
\end{equation}

The transfer variables $x$ and $y$ are strongly correlated, so we make
a variable transformation from $(x,y)$ to $(\xi,Y)$ as
\begin{equation}
\label{eq:xieta}
\xi = \frac{x-y}{x},\;\;\;
Y = y. 
\end{equation}
The TF for variables $(\xi,Y)$ is defined by
\begin{equation}
f(\xi,Y)\,d\xi\,dY=\frac{1}{n_{x}}\,D(x,y)dx dy.
\end{equation}
In practice, $f(\xi,Y)$ is obtained by filling a $(\xi,Y)$ histogram
with weight $1/n_{x}$ for each Monte Carlo event.  We call the
variable $\xi$ a ``response variable'' in this paper.  In the function
$f(\xi,Y)$, $\xi$ and $Y$ are much less correlated than $x$ and $y$ in
$w(y|x)$, so wider bins can be used in $Y$.  In the reconstruction of
parton kinematics, $\xi$ is generated from the observed value of
$Y(=y)$ according to $f(\xi,Y)$, and $x$ is then determined by
Eq. (\ref{eq:xieta}).

An advantage of deriving the TF from Monte Carlo events is that the 
effect of the detection efficiency and acceptance is automatically included 
in the determination of the TF.

\begin{figure}[htbp]
\begin{center}
\includegraphics[width=1\columnwidth]{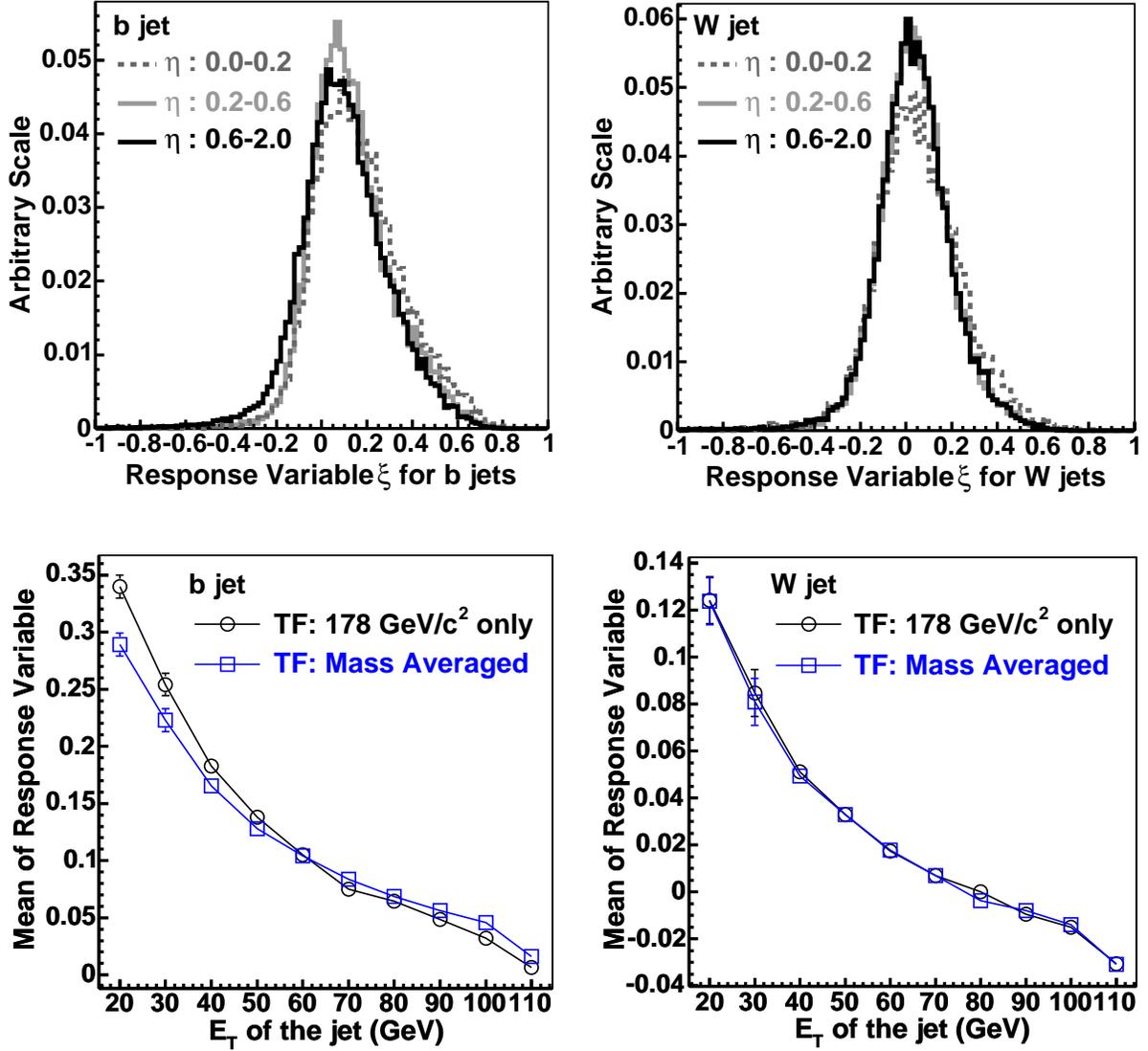}
\end{center}
\caption{The distributions of the response variable $\xi$ for $b$ jets
(upper-left) and $W$ jets (upper-right) for different ranges of jet
$\eta$.  Each distribution is normalized to unit area.  The jet $E_T$
dependence is shown in the lower-left and lower-right for $b$ and $W$
jets, respectively, by plotting the mean of the response variables as
a function of jet $E_T$.}
\label{Fig:TF_bandW}
\end{figure}

As illustrated in Fig.~\ref{Fig:TF_bandW}, the TF's strongly depend on
the $E_T$ and slightly depend on pseudorapidity $\eta$ of the jets.
Therefore we calculate TF's in 10 bins of jet $E_T$ (15 to $>$105 GeV
in 10 GeV steps) and 3 bins of $|\eta|$ (0.0-0.2-0.6-2.0) that
correspond to different regions of the
calorimeter~\cite{Ref:CDF}. Thus separately for $b$ and $W$ jets, we
make thirty histograms. In each bin, the mass averaged TF contains
5,000 jets on average, while if we use $M_{top}$-dependent TF, it is
about 250 which is not enough to get smooth functions.  In the figure,
the means of the response variable as a function of $E_T$ are compared
with the transfer functions extracted from only a single mass sample
($M_{top}$ = 178 GeV/$c^2$). The $b$ jet response is lower (higher) at
lower (higher) $E_T$ for the mass averaged TF, while for $W$ jets the
response is almost identical. This is because the $b$ jets, being the
direct daughters of the top quarks, carry more of the mass
information.  By averaging over the samples in a wide mass range, the
top quark mass dependence is reduced without needing the enormous
statistics for making mass dependent TF's.  The response distributions
are asymmetric due to the finite size of the clustering cone.
Consequently we do not fit the distributions with a functional form,
but rather generate random numbers to accurately sample the full
distributions.

To validate the transfer function performance using $t\bar t$ Monte
Carlo samples with different masses, we investigate the invariant
mass of the jet pair from the $W$ and the three jets from the
hadronically decaying top quark using the following procedure.
\begin{enumerate}
\item Jet-parton matching \\To ensure proper assignment of jets to
partons, we require the distance ($\Delta$R) between a jet's direction
and a parton's direction to be less than 0.4. Moreover, if two or more
jets are within $\Delta$R$<$0.4 of a parton direction, we discard the
event.
\item Applying the transfer function
\\ This is performed by random generation of the response variable 
$\xi$ from the given $Y=y$.  Explicitly, the transverse energy of the
parton is obtained by
\begin{equation}
E_{T}(\mbox{parton}) = \frac {E_{T}(\mbox{jet})}{(1-\xi)}
\label{eq:TF1}
\end{equation}
Then the dijet ($W$) and trijet (top) invariant masses are calculated.
The random number generation is repeated more than 50,000 times (we
call this ``scanning'').  After scanning, distributions of the dijet
and trijet invariant masses are obtained for each event.
\item Extracting the invariant mass
\\We calculate the mean of the distribution obtained in step 2 by
fitting the distribution from each event with a Gaussian function and
storing the fitted mean value in a histogram.
\end{enumerate}

\begin{figure}[htbp]
\begin{center}
\includegraphics[width=1\columnwidth]{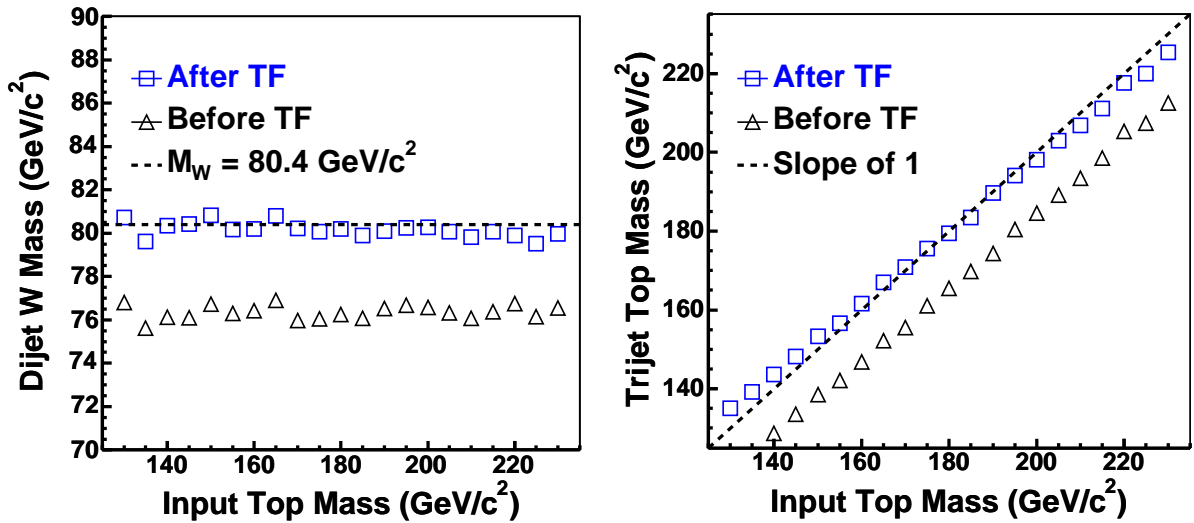}
\end{center}
\caption{Comparisons of reconstructed invariant masses of $W$ dijets
  (left) and top trijets (right) in HERWIG Monte Carlo samples, as a
  function of input top quark mass, before and after the transfer
  function is applied. Dashed lines correspond to the input masses of
  the $W$ and top quark.}
\label{Fig:TFMassWT}
\end{figure}

\begin{figure}[htbp]
\begin{center}
\includegraphics[width=1\columnwidth]{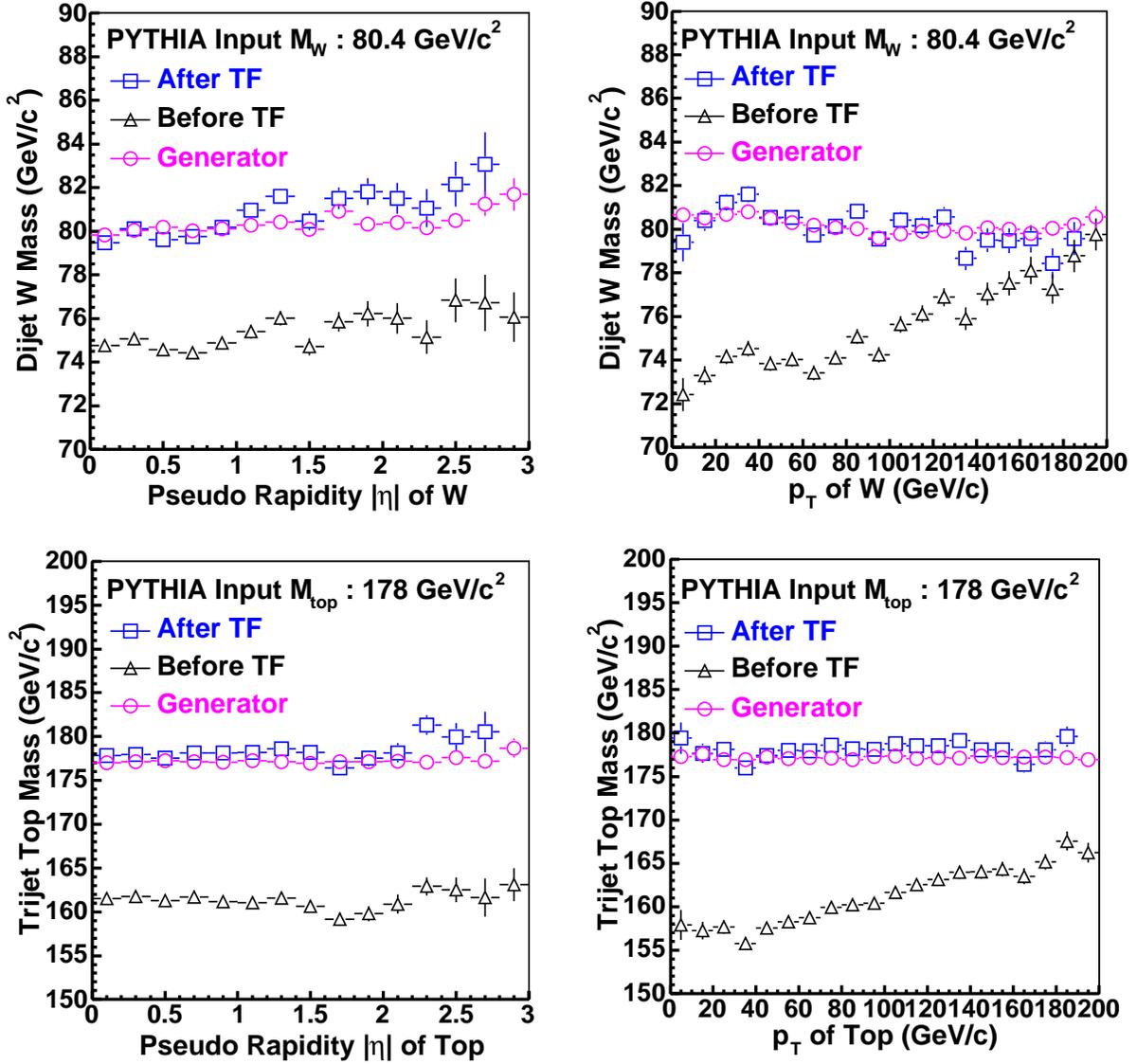}
\end{center}
\caption{$\eta$ and $p_T$ dependence of dijet $W$ (upper) and trijet
top (lower) invariant masses, with the generated masses in PYTHIA Monte
Carlo shown as open circles. Masses with only generic
corrections are shown in open triangles, and the open squares
show the results after TF application.}
\label{Fig:TFMassWTdep}
\end{figure}

The invariant masses of the dijets and trijets before and after
applying the transfer function are shown in Fig.~\ref{Fig:TFMassWT}.
Since the out-of-cone correction is not applied to the masses before
the transfer function, (we start with hadrons within the jet cone and
apply the transfer function to obtain the parton energy), lower masses
are observed, while after the transfer function is applied, the final
values of the mean agree with the generated input masses.  The left
plots in Fig.~\ref{Fig:TFMassWTdep} show the $\eta$ dependence of the
invariant masses, while the right plots show the $p_T$
dependence. There is a large $p_T$ dependence in the plots before the
transfer functions are applied.  The transfer functions, however,
largely eliminate this dependence.

We investigate the Monte Carlo generator dependence by comparing
PYTHIA and HERWIG, which have different fragmentation modeling, but no
significant discrepancies are observed. Since the top quark mass
samples are produced with HERWIG, we make the transfer functions with
HERWIG and examine the generator bias in Section~\ref{sec:chap11}.  We
also check alternative variables that could be used in the response
function: jet $E$, $p$, or $p_T$.  As with the generator dependence,
no differences are found in the shape or mean of the response
functions or in the reconstructed invariant masses of $W$ dijets or
top trijets.

\subsection{\label{sec:met}Missing Transverse Energy}
The ``raw $\not \!\! E_T$'', which is defined in
Eq.~(\ref{eq:metdef}), is corrected by applying generic jet energy
corrections and then the transfer functions. First, the definition of
missing transverse energy is rewritten using the observed objects in
the sample to take into account the generic jet corrections,
\begin{equation}
- \bm{\not \!\! E_T} = \bm{E_T}({\rm lepton})+\sum_{i=1}^{4}\bm{E_T^i}({\rm jet})+\bm{X_T}
\end{equation}
where $\bm{E_T^i}({\rm jet})$ is the $E_T$ of the jet after generic
corrections, and $\bm{X_T}$ corresponds to all other
calorimeter-deposited energies. (Within $\bm{X_T}$, the generic
corrections are also applied to all jets with $E_T>$ 8 GeV and
$|\eta|<$ 2.4.) The above expression shows that the \mbox{$\protect
\raisebox{.3ex}{$\not$}E_T$~} measurement is highly correlated with
the jet energy measurements and corrections. Therefore, it is not
considered to be an independent observable in this analysis. We
calculate the transverse component of the neutrino momentum,
$\bm{\nu_T}$, from the leptonic $W$ decay as
\begin{equation}
\bm{\nu}_{T}  =  \not \!\!\bm{E}_T + \sum_{j=1}^{4} (\bm{E_T}^{j}({\rm jet}) - 
\bm{E_T}^{j}({\rm corr}))
\end{equation}
where $\bm{E_T}^{j}({\rm corr})$ is the jet $E_T$ after generic and
transfer function corrections are applied to each of the leading four
jets.

%===============================================================================
\section{\label{sec:chap8}Top Quark Mass Reconstruction}
This section describes how we extract the top quark mass and checks of
the top quark mass reconstruction using Monte Carlo simulation. In
this analysis, all events are assumed to be signal when the likelihood
is calculated.  The result is then corrected for the presence of
background. Therefore, we first present the behavior of the background
and its effects on signal reconstruction. Based on large sets of
pseudo-experiments with varying background fractions, we derive the
background correction function (``mapping function'') for the top
quark mass parameterized as a function of the background fraction. At
this point this method is fully calibrated with the Monte Carlo sample.
     
\subsection{Background Effect on the Likelihood}
As described in Table~\ref{tab:expectedSummary} in
Section~\ref{sec:chap5}, there are various background processes that
may affect this measurement.  We use the ALPGEN
Monte Carlo with the CDF detector simulation to model mistags and $W$
+ heavy flavor events. The $W$ + four light-flavor partons ($W$4p)
process can be used to investigate mistags, since mistags come from a
false secondary vertex, which is mainly due to track and vertex
resolution effects. For non-$W$ (QCD) background, we use a
non-isolated lepton sample (isolation $I>$ 0.2, but $\not \!\!E_T >$
20 GeV) from real data. Other electroweak processes, diboson and
single top production, are modeled by PYTHIA Monte
Carlo samples.  All events are subject to the event selection
described in Section~\ref{sec:chap4}.

The likelihood distribution and the mass-likelihood peak are expected
to be changed by the existence of background events. To understand the
background effects more fully, we first calculate the dynamical
likelihood defined by Eq.~(\ref{eq:evlikedef1}) for each background
sample, and the average joint maximum likelihood masses are estimated
from pseudo-experiments with $\sim$100--1000 events, depending on
background source.  Their values mainly result from the lepton ($E_T
>$ 20 GeV) and jet energy ($E_T >$ 15 GeV) cut thresholds.  The
mistag, $W$+HF, and non-$W$ samples produce almost the same maximum
location in the range of 155--160 GeV/$c^{2}$, while the single top
sample has 170 GeV/$c^{2}$, a slightly higher mass. The diboson
background has a slightly lower mass, around 155 GeV/$c^{2}$, near the
lower limit of the search region (155--195 GeV/$c^2$).  For each
background, the peak width of maximum likelihood masses per event is
much larger than for signal events, and its peak is relatively lower
compared to the top quark mass search range (as shown in
Fig.~\ref{Fig:evMassdataMC} in Section~\ref{sec:chap10}).

\begin{figure}[htbp]
\includegraphics[width=1\columnwidth]{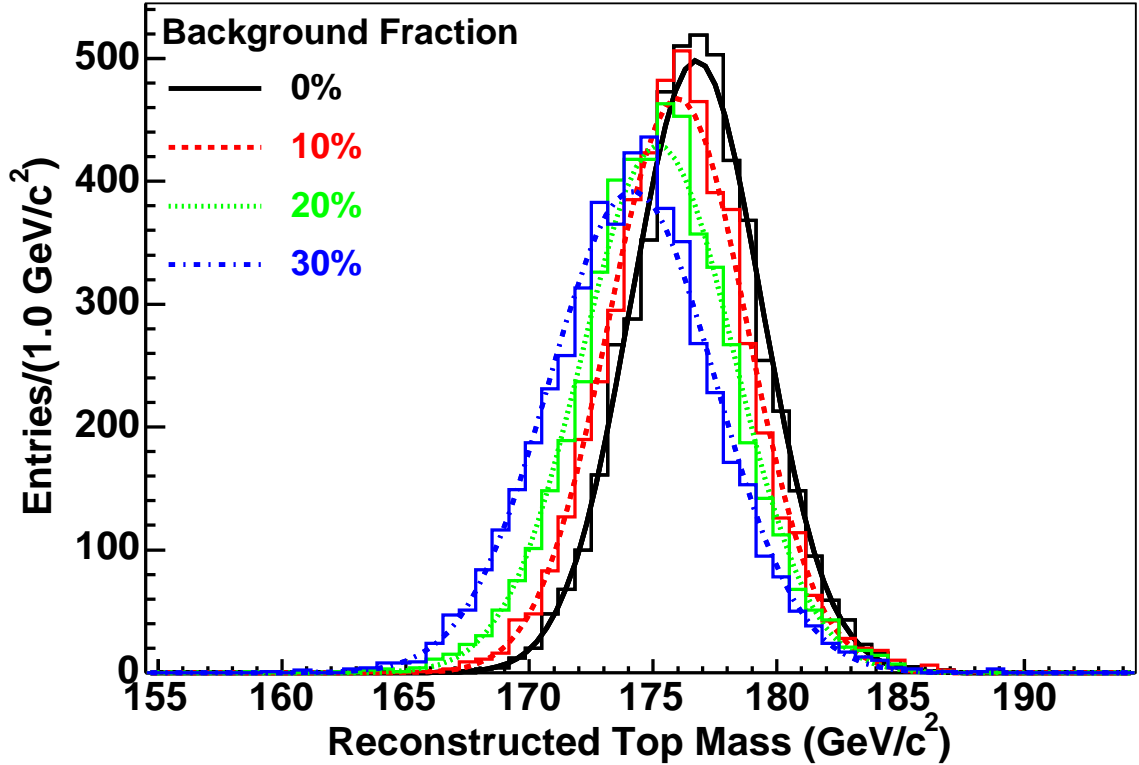}
\caption{An example of mass shift due to background events. The plot shows the 
reconstructed mass (input $M_{top}=178$ GeV/$c^2$), varying the
background fraction from 0$\%$ to 30$\%$ (expected fraction is
14.5$\%$).  For each distribution, 5000 sets of pseudo-experiments
that contain 63 events each are performed. Each distribution is fitted
with a Gaussian function.}
\label{Fig:bkgshiftill}
\end{figure}

The effect of background on top quark mass extraction is seen in
Fig.~\ref{Fig:bkgshiftill}, which shows the reconstructed top quark
mass from 63-event pseudo-experiments as a function of the background
fraction. The peak mass is shifted lower and the width broadens as the
background fraction increases.

\begin{figure}[htbp]
\includegraphics[width=1\columnwidth]{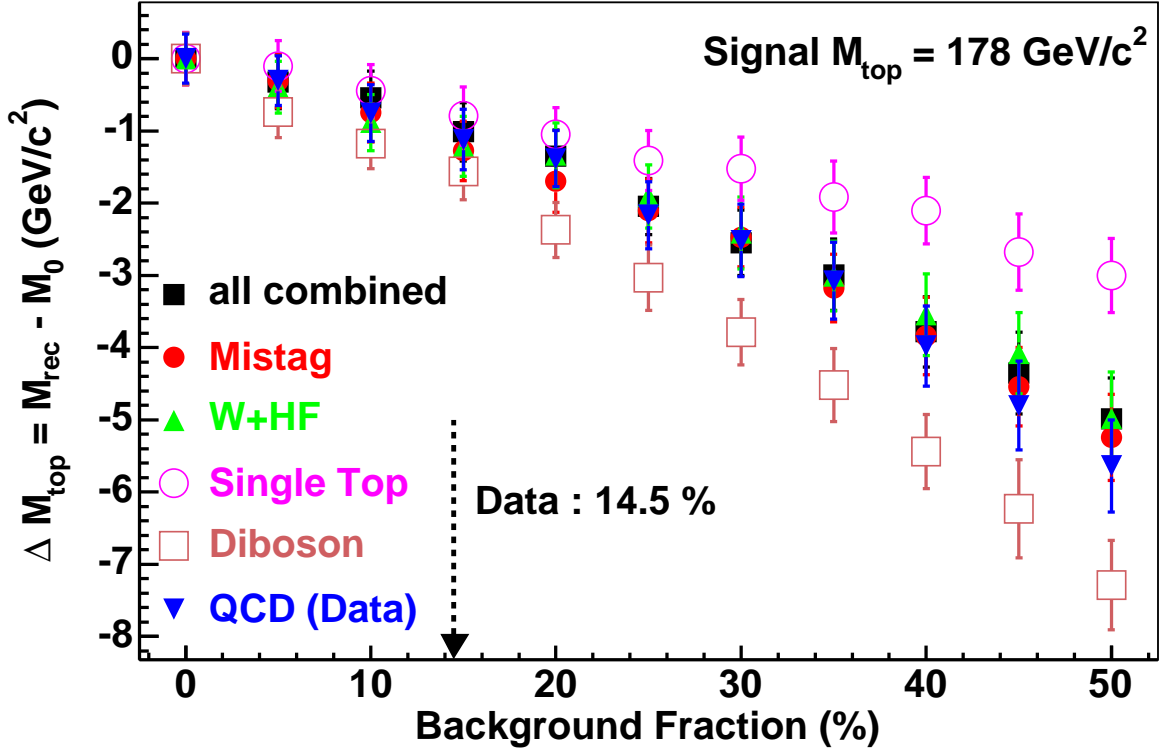}
\caption{The difference between the reconstructed mass
  ($M_{\mathrm{rec}}$) and $M_{\mathrm{0}}$, the mass at 0$\%$
  background ($\sim$ 177.5 GeV/$c^2$), due to individual background
  sources, using a signal sample of $M_{top}$=178 GeV/$c^2$, as a
  function of the background fraction. The closed squares represent
  the combined background using the expected composition from
  Table~\ref{tab:expectedSummary}. The expected background fraction of
  14.5$\%$ is shown as the dashed line.}
\label{Fig:BkgEachEff}
\end{figure}

It is important to know the effect of each of the backgrounds on the
mass determination in order to properly account for the background
composition uncertainty. Figure~\ref{Fig:BkgEachEff} shows, for 178
GeV/$c^2$ $t\bar t$ Monte Carlo, how the reconstructed mass is shifted
from the input mass by individual background sources as the background
fraction is varied over the range 0--50$\%$. This is done with
pseudo-experiments having 63 total events, where the number of
background events is fluctuated using Poisson statistics. We do not
see significant differences among the $W$+HF, mistag, and non-$W$
(QCD) samples, which in sum account for more than 90$\%$ of the
background and hence dominate the total background (the solid squares
in Fig.~\ref{Fig:BkgEachEff}).  Thus the size of the mass-shift
produced by the background is not sensitive to the relative fractions
of $W$+HF, mistag and non-$W$. On the other hand, the single top
sample produces a smaller negative shift and diboson events a slightly
larger negative shift compared to the dominant sources of
mistag/$W$+HF/non-$W$.  Each of these two sources is responsible for
approximately 5$\%$ of the total background.

In summary, background reduces the likelihood peak mass. We evaluate
the size of these mass shifts and derive a correction, the ``mapping
function'' discussed in the next section.

\subsection{\label{sec:mapping} The mapping function}
\begin{figure}[htp]
\includegraphics[width=1\columnwidth]{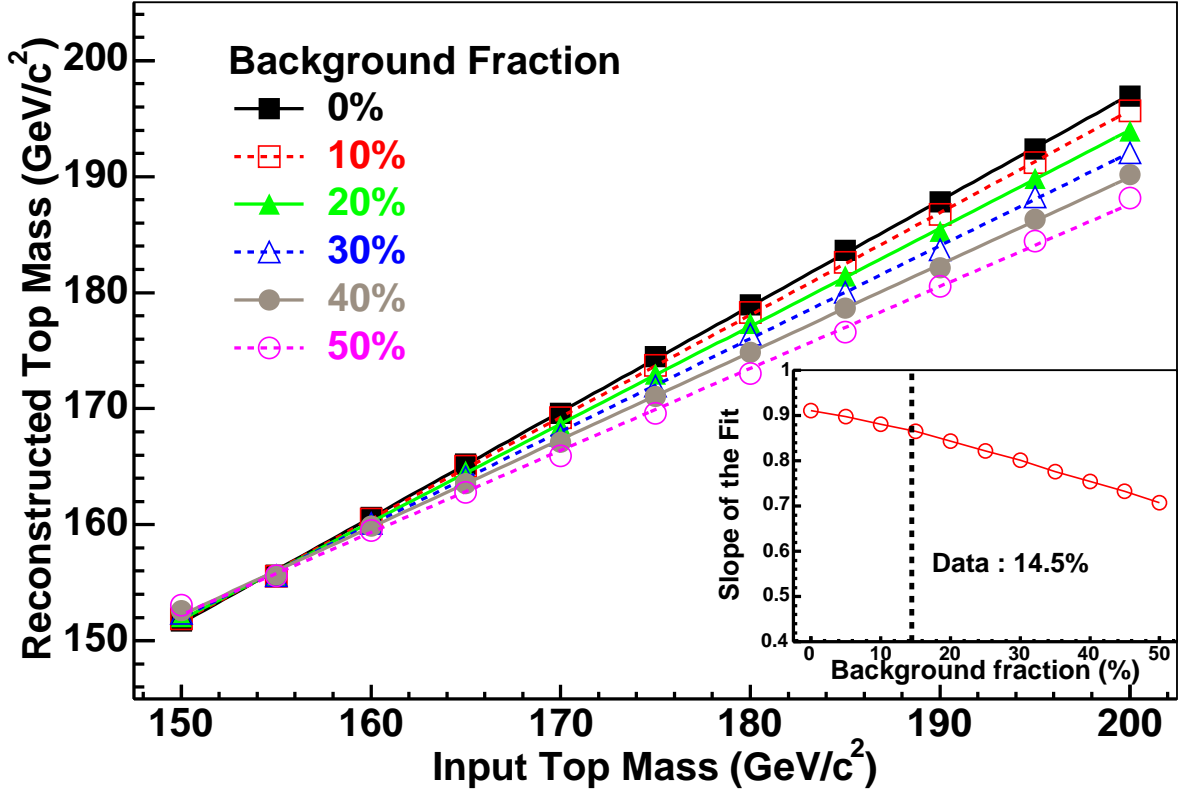}
\caption{The reconstructed mass obtained from the mean of the pseudo-experiments, 
as a function of input mass with the background fraction varying from
0$\%$ to 50$\%$.  The inset shows the slope of the linear fit to the
mapping function as a function of the background fraction. The
expected background fraction of 14.5$\%$ in this data sample is shown
as the dashed line.}
\label{Fig:Mapbkg}
\end{figure}

There are two sources that cause the input top quark mass and the
reconstructed top quark mass to differ. One is the top quark mass
dependence of the transfer function, and the other is the effect of
background. We combine the two effects into a single mass-dependent
correction factor, the mapping function, which is obtained from many
sets of pseudo-experiments.  Figure~\ref{Fig:Mapbkg} shows the
reconstructed top quark mass as a function of its input mass for
various background fractions. The background fraction ranges from
0$\%$ to 50$\%$, where the relative fraction of each background is
that given in Table~\ref{tab:expectedSummary}.  In each
pseudo-experiment, the number of events from each background source
and the total number of events are Poisson fluctuated.  As one can see
in the figure, even with 0$\%$ background the reconstructed top quark
mass does not have unit slope.  This is due to a small top quark mass
dependence of the transfer function as well as to the effect of gluon
radiation and the contamination of the data sample from other top
quark decay modes.  As expected from the background study, the
reconstructed top quark mass is shifted lower as the background
fraction increases. The inset of Fig.~\ref{Fig:Mapbkg} shows the slope
of the linear fit ($p_0$ of $p_0\cdot x+p_1$) to the mapping functions
as a function of background fraction. One can see very stable behavior
up to background fraction of 50$\%$. The estimated background fraction
of 14.5$\%$ is used to extract the top quark mass.

\subsection{\label{sec:methodcheck} Method Check}
\begin{figure}[htbp]
\includegraphics[width=1\columnwidth]{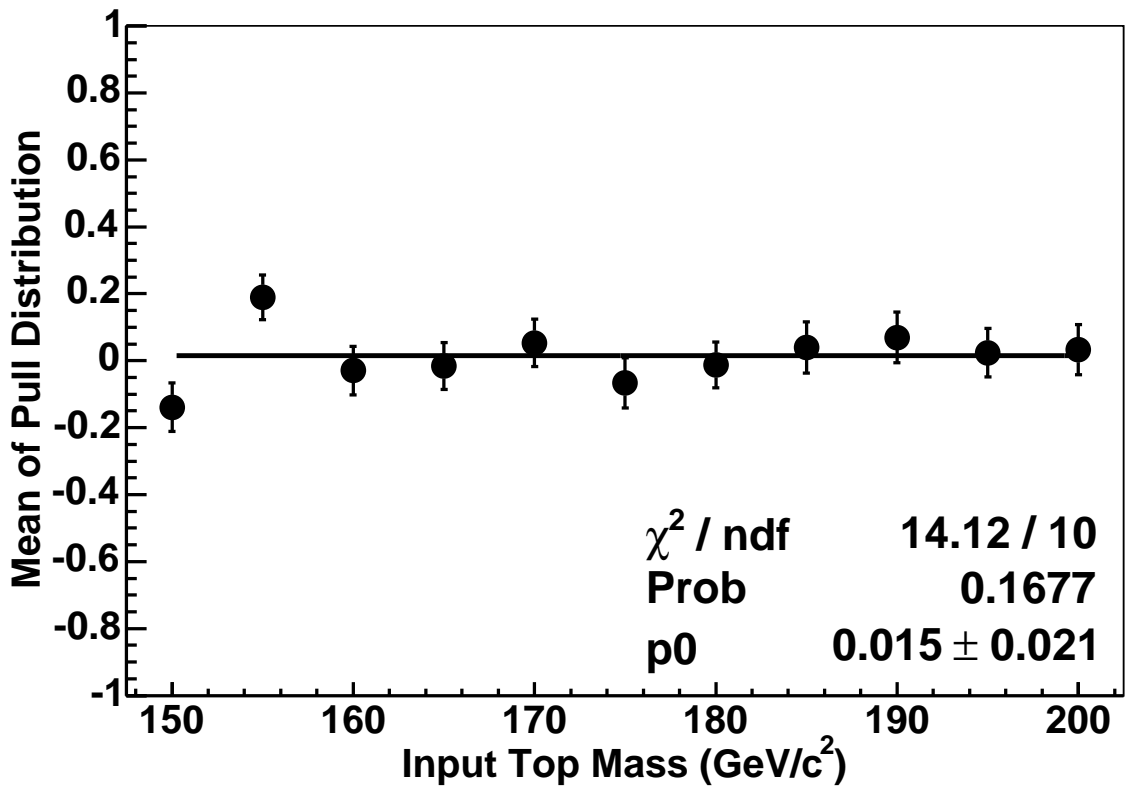}
\caption{The mean (center) of the pull distribution as a function of the input top quark mass 
is consistent with zero, as shown in the figure.}
\label{Fig:SigBkgSanity}
\end{figure}

\begin{figure}[htbp]
\includegraphics[width=1\columnwidth]{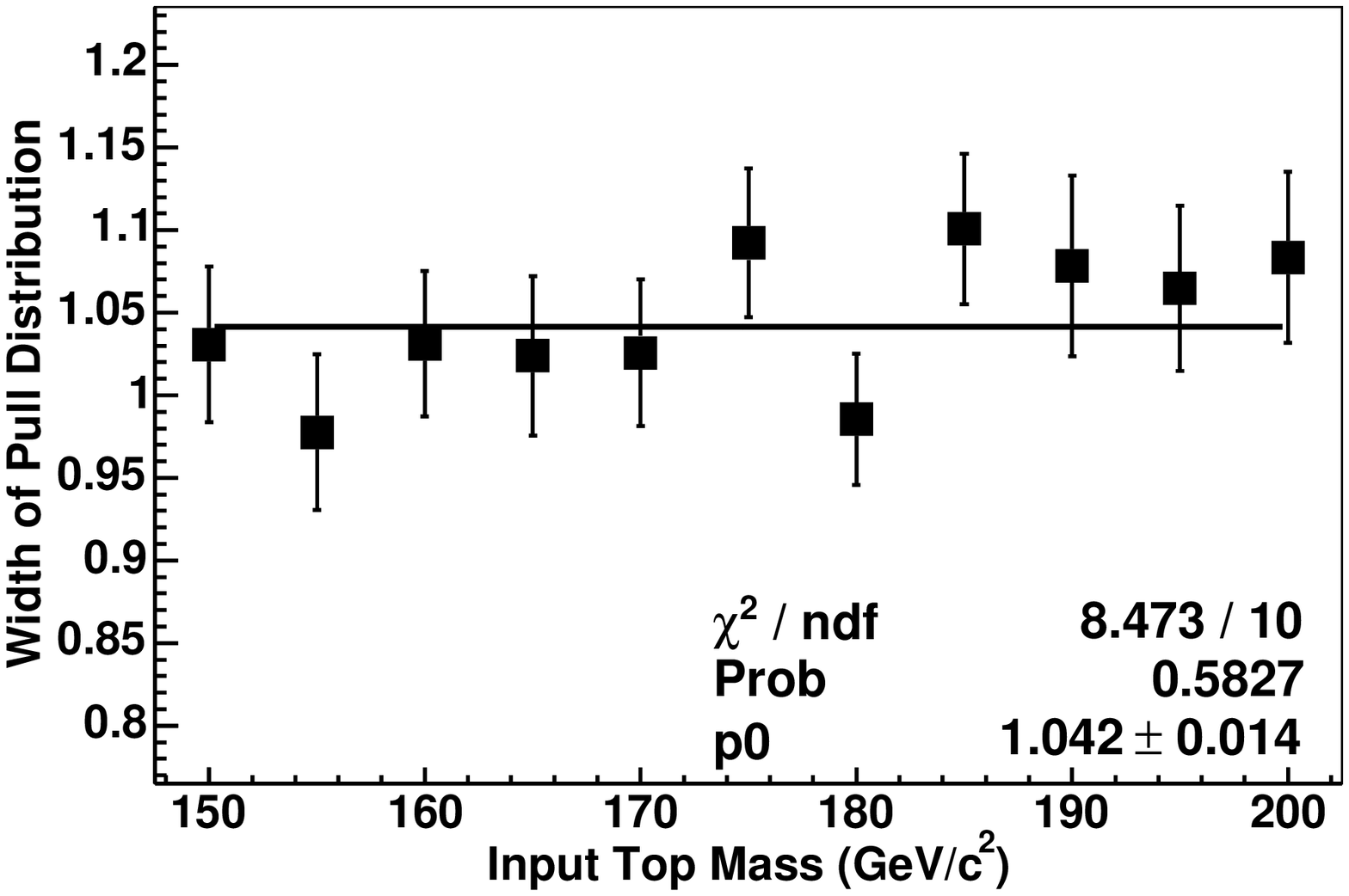}
\caption{The width of the pull distribution as a function of the input top quark mass 
is consistent with a horizontal line fit ($p0$ = 1.042), as shown in the figure.}
\label{Fig:RecSigBkgCW}
\end{figure}

The method described above is tested for possible systematic bias by
running large numbers of pseudo-experiments using Monte Carlo samples.
Each set of 63 events (mean) in a pseudo-experiment consists of on
average 53.8 signal events and 9.2 background events, with each source
Poisson fluctuated.  For each pseudo-experiment, the fit of the $-2\ln
L$ distribution provides a measured top quark mass as well as the
positive and negative uncertainties by fitting with a second order
polynomial with different curvature on the positive and negative sides
(four parameters). After applying mapping functions for a 14.5$\%$
background fraction to each pseudo-experiment, we obtain a slope
consistent with unity (0.997 $\pm$ 0.006) between the input and
reconstructed masses. 
A pull distribution, defined as the input top quark mass minus the 
reconstructed mass divided by its estimated uncertainty, is generated for
each of 11 different input top quark mass samples, where each mass
point is generated from 1000 pseudo-experiments and then is fitted
with a Gaussian function to extract the center and the width of the
pull distribution. The center of the pull distribution is consistent
with zero (0.015 $\pm$ 0.021) as illustrated in Fig.~\ref{Fig:SigBkgSanity}.  
The width of the pull distribution as
a function of the top quark mass is shown in
Fig.~\ref{Fig:RecSigBkgCW}. It is seen that the pull widths are
slightly larger than one (1.042 $\pm$ 0.014). This is because this technique 
assumes that all events are from $t\bar t$ signal. When backgrounds, other decay
channels or extra gluon radiation contaminate the data sample, our
assumption is violated and the reported uncertainties will not
necessarily be correct.  This effect is observed in the pull
width. Therefore we correct the final statistical uncertainty in order
to have a pull width equal to one, corresponding to 68$\%$ coverage in
Gaussian statistics, by scaling the reported uncertainties. The scale
factor of 1.04 is extracted by fitting the pull width over the full
range of true top quark mass.  After applying the mapping function and
scaling the statistical uncertainty, we conclude that the top quark
mass is reconstructed without bias, over a wide range of input masses.

%=======================================================================
\section{\label{sec:chap9}The results from the data}
\begin{figure}[htbp]
\includegraphics[width=1\columnwidth]{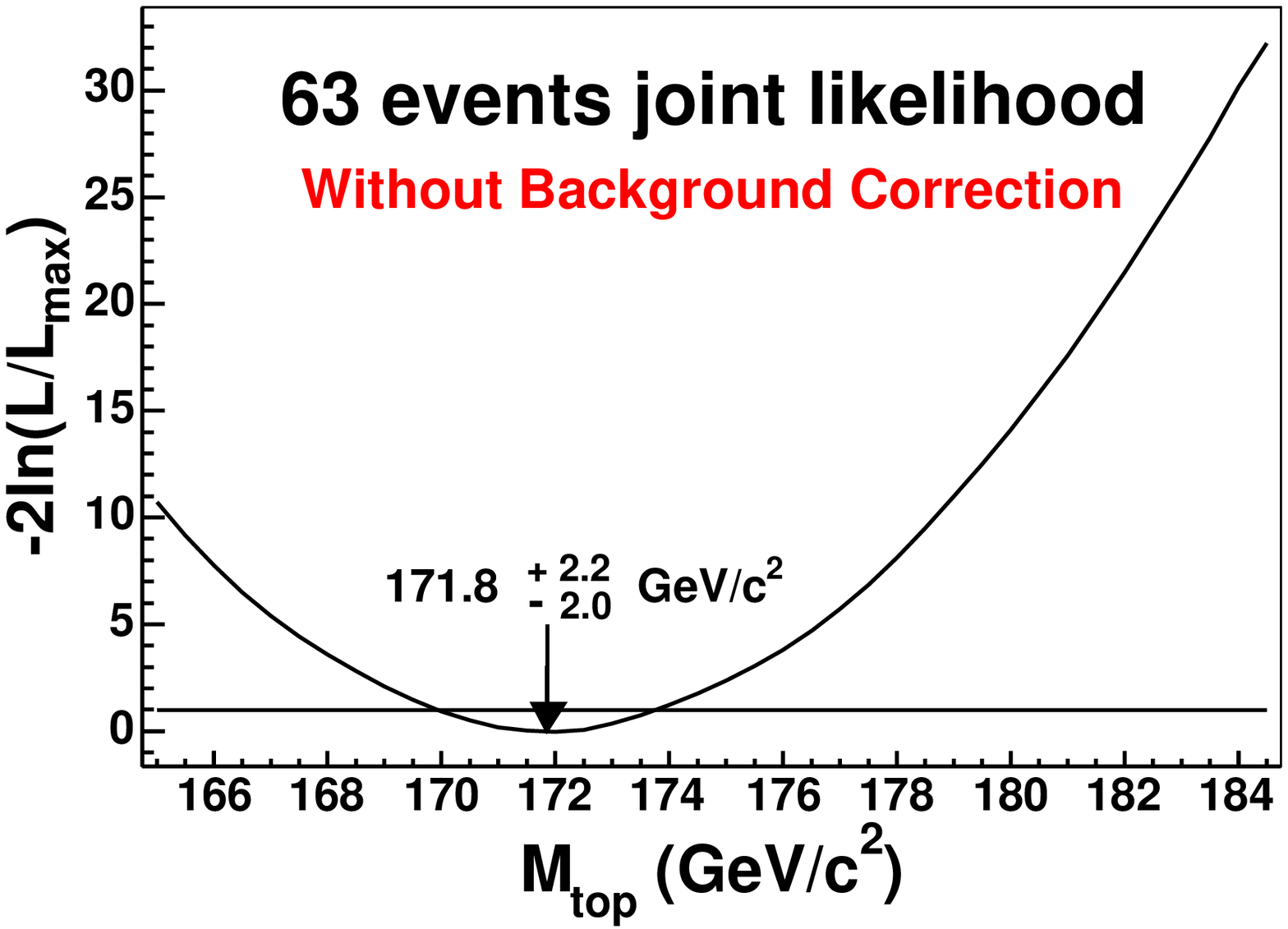}
\caption{ The joint negative log likelihood distribution of the 63
events observed in the data. The fit gives $M_{top} = 171.8$
$^{+2.2}_{-2.0}$ GeV/$c^{2}$, before any corrections.}
\label{Fig:Datafit}
\end{figure}

\begin{figure}[htbp]
\includegraphics[width=1\columnwidth]{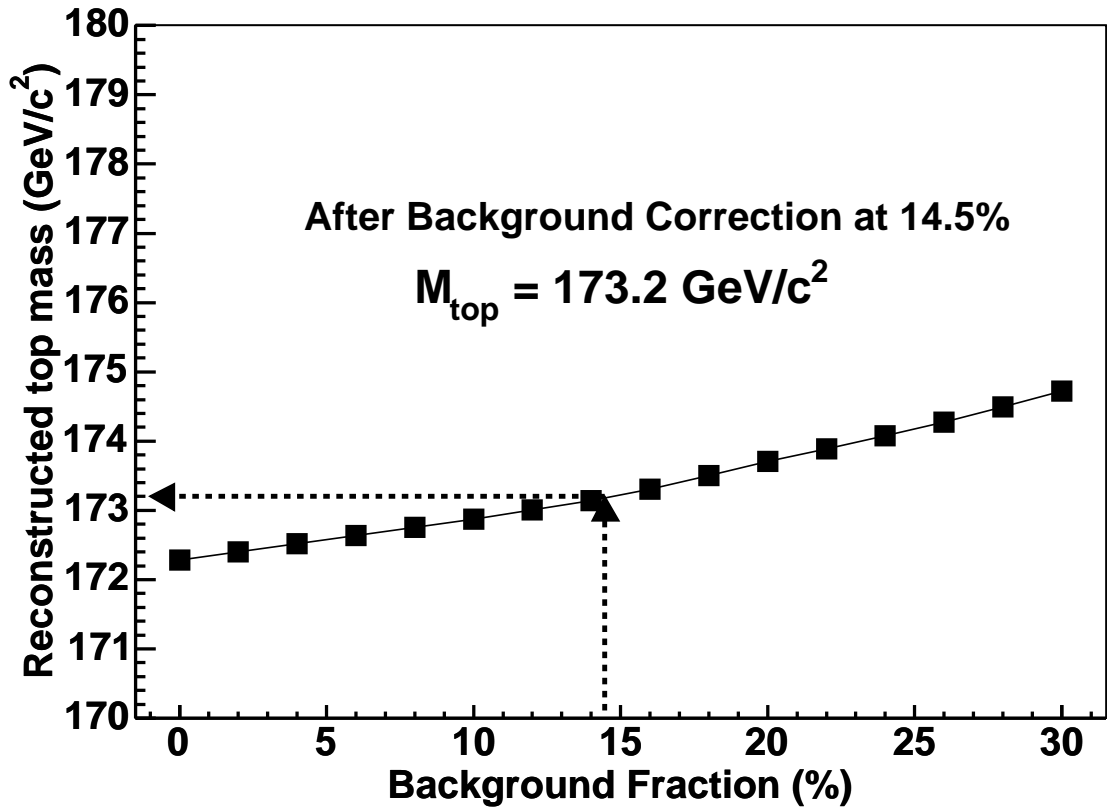}
\caption{Extracted top quark mass using the mapping function as a function of the
background fraction.}
\label{Fig:DataFrac}
\end{figure}

We have 63 $t\bar{t}$ candidate events passing the event selection
criteria.  The joint likelihood of these events is shown in
Fig.~\ref{Fig:Datafit}. From the fit, we obtain $M_{top} = 171.8$
$^{+2.2}_{-2.0}\hspace{0.1cm}
\mathrm{(stat.~only)}$ GeV/$c^{2}$, assuming there is no background. We 
then apply the mapping function to remove the mass-pulling effect of 
the background. Figure~\ref{Fig:DataFrac} shows the extracted top
mass as a function of the background fraction. The top quark mass changes 
by $+$1.4 GeV/$c^2$ for a background fraction of 14.5$\%$.

\begin{figure}[htbp]
\includegraphics[width=1\columnwidth]{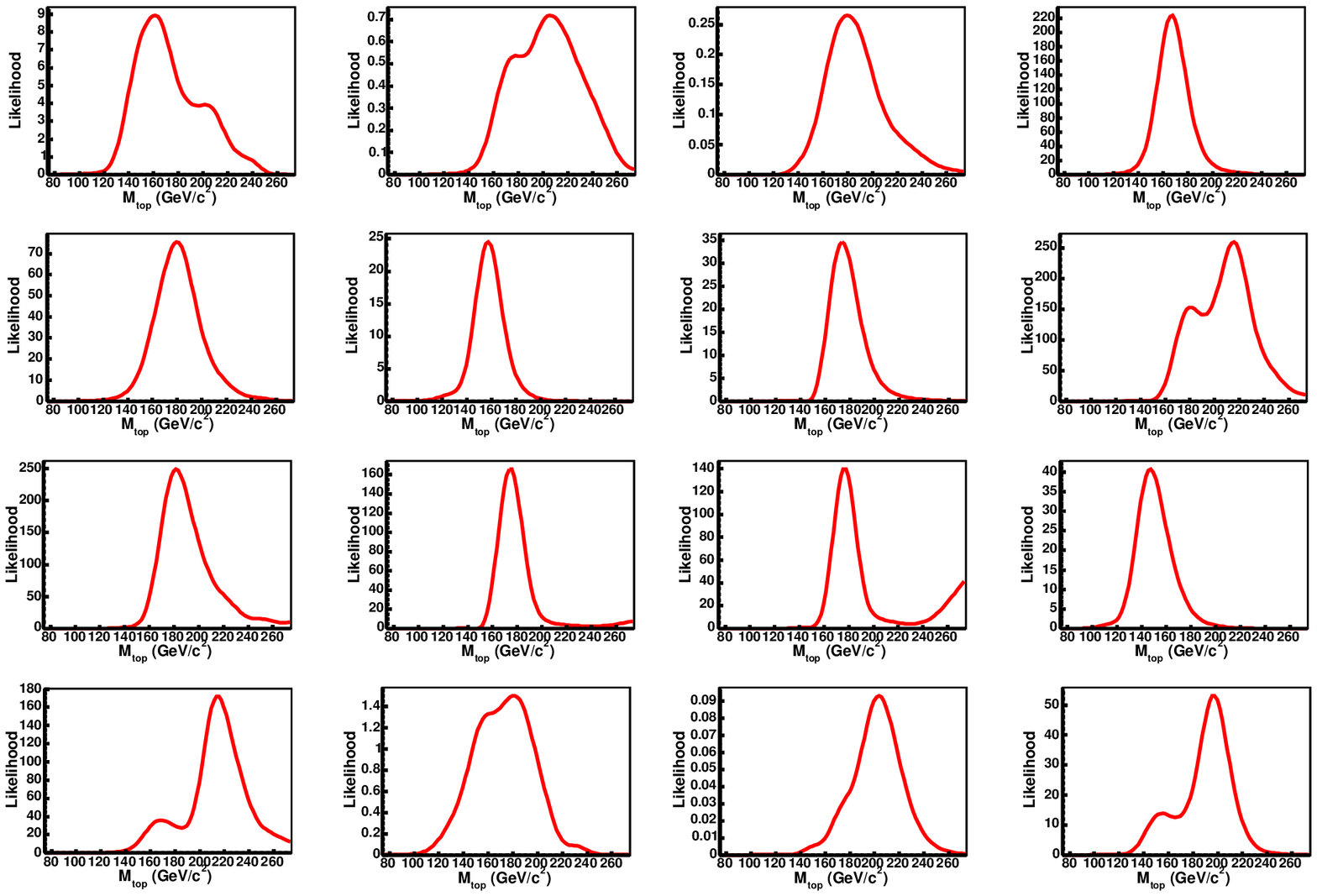}
\caption{Event likelihood distributions as a function of the top quark mass 
for the 16 double $b$-tagged events in the data.}
\label{Fig:Data2tag}
\end{figure}

For the final result, we use the estimated 14.5$\%$ background
fraction, which gives $M_{top} =173.2$
$^{+2.6}_{-2.4}\hspace{0.1cm}\mathrm{(stat.~only)}$ GeV/$c^{2}$. The
statistical uncertainty is also scaled by the slope of the mapping
function mass shift extracted from Fig.~\ref{Fig:Mapbkg} and by 1.04
from the pull width in Fig.~\ref{Fig:RecSigBkgCW}.
Figure~\ref{Fig:Data2tag} shows the likelihood distribution for each
of the 16 data events containing two $b$-tagged jets. Some of these
events have two or three peaks because we sum up all combinations,
each of which could produce a different maximum likelihood point.  For
these 16 events, backgrounds are expected to be small ($\sim$1.4
events) since two $b$ jets are tagged.

\begin{figure}[htbp]
\begin{center}
\includegraphics[width=1\columnwidth]{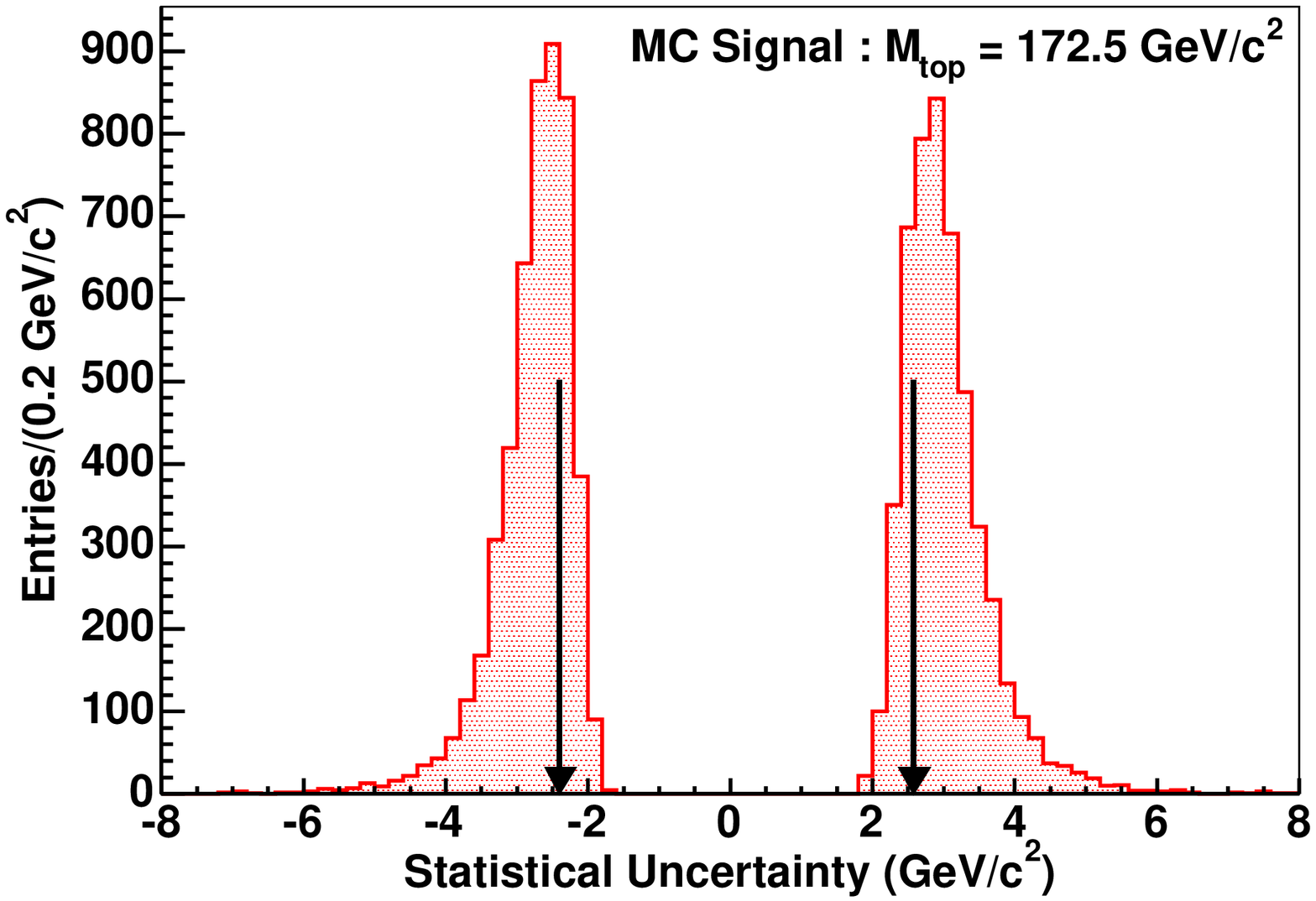}
\caption{The expected positive and negative statistical uncertainties
from pseudo-experiments using 172.5 GeV/$c^2$ $t\bar{t}$ Monte Carlo
samples with the same number of events as in the data.  The arrows
indicate the positive and negative uncertainties for the data.  19
$\%$ of the pseudo-experiments have smaller uncertainties than those
in the data.}
\label{Fig:PseExpAmap}
\end{center}
\end{figure}

To test how likely the reported statistical uncertainty is, we
generated a set of Monte Carlo pseudo-experiments at a top quark mass
of 172.5 GeV/$c^2$ (the closest mass sample to the measured mass),
with the number of events in each subsample equal to that observed in
the data.  Figure~\ref{Fig:PseExpAmap} shows the expected negative and
positive statistical uncertainties. The arrows indicate the
statistical uncertainties for the fit to the data.  The probability of
having a smaller uncertainty than that from data is estimated to be
19$\%$.

\begin{figure}[htbp]
\includegraphics[width=1\columnwidth]{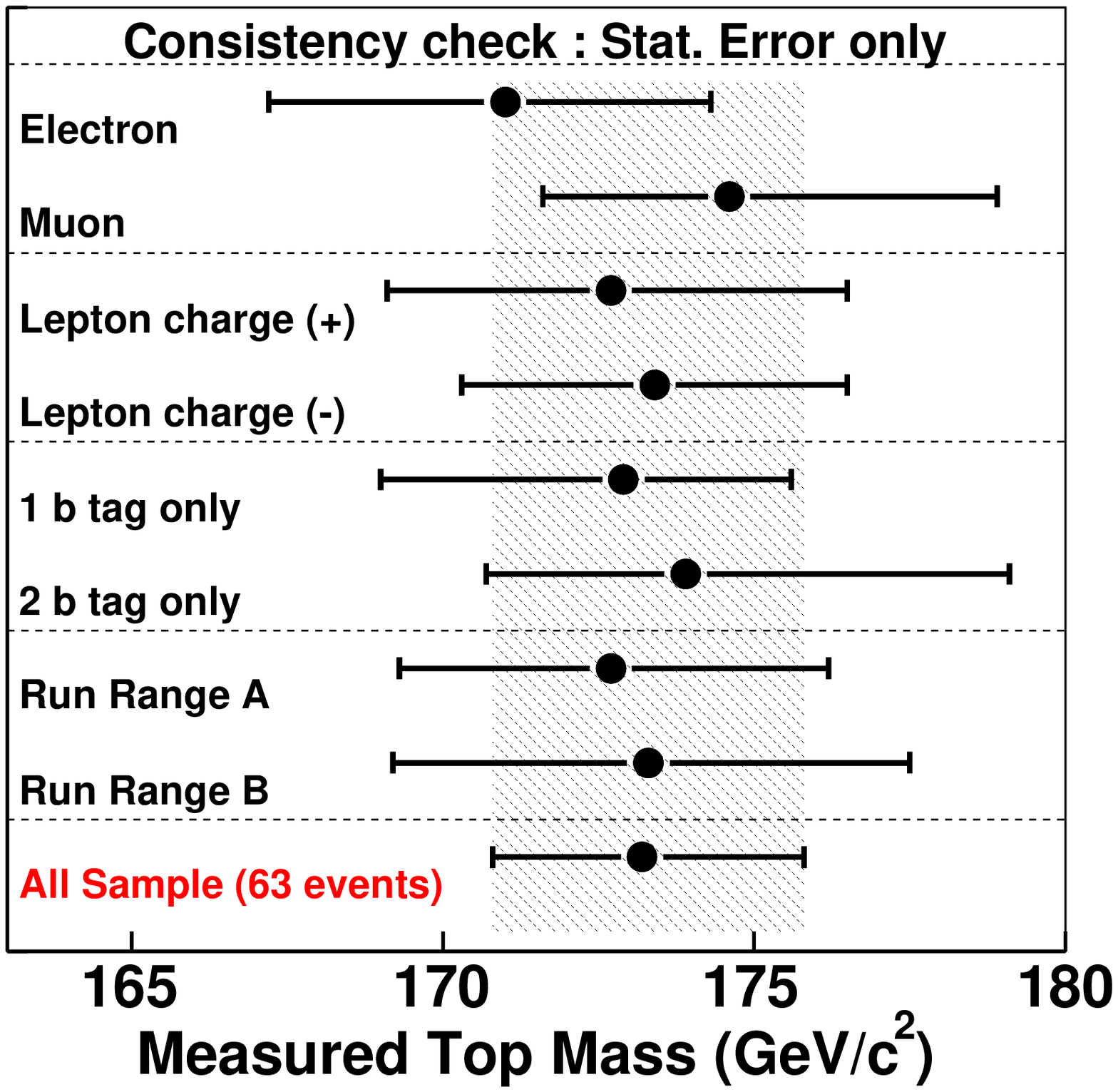}
\caption{Consistency checks: Comparisons between (1) electron and muon
channel, (2) lepton charge ($\pm$), (3) one $b$-tag and two $b$-tag
events, (4) run period A which collected data until September 2003 and
run range B which is after September 2003. The corresponding
integrated luminosities are roughly the same between the two run
ranges. Each point includes the statistical uncertainty only.}
\label{Fig:crosschecks}
\end{figure}

As a consistency check, the top quark mass is measured using different
subsamples to ensure the robustness of the final result. The analysis
procedure applied to these measurements is the same as the one used
for the entire data sample. Figure~\ref{Fig:crosschecks} shows the
resulting top quark mass for the various categories. Comparisons are
made by splitting the events into (1) electron and muon channel, (2)
lepton charge ($\pm$), (3) 1 $b$-tag and 2 $b$-tag events, (4) run
period A which collected data until September 2003 and run range B
with data accumulated after that date. The corresponding integrated
luminosities are roughly the same for the two run ranges.  The same
mapping function is used to estimate the mass in each category using
the expected background fraction of 14.5$\%$ except that a background
fraction of 9 $\%$ (1.4/16) is used for 2 $b$-tag events in category
(3).  1.4 and 16 are the expected number of background 2 $b$-tag
events and the number of 2 $b$-tag events observed in the data,
respectively.  Although inconsistencies would indicate the presence of
new physics in this mode, or perhaps problems with the analysis
method, the Monte Carlo modeling, or detector performance, all results
are consistent with each other and with the default measurement.

%===========================================================================
\section{\label{sec:chap10}Cross checks}
In order to ensure that the method, calibrated by Monte Carlo
samples, describes the data correctly as well as to check how well the
Monte Carlo itself models the data, we compare various variables for
the data with the Monte Carlo predictions for combined signal and
background with regard to (1) the absolute likelihood, (2) the
maximum-likelihood top quark mass, (3) the maximum-likelihood hadronic
$W$ mass, and (4) transfer functions. The normalization of these
comparisons is done in the same way, using the expected numbers of
events of 9.2 for background and 53.8 for the signal,
giving the observed 63 events in total.

\subsection{Absolute Likelihood Value}
\begin{figure}[htbp]
\includegraphics[width=1\columnwidth]{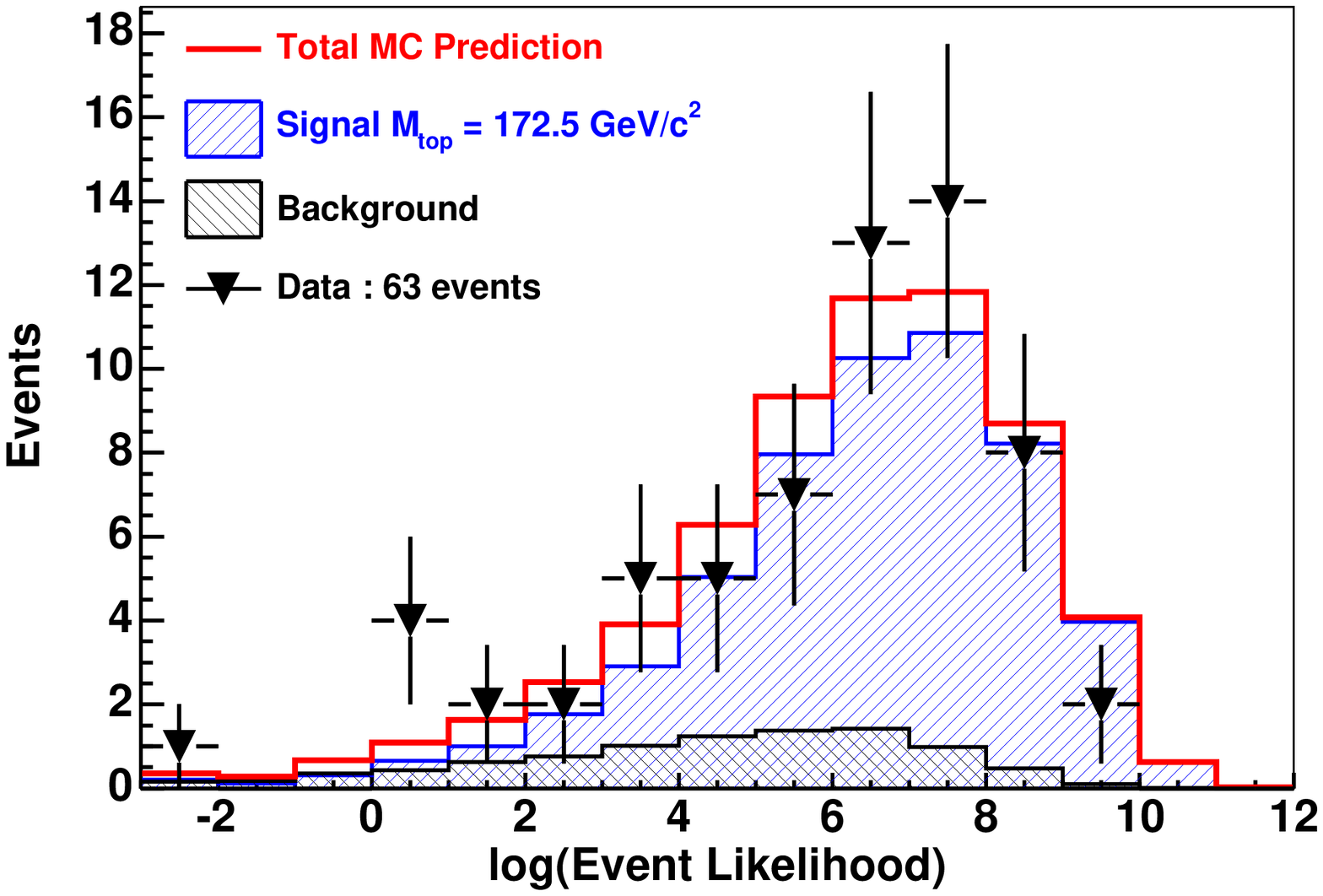}
\caption{Event likelihood distribution.  The number of signal and background 
events is normalized to 63, the number of observed events. The Monte
Carlo signal, background, and the combined predictions are shown as
histograms.  The triangles are the 63 data events.}
\label{Fig:evLikedataMC}
\end{figure}

Although the absolute value of the likelihood in DLM is arbitrary, we
can compare the Monte Carlo with the data.  The signal likelihood for
the $i$-th event is defined as
\begin{equation}
\label{eq:evlikeabs}
L^{i}_{event} = \int L^{i}(M)dM,
\end{equation}
where the integration is over the search region 155--195
GeV/$c^{2}$. Figure~\ref{Fig:evLikedataMC} shows the comparison of the
log of the event likelihoods in the data and the Monte Carlo
samples. We find good agreement between the data and Monte Carlo.

\subsection{Maximum Likelihood Top Quark Mass}
\begin{figure}[htbp]
\includegraphics[width=1\columnwidth]{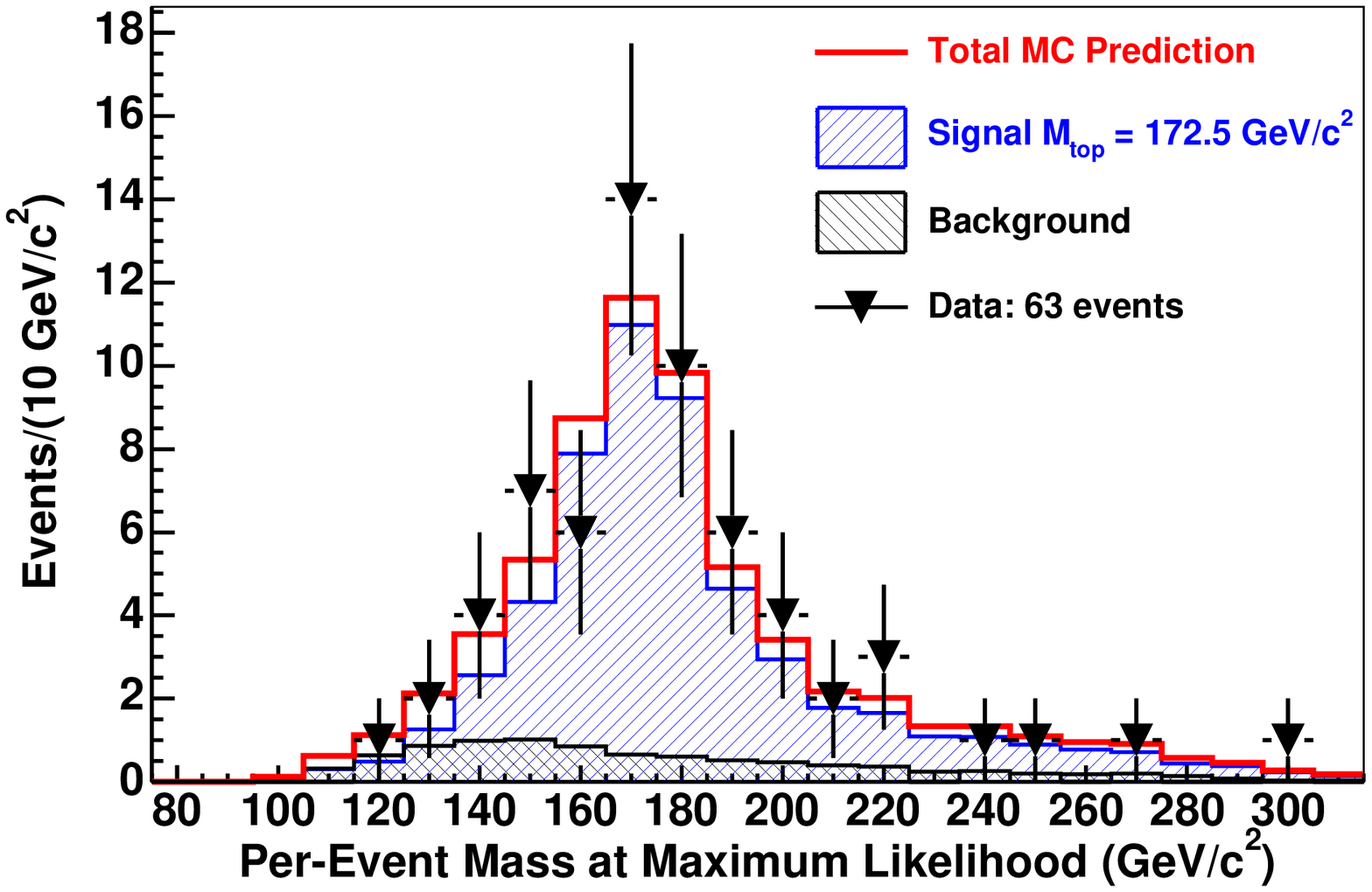}
\caption{The maximum likelihood mass for each event in the data
  compared to Monte Carlo.  The signal Monte Carlo sample is for
  $M_{top}$=172.5 GeV/$c^{2}$.}
\label{Fig:evMassdataMC}
\end{figure}

A second check uses the event-by-event maximum likelihood mass. We
show this quantity for each event in Fig.~\ref{Fig:evMassdataMC}.  The
signal Monte Carlo sample used for the comparison is generated with
$M_{top}$=172.5 GeV/$c^{2}$, close to the central value from the data.
The combined background distribution has a peak around 150--160
GeV/$c^2$, while the signal events peak at the input value of 172.5
GeV/$c^2$. The Monte Carlo prediction agrees well with the data.

\subsection{Hadronic $W$ Mass ($W\to jj$)}
\begin{figure}[htbp]
\includegraphics[width=1\columnwidth]{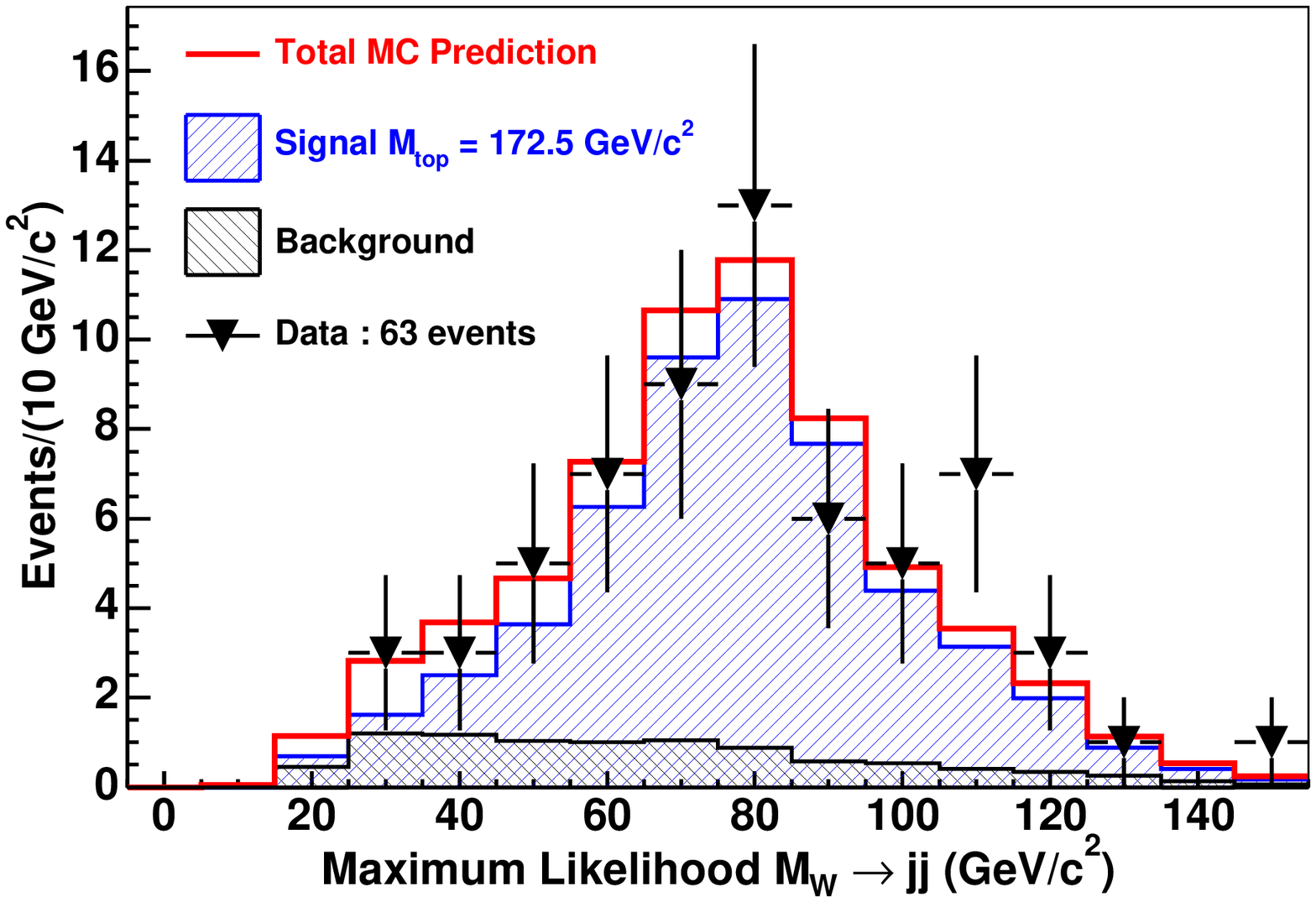}
\caption{Dijet ($W\to jj$) invariant mass distribution for the maximum
likelihood solution for signal ($M_{top}$=172.5 GeV/$c^{2}$) and
background, normalized to the expected number of events.  The
triangles show the 63 data events.}
\label{Fig:evWMassdataMC}
\end{figure}

We assume that the top quark always decays to a $b$ quark and a real
$W$ boson. Therefore in the top quark mass likelihood, we fix the $W$
mass at 80.4 GeV/$c^{2}$.  To check this, we remove the constraint in
the likelihood on the mass of the $W$ that decays into two jets and
instead constrain the top quark mass to 172.5 GeV/$c^{2}$.  Then in
each event, the invariant mass of the two jets assigned to the $W$ at
the maximum likelihood point is plotted.
Figure~\ref{Fig:evWMassdataMC} shows the comparison between the data
and Monte Carlo. We conclude that the dijet mass is consistent with
that expected from Monte Carlo $t\bar t$ events.

\subsection{Validation of Transfer Functions}
\begin{figure}[htbp]
\includegraphics[width=1\columnwidth]{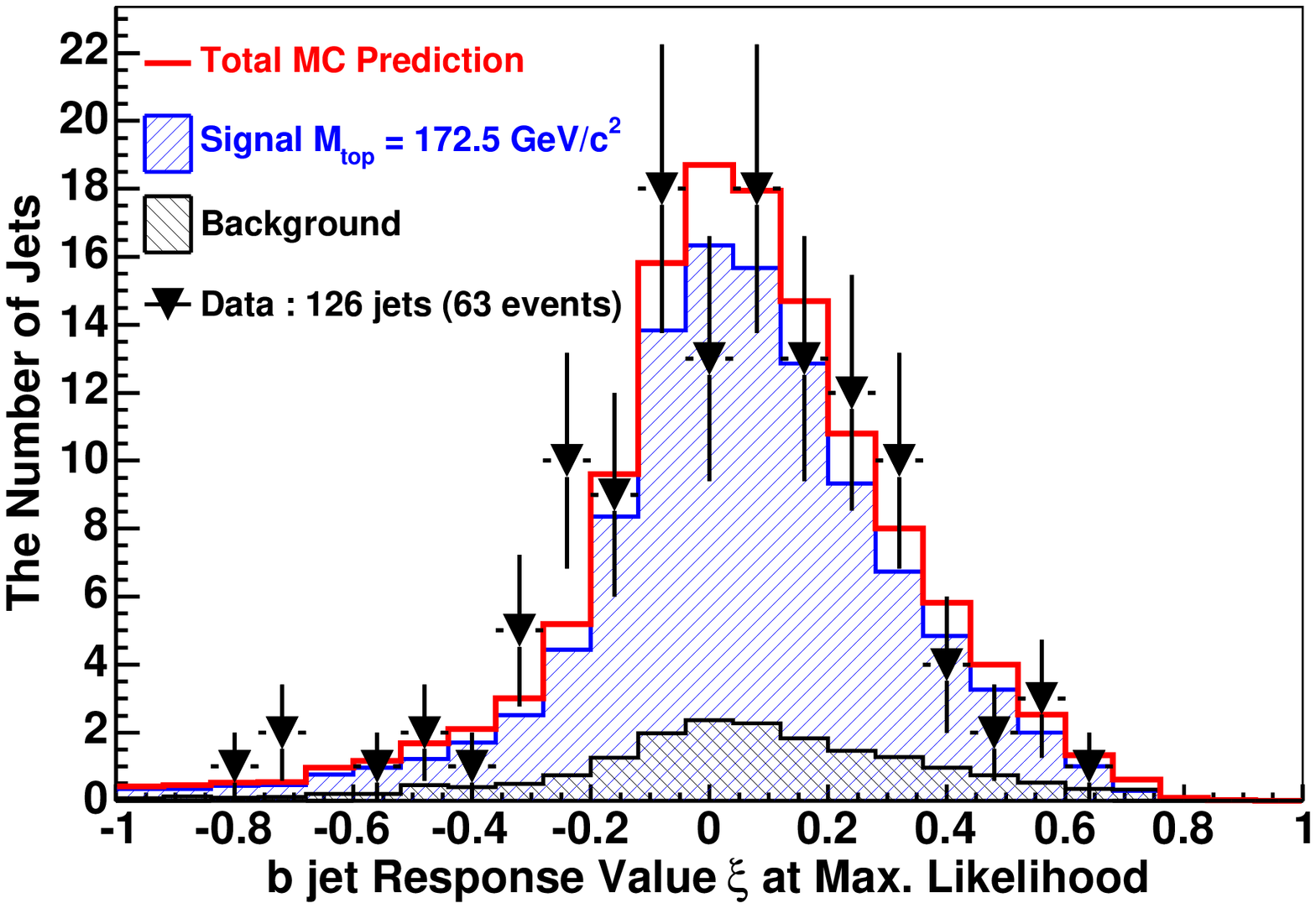}
\caption{Comparison of the $b$-jet response variable $\xi$ between the
data (triangles) and the simulation (histograms show signal,
background, and total($=$ signal + background)). The means and
resolutions are summarized in Table~\ref{tab:TFsummary_xi}.}
\label{Fig:TFBjetdataMC}
\end{figure}

\begin{figure}[htbp]
\includegraphics[width=1\columnwidth]{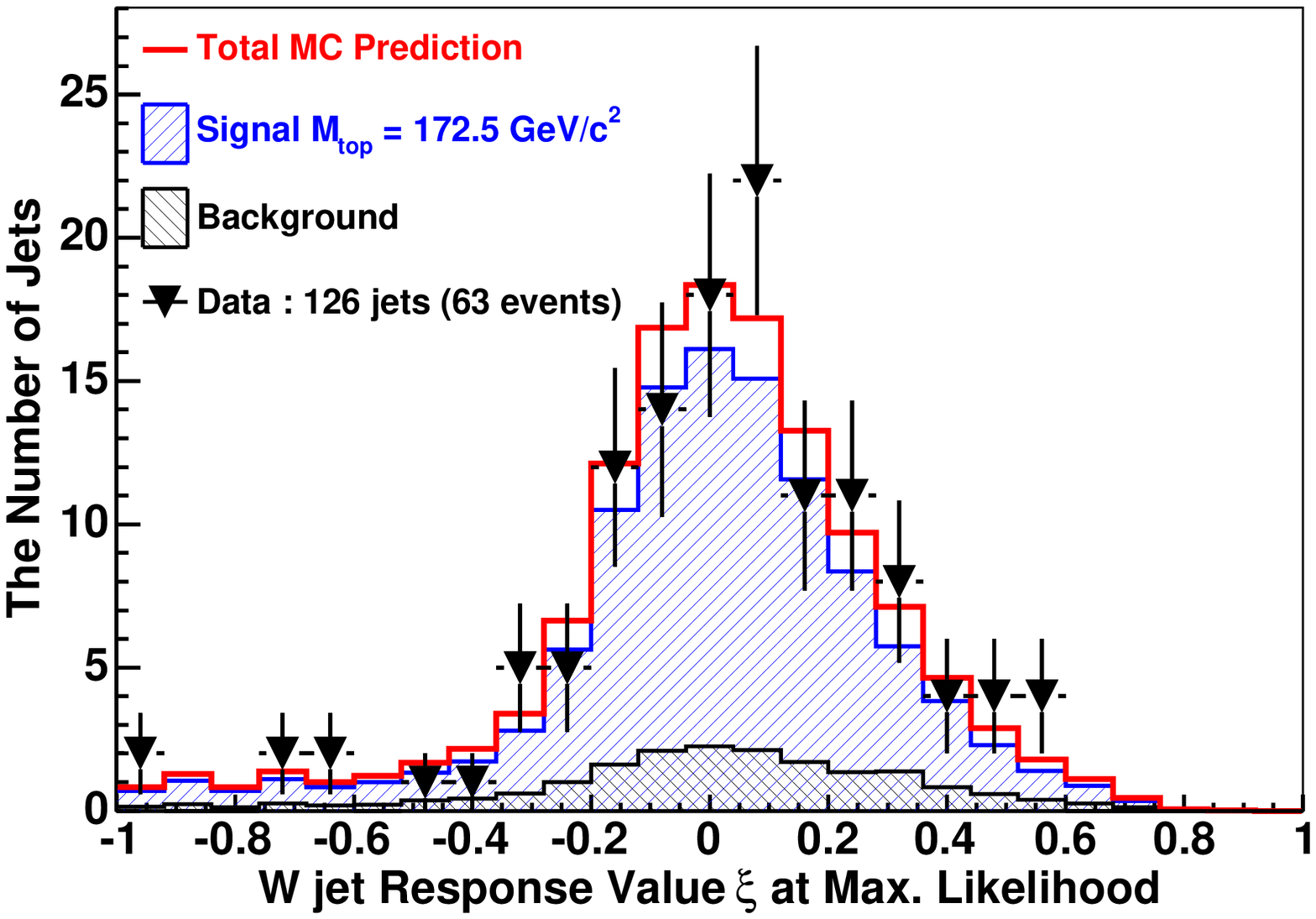}
\caption{Comparison of the $W$-jet response variable $\xi$ between the
data (triangles) and the simulation (histograms show signal,
background, and total($=$ signal + background)). The means and resolutions are summarized in
Table~\ref{tab:TFsummary_xi}.}
\label{Fig:TFWjetdataMC}
\end{figure}

The transfer function is checked by comparing the data and the
simulation directly. This is important because we rely on the Monte
Carlo simulation for the relation between partons and jets.  The
energy scale of the jets is understood to $\sim 3\%$, with possible
biases taken into account through the systematic uncertainty on the
top quark mass. However the resolution and even the scale itself for
this specific physics process should be checked. To do this, the
response variable $\xi$ is selected at the maximum likelihood point
for each event. Since each time the likelihood is calculated, we
assign which jet corresponds to which parton, we can extract the
response variables for ``jets assigned as $b$ quarks'' and ``jets
assigned as $W$ daughter jets''.  These distributions will of course
include mis-assignments and gluon contamination, but by comparing the
Monte Carlo and the data directly, it is possible to check whether the
transfer functions are well modeled.  Monte Carlo studies have shown
that the mean value of the $\xi$ distribution is slightly different
for signal and background, and the resolution of the background is
much wider than for the signal sample.  The direct comparisons between
data and MC are shown in Figs.~\ref{Fig:TFBjetdataMC} and
\ref{Fig:TFWjetdataMC} for $b$ jets and $W$ jets respectively.  Since
in each event there are two $b$ jets and two $W$ jets, the number of
data entries in these plots is twice the number of events (63).  As a
summary, the mean and RMS are listed in
Table~\ref{tab:TFsummary_xi}. The good agreement indicates that the
jet energy scale is well calibrated and no additional systematic
uncertainty is needed beyond those from generic jet energy
corrections. This test has the potential to further constrain the jet
energy scale. In the future, as the integrated luminosity increases,
we can use this together with the hadronic $W\to jj$ mass to reduce
the jet energy scale uncertainty. Indeed, CDF has recently used the
dijet mass (hadronic $W$ mass) to reduce the jet energy scale
systematic uncertainty in the template top quark mass
analysis~\cite{Ref:TMassTempII}.

\begin{table}
\begin{ruledtabular}
\caption{\label{tab:TFsummary_xi} Summary of the mean and RMS of the response variables $\xi$
for the data and the Monte Carlo in Fig.~\ref{Fig:TFBjetdataMC} and
\ref{Fig:TFWjetdataMC}.}
\begin{tabular}{lcccc}
		& \multicolumn{2}{c}{$b$ jet} & \multicolumn{2}{c}{$W$ jet}\\
		& Mean & RMS & Mean & RMS \\
\hline
MC    &	0.044$\pm$0.002 & 0.264$\pm$0.002 & 0.012$\pm$0.002 & 0.280$\pm$0.002 \\
Data  &	0.039$\pm$0.022 & 0.263$\pm$0.018 & 0.022$\pm$0.026 & 0.281$\pm$0.020 \\
\end{tabular}
\end{ruledtabular}
\end{table}

%===========================================================================
\section{\label{sec:chap11}The Systematic Uncertainty}

We have performed a number of studies of systematic uncertainties.
For each source of uncertainty, we change the input sample and
estimate the impact on the reconstructed top quark mass based on a
number of pseudo-experiments using Monte Carlo simulations where the
input top quark mass is the Run I Tevatron average, 178
GeV/$c^2$~\cite{Ref:PDG}. The reconstructed mass from each input
sample for the various systematic sources is calculated by the same
procedure as applied to the data sample; i.e., likelihood
computations, followed by the mapping function for a background
fraction of 14.5$\%$. These masses are compared to the nominal mass
from HERWIG or PYTHIA, depending on the source. The shift in the mean
from a Gaussian fit over a large number of pseudo-experiments is taken
as the systematic uncertainty.
 
\subsection{Jet Energy Scale}
With regard to the jet energy corrections, we consider three
systematic sources: first the generic corrections calibrated by the
QCD dijet process, second the transfer functions for $b$ and $W$
daughter jets from top decay, and third the $b$-jet energy scale.

First we evaluate the impact on the top quark mass from systematic
uncertainties in the generic jet energy corrections.
The details of the generic jet energy corrections are described in 
Section~\ref{sec:JetIDCorr}. The relative, absolute energy scale (hadron jet modeling), 
and out-of-cone corrections have uncertainties of roughly $1\%$, $2\%$, and $2.5\%$, respectively. 
We apply a $\pm 1 \sigma$ shift to both signal and background events 
and make event selection cuts on the shifted samples. The reconstructed masses 
are then calculated by the DLM procedure. 
We take half the difference between the means of the
$\pm 1 \sigma$ distributions. 
Table~\ref{tab:SysJESsum} lists the uncertainties from individual
corrections. The total uncertainty is taken to be the quadrature sum of these 
uncertainties and is found to be $\pm$3.0 GeV/$c^2$.
\begin{table}
\begin{ruledtabular}
\caption{\label{tab:SysJESsum} The systematic uncertainties on the top quark 
mass for each jet energy systematic source.}
\begin{tabular}{lc}
Jet Energy Systematic &  $\Delta$ $M_{top}$ GeV/$c^{2}$\\
\hline
Response relative to central scale  		       & 0.6 \\
Modeling of hadron jets (absolute scale)               & 2.0 \\
Modeling of parton showers (out-of-cone)               & 2.2 \\
\hline
Total systematic due to jet energy scale               & 3.0  \\
\end{tabular}
\end{ruledtabular}
\end{table}

Second is the systematic uncertainty from modeling of the transfer
functions.  In Section~\ref{sec:chap10}, the TF is checked by
comparing the Monte Carlo simulation with the data and found to be
consistent. Therefore we only account for the difference of TF's
between PYTHIA and HERWIG. We make two sets of TF's, one each from
PYTHIA and HERWIG. They are applied to the same Monte Carlo sample,
HERWIG with $M_{top}$ = 178 GeV/$c^2$. The difference between the two
is found to be 0.2 GeV/c$^2$.

The last systematic related to the jet energy scale arises from the
$b$-jet specific energy scale. The light quark jet scale is set by the
generic corrections which are deduced using samples that are mainly
light quark and gluon jets. In addition, the sensitivity of the top
mass to the light quark jet energy scale is reduced by the $W$ mass
constraint in the likelihood.  On the other hand, the top quark mass
is very sensitive to the $b$-jet energy scale, so its additional
uncertainty has to be estimated. We consider three possible sources:
(1) $b$-quark decay properties, (2) fragmentation properties, and (3)
different color flow.

The $B$ meson semi-leptonic branching ratios are varied in the
simulation by 3$\%$ (30 $\pm$ 3$\%$), corresponding approximately to
the uncertainty in the current world average~\cite{Ref:PDG}, to
estimate its impact on the $b$-jet energy scale. We find that the
total uncertainty on $b$-jet response is 0.4$\%$, which translates to
a top quark mass difference of 0.4 GeV/$c^2$.  Using the
LEP~\cite{Ref:ALEPH,Ref:OPAL} and SLC~\cite{Ref:SLD} results from
large $Z\to b\bar b$ datasets, we constrain the possible fragmentation
models in Monte Carlo calculations by changing the Peterson
parameter~\cite{Ref:Peterson} to match the experimental results within
their uncertainties.  The variations introduce an additional
uncertainty of $\pm$0.4 GeV/$c^2$.  For color flow modeling, we vary
the parameters of the algorithms used to generate color flow in both
PYTHIA and HERWIG. The amount of ambiguous energy, i.e.,  energy that
cannot be assigned to the $b$ jet or the initial state parton due to
the color connection, is estimated to be 3$\%$ of the $b$-jet energy
scale.  By considering large variations of the parameter related to
color flow modeling, the amount of ambiguous energy changes by 0.3$\%$
of the total $b$-jet energy, corresponding to $\pm$0.3 GeV/$c^2$ in
the top quark mass.
 
These three contributions are added in quadrature, and the resulting
$\pm$0.6 GeV/$c^2$ is assigned as an additional systematic uncertainty
due to the modeling of the $b$-quark energy scale.

\subsection{Initial and final state hard radiation}
\begin{figure}[htbp]
\includegraphics[width=1\columnwidth]{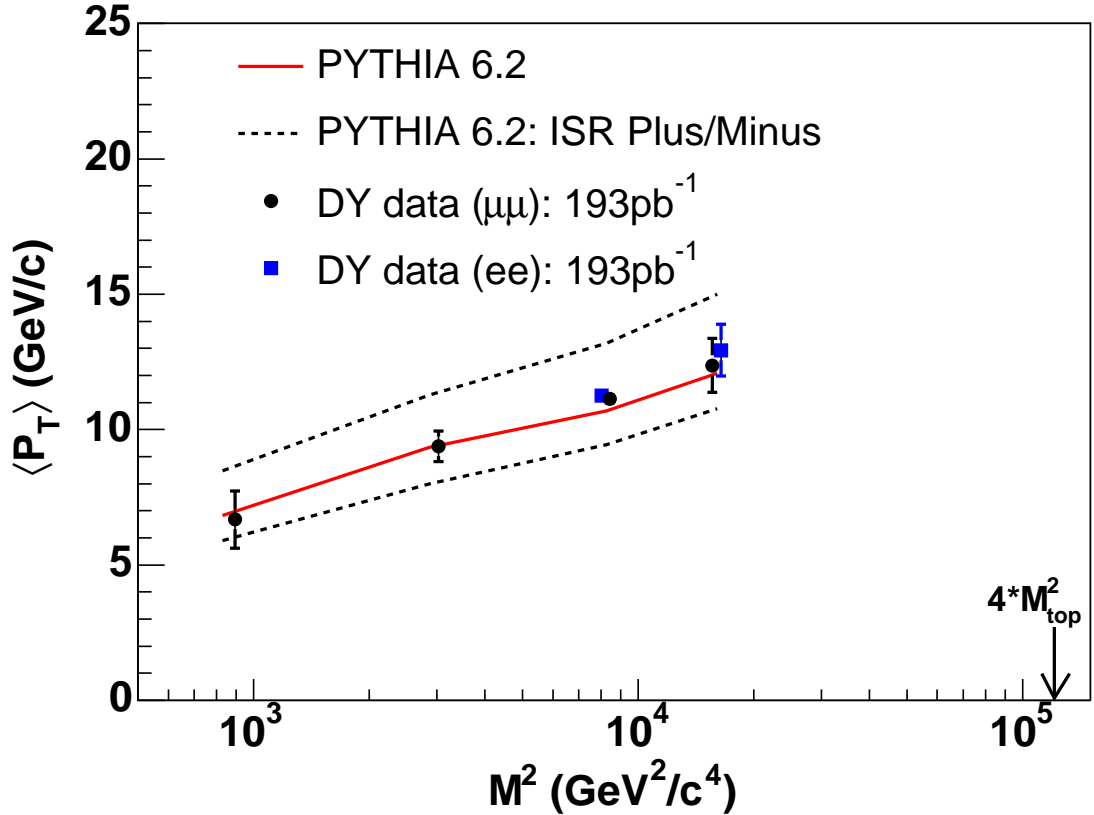}
\caption{The average $p_T$ of dilepton events shows a logarithmic dependence
on the dilepton invariant mass squared. The data are compared with
PYTHIA samples created with nominal settings as well as those with
increased and decreased ISR activity.}
\label{Fig:sysISR}
\end{figure}

Initial and final state gluon radiation (ISR and FSR) affect the top
quark mass measurement.  ISR produces extra jets that can be
misidentified as a $t\bar{t}$ daughter, while FSR can cause a final
state quark jet energy to be measured low. To evaluate the level of
ISR, Drell-Yan dilepton events ($ee$ and $\mu\mu$) are used since
there is no FSR and they are produced via $q\bar q$ annihilation, the
dominant production mechanism for $t\bar t$ at the Tevatron (85$\%$ at
NLO). The average dilepton $p_T$, $\langle p_T \rangle$, which
reflects the size of ISR activity, is shown in Fig.~\ref{Fig:sysISR}
as a function of the dilepton mass squared.  A logarithmic dependence
is seen between the two. By extrapolating to the energy scale of
$t\bar t$ production, we find the allowed range for $\langle p_T
\rangle$.  Two PYTHIA Monte Carlo samples are made with parameters
adjusted to cover the range: one with $\Lambda_{QCD}$ = 73 MeV, $K$ =
2.0 and the other with $\Lambda_{QCD}$ = 292 MeV, $K$ = 0.5 for $-1$
$\sigma_{ISR}$ and $+1$ $\sigma_{ISR}$, respectively, where $K$ is a
scale factor applied to the transverse momentum scale. Corresponding
curves are also shown in Fig.
\ref{Fig:sysISR}. This yields an uncertainty of $\pm$0.4 GeV/$c^2$.
Since both ISR and FSR are controlled by the same DGLAP evolution
equation, the same variations of $\Lambda_{QCD}$ and $K$ are used to
generate FSR systematic samples by varying only FSR modeling.  This
results in a $\pm$0.5 GeV/$c^2$ variation in the top quark mass.

\subsection{Parton Distribution Functions}
 For the parton distribution functions, we add in quadrature
 uncertainties derived from two sources: differences from 20 pairs of
 CTEQ6M~\cite{Ref:CTEQ6M} uncertainty eigenvectors ($\pm1\sigma$), and
 MRST~\cite{Ref:MRST} with two different $\Lambda_{QCD}$ values (300
 and 228 MeV).  The result is an uncertainty of $\pm$0.5 GeV/$c^2$, of
 which 0.45 GeV/$c^2$ comes from the 20 eigenvectors.

 Both the PDF and ISR systematics, which impact the $p_T$ of the
 $t\bar{t}$ system, reflect the sensitivity of the DLM method to the
 production mechanism.  To make an extreme test of this, we created a
 signal Monte Carlo sample of $t\bar t$ resonance production in which
 175 GeV/$c^2$ top quarks are produced from the decay of a 700
 GeV/$c^2$ resonance. In this sample, the top quark decay properties
 are the same as those in the SM. The shifted mass is
 found to be 2.0 GeV/$c^2$, demonstrating that the method is
 relatively insensitive to major variations in the production
 mechanism even though we use the SM $t\bar t$ production matrix
 element in the event likelihood.  This is because the most sensitive
 factor in the likelihood is the top quark propagator rather than the
 production and decay matrix elements.

\subsection{Other Systematic Uncertainties}
Possible bias in the Monte Carlo generator is estimated by comparing
PYTHIA and HERWIG. HERWIG deals with spin correlations in the
production and decay of $t\bar{t}$, while PYTHIA does not.  Another
difference between the two is the fragmentation model, where PYTHIA
uses the string model while HERWIG adopts the cluster model. We
estimate the associated uncertainty to be $\pm$0.3 GeV/$c^2$ by taking
the difference of the reconstructed masses between PYTHIA and HERWIG
using the same mapping function extracted from HERWIG.  Another
systematic uncertainty comes from the mapping function, for which we
use a background fraction of 14.5$\%$. The uncertainty on this
fraction is $\pm$2.9$\%$ from the uncertainty in the mean expected
background of $\pm$1.8 events as shown in
Table~\ref{tab:expectedSummary}. From a series of pseudo-experiments
by changing background fraction by $\pm$2.9$\%$, we estimate this
uncertainty to be $\pm$0.2 GeV/$c^2$.  The statistical uncertainty on
the expected number of background events (9.2) is already taken into
account by the correction obtained from the width of pull distribution
discussed in Section~\ref{sec:methodcheck} because the expected number
of background events has been Poisson fluctuated in the
pseudo-experiments.  

The uncertainty due to background modeling, $\pm$0.4 GeV/$c^2$, comes
from two sources: We evaluate the difference between the reconstructed
masses obtained by using only one of the individual background
process, rather than using combined background.  Then the maximum
difference among the major background sources ($W+$ heavy flavor
quarks, $W+$ mistagged jets, non-$W$ background) is used.  The other
source is the variation with different choices of the $Q^2$ scale
($4M_W^2,M_W^2,M_W^2/4$, and $M_W^2+P_{TW}^2$) which is the
characteristic energy scale of the hard scattering process using the
ALPGEN Monte Carlo program. This takes into account possible
variations in the background composition.  Finally, as described in
Section~\ref{sec:chap4}, the $b$-tagging efficiency is different in
data and Monte Carlo. Only the jet $E_T$ dependence of the tagging
efficiency is important in the mass analysis. By varying the slope of
the efficiency as a function of $E_T$ by $\pm$1$\sigma$, we find the
top quark mass shifts by $\pm$0.2 GeV/$c^2$. The uncertainty due to
the finite statistics of the non-$W$ data sample and the Monte Carlo
samples used to make the mapping functions are negligible.

\subsection{Summary of systematic uncertainties}
The systematic uncertainties are summarized in
Table~\ref{tab:Syssummary}. The largest one comes from the uncertainty
in the jet energy measurement.  The sum in quadrature of all the
systematic uncertainties is 3.2 GeV/$c^{2}$.
\begin{table}
\begin{ruledtabular}
\caption{\label{tab:Syssummary}The summary of systematic uncertainties.}
\begin{tabular}{lc}
Source &  $\Delta$ $M_{top}$ GeV/$c^{2}$\\
\hline
Jet Energy Corrections          & 3.0 \\
Transfer Function               & 0.2 \\
ISR                             & 0.4 \\
FSR                             & 0.5 \\
PDFs                            & 0.5 \\
Generator                       & 0.3 \\
Background Fraction             & 0.2 \\
Background Modeling             & 0.4 \\
$b$ Jet Energy Modeling         & 0.6 \\
$b$ Tagging                     & 0.2 \\
\hline
Total                           & 3.2  \\
\end{tabular}
\end{ruledtabular}
\end{table}

%===========================================================================
\section{\label{sec:chap12}Conclusion}  
Using the dynamical likelihood method, we measure the top quark mass
to be\\
\vspace{-0.5cm}
\begin{center}
$M_{top}$ = 173.2 $^{+2.6}_{-2.4}$ (stat.) $\pm$ 3.2 (syst.) GeV/$c^2$\\
\vspace{0.2cm}
= 173.2 $^{+4.1}_{-4.0}$ GeV/$c^2$
\end{center}
from 63 events, corresponding to an integrated luminosity of
318~pb$^{-1}$ accumulated in the CDF Run II experiment.  By using the
maximal information from the $t\bar t$ production mechanism and
assuming the validity of the SM, a reduction of the
statistical uncertainty is obtained.  The precision of this single
measurement in fact is slightly better than the Run I world average,
and the result is consistent with other recent measurement by
CDF~\cite{Ref:TMassTempII}, which provided the best single measurement
(173.5 $^{+3.9}_{-3.8}$ GeV/$c^2$) using the template technique with a
dijet $W$ mass constraint to reduce the jet energy scale uncertainty.
The current DLM analysis technique uses the jet energy scale
determined with generic jet samples.  However as the luminosity
increases, a reduction of the dominant systematic uncertainty, due to
the jet energy scale, is crucial.  DLM will be able to further
constrain the jet energy scale using the hadronic $W \to jj$ mass in
$t\bar{t}$ events as done in~\cite{Ref:TMassTempII}. We expect that
other systematics also can be improved as the size of control samples
increase. A reduced top quark mass uncertainty with increased data
sample size will contribute to the detailed understanding of the
electroweak interaction as well as to the search for physics beyond
the standard model.

\begin{acknowledgments}
We thank the Fermilab staff and the technical staffs of the
participating institutions for their vital contributions.  This work
was supported by the U.S. Department of Energy and National Science
Foundation; the Italian Istituto Nazionale di Fisica Nucleare; the
Ministry of Education, Culture, Sports, Science and Technology of
Japan; the Natural Sciences and Engineering Research Council of
Canada; the National Science Council of the Republic of China; the
Swiss National Science Foundation; the A.P. Sloan Foundation; the
Bundesministerium f\"ur Bildung und Forschung, Germany; the Korean
Science and Engineering Foundation and the Korean Research Foundation;
the Particle Physics and Astronomy Research Council and the Royal
Society, UK; the Russian Foundation for Basic Research; the Comisi\'on
Interministerial de Ciencia y Tecnolog\'{\i}a, Spain; in part by the
European Community's Human Potential Programme under contract
HPRN-CT-2002-00292; and the Academy of Finland.
\end{acknowledgments}


\begin{thebibliography}{unsrt}
\bibitem{Ref:RunIdilCDF} F. Abe {\it et al.} (CDF Collaboration), Phys. Rev. Lett. {\bf 82}, 2808 (1999).
\bibitem{Ref:RunIdilD0}  B. Abbott {\it et al.} (D$\O$ Collaboration), Phys. Rev. {\bf D60}, 052001 (1999).
\bibitem{Ref:RunIljCDF}  T. Affolder {\it et al.} (CDF Collaboration), Phys. Rev. {\bf D63}, 032003 (2001).
\bibitem{Ref:RunIljD0}   B. Abbott {\it et al.} (D$\O$ Collaboration), Phys. Rev. {\bf D58}, 052001 (1998).
\bibitem{Ref:RunIallj}   F. Abe {\it et al.} (CDF Collaboration), Phys. Rev. Lett. {\bf 79}, 1992 (1997).
\bibitem{Ref:RunIalljD0} V.M. Abazov {\it et al.} (D$\O$ Collaboration),  Phys. Lett. {\bf B606}, 25 (2005).
\bibitem{Ref:D0runIana}  V.M. Abazov {\it et al.} (D$\O$ Collaboration), Nature {\bf 429}, 638 (2004).
\bibitem{Ref:RunIcomb} P. Azzi {\it et al.} (2004), hep-ex/0404010, http://tevewwg.fnal.gov/.
\bibitem{Ref:HiggsLEP} The LEP ElectroWeak Working Group, hep-ex/0509008.
\bibitem{Ref:dlmKK1} K. Kondo, J.Phys. Soc. Jpn. {\bf 57}, 4126 (1988).
\bibitem{Ref:dlmKK2} K. Kondo, J.Phys. Soc. Jpn. {\bf 60}, 836 (1991).
\bibitem{Ref:dlmKK3} K. Kondo, T. Chikamatsu, and S.H. Kim, J.Phys. Soc. Jpn. {\bf 62}, 1177 (1993).
\bibitem{Ref:dlmKK4} K. Kondo, RISE Technical Report 05-01(2005), Waseda University, hep-ex/0508035.
\bibitem{Ref:RunIWhelD0}  V.M. Abazov {\it et al.} (D$\O$ Collaboration), Phys. Lett. {\bf B617}, 1 (2005).
\bibitem{Ref:DGM1} R.H. Dalitz and G.R. Goldstein, Phys. Rev. {\bf D45}, 1531 (1992).
\bibitem{Ref:TMassTempII} A. Abulencia {\it et al.} (CDF Collaboration),  Phys. Rev. {\bf D73}, 032003 (2006).
\bibitem{Ref:TMassPRL}  A. Abulencia {\it et al.} (CDF Collaboration), Phys. Rev. Lett. {\bf 96}, 022004 (2006).
\bibitem{Ref:topdiscov}  F. Abe  {\it et al.} (CDF Collaboration), Phys. Rev. Lett. {\bf 74}, 2626 (1995); 
                         S. Abachi {\it et al.} (D$\O$ Collaboration), Phys. Rev. Lett. {\bf 74}, 2632 (1995). 
\bibitem{Ref:RunIIXsecKin} D. Acosta {\it et al.} (CDF Collaboration),  Phys. Rev. {\bf D72}, 052003 (2005);
                           V. Abazov {\it et al.} (D$\O$ Collaboration),  Phys. Lett. {\bf B626}, 35 (2005).
\bibitem{Ref:RunIIWhel} A. Abulencia {\it et al.} (CDF Collaboration), submitted to  Phys. Rev. Lett (2005), hep-ex/0511023; 
                        V. Abazov {\it et al.} (D$\O$ Collaboration),  Phys. Rev. {\bf D72}, 011104 (2005).
\bibitem{Ref:RunIIBR}  D. Acosta {\it et al.} (CDF Collaboration),  Phys. Rev. Lett. {\bf 95}, 102002 (2005).
\bibitem{Ref:CDF} D. Acosta {\it et al.} (CDF Collaboration), Phys. Rev. {\bf D71}, 032001 (2005); 
                  R. Blair {\it et al.} (CDF Collaboration), Fermilab Report No. FERMILAB-PUB-96/390-E, (1996).
\bibitem{Ref:COT} T. Affolder {\it et al.}, Nucl. Instrum. Meth. {\bf A526}, 249 (2004).
\bibitem{Ref:L00} C.S. Hill {\it et al.}, Nucl. Instrum. Meth. {\bf A530}, 1 (2004).
\bibitem{Ref:SVX} A. Sill {\it et al.}, Nucl. Instrum. Meth. {\bf A447}, 1 (2000).
\bibitem{Ref:ISL} T. Affolder {\it et al.},  Nucl. Instrum. Meth. {\bf A453}, 84 (2000).
\bibitem{Ref:elemagCal} L. Balka {\it et al.}, Nucl. Instrum. Meth. {\bf A267}, 272 (1988). 
\bibitem{Ref:hadronCal} S. Bertolucci {\it et al.}, Nucl. Instrum. Meth. {\bf A267}, 301 (1988). 
\bibitem{Ref:Cal} M. G. Albrow {\it et al.}, Nucl. Instrum. Meth. {\bf A480}, 524 (2002). 
\bibitem{Ref:CMU} G. Ascoli {\it et al.}, Nucl. Instrum. Meth. {\bf A268}, 33 (1988).
\bibitem{Ref:CMP} T. Dorigo {\it et al.}, Nucl. Instrum. Meth. {\bf A461}, 560 (2001).
\bibitem{Ref:Herwig} G. Marchesini {\it et al.}, Comput. Phys. Commun. {\bf 67}, 465 (1992); 
                     G. Corcella {\it et al.}, JHEP {\bf 0101}, 010 (2001). 
\bibitem{Ref:Pythia} T. Sjostrand {\it et al.}, Comput. Phys. Commun. {\bf 135}, 238 (2001).
\bibitem{Ref:CTEQ5L} J. Pumplin {\it et al.}, JHEP {\bf 0207}, 012 (2002).
\bibitem{Ref:QQv91} P. Avery, K. Read, and G. Trahern, Cornell Internal Report No. CSN-212, 1985 (unpublished).
\bibitem{Ref:Alpgen} M. L. Mangano {\it et al.}, JHEP {\bf 0307}, 001 (2003).
\bibitem{Ref:CDFsim}  E. Gerchtein and M. Paulini, ECONF {\bf C0303241}, TUMT005 (2003), physics/0306031.
\bibitem{Ref:Geant3} R. Brun and F. Carminati, CERN Programming Library Long Writeup {\bf W5013}, (1993).
\bibitem{Ref:Gflash} G. Grindhammer, M. Rudowicz, and S. Peters, Nucl. Instrum. Meth. {\bf A290}, 469 (1990).
\bibitem{Ref:WZcross}  A. Abulencia {\it et al.} (CDF Collaboration), submitted to Phys. Rev. {\bf D}, (2005), hep-ex/0508029. 
\bibitem{Ref:JetCone}  F. Abe {\it et al.} (CDF Collaboration), Phys. Rev. {\bf D45}, 1448 (1992).
\bibitem{Ref:JetRes}  F. Abe {\it et al.} (CDF Collaboration), Phys. Rev. Lett. {\bf 68}, 1104 (1992).
\bibitem{Ref:JESnim} A. Bhatti {\it et al.}, to be submitted to Nucl. Instrum. Methods {\bf A}, hep-ex/0510047.
\bibitem{Ref:btagRunI}  T. Affolder {\it et al.} (CDF Collaboration), Phys. Rev. {\bf D64}, 032002 (2001).
\bibitem{Ref:RunIIXsecCount} D. Acosta {\it et al.} (CDF Collaboration),  Phys. Rev. {\bf D71}, 052003 (2005).
\bibitem{Ref:BkgXsec1} J. M. Campbell and R. K. Ellis, Phys. Rev. {\bf D60}, 113006 (1999).
\bibitem{Ref:BkgXsec2} B. W. Harris, E. Laenen, L. Phaf, Z. Sullivan and S. Weinzierl, Phys. Rev. {\bf D66}, 054024 (2002).
\bibitem{Ref:ttXsecRef} M. Cacciari {\it et al.}, JHEP {\bf 0404}, 68 (2004).
\bibitem{Ref:Hagi} K. Hagiwara {\it et al.}, Phys. Rev. {\bf D66}, 010001 (2002).
\bibitem{Ref:ME1} R. K. Ellis, W. J. Stirling, and B. R. Webber, {\it ``QCD and Collider Physics''},
                  Cambridge Monographs on Particle Physics, Nuclear Physics and Cosmology (1996), p.348;
                  R. K. Ellis, {\it ``Strong Interactions and Gauge Theories''},
                  edited by J. Tran Thanh Van, Editions Fronti\'eres, Gif-sur-Yvette (1986), p.339.
\bibitem{Ref:ME2} M. Gluck, J. F. Owens, and E. Reya, Phys. Rev. {\bf D17}, 2324 (1978);
                  B. Combridge, Nucl. Phys. {\bf 151}, 429 (1979).
\bibitem{Ref:ME3} G. Mahlon and S. Parke,  Phys. Rev. {\bf D53}, 4886 (1996);
                  G. Mahlon and S. Parke,  Phys. Lett. {\bf B411}, 173 (1997).	
\bibitem{Ref:PDG} S. Eidelman {\it et al.}, Phys. Lett. {\bf B592}, 1 (2004).
\bibitem{Ref:ALEPH} A. Heister {\it et al.} (ALEPH Collaboration),  Phys. Lett. {\bf B512}, 30 (2001).
\bibitem{Ref:OPAL} G. Abbiendi {\it et al.} (OPAL Collaboration),  Eur. Phys. J. {\bf C29}, 463 (2003).
\bibitem{Ref:SLD}  K. Abe {\it et al.} (SLD Collaboration), Phys. Rev. {\bf D65}, 092006 (2002).
\bibitem{Ref:Peterson}  C. Peterson, D. Schlatter, I. Schmitt and  P. M. Zerwas, Phys. Rev. {\bf D27}, 105 (1983).
\bibitem{Ref:CTEQ6M} D. Stump {\it et al.}, JHEP {\bf 0310}, 046 (2003).
\bibitem{Ref:MRST} A. D. Martin {\it et al.}, Eur. Phys. J. {\bf C14}, 133 (2000).
\end{thebibliography}
\end{document}